\documentclass[twocolumn,nofootinbib,superscriptaddress,preprintnumbers,floatfix]{revtex4}

\usepackage{graphicx}
\usepackage{amsmath}
\usepackage{amssymb}
\usepackage{color}
\usepackage{xspace}
\usepackage[small]{subfigure}
\usepackage[hyperfootnotes=false]{hyperref}

\newcommand{\eq}[1]{Eq.~\eqref{eq:#1}}
\newcommand{\eqs}[2]{Eqs.~\eqref{eq:#1} and \eqref{eq:#2}}
\renewcommand{\sec}[1]{Sec.~\ref{sec:#1}}
\newcommand{\subsec}[1]{Sec.~\ref{subsec:#1}}
\newcommand{\app}[1]{App.~\ref{app:#1}}

\newcommand{\fig}[1]{Fig.~\ref{fig:#1}}

\newcommand{\head}[1]{
\vspace{0.3cm}
\noindent {\bf \underline {#1}}
\vspace{0.1cm}
}

\newcommand{\OMIT}[1]{}

\newcommand{\e}{\epsilon}

\newcommand{\V}{V}

\newcommand{\df}{\mathrm{d}}

\newcommand{\cL}{\mathcal{L}}

\newcommand{\beq}{\begin{equation}}
\newcommand{\eeq}{\end{equation}}
\newcommand{\beqa}{\begin{eqnarray}}
\newcommand{\eeqa}{\end{eqnarray}}

\newcommand{\lqcd}{\Lambda_\mathrm{QCD}}
\newcommand{\GeV}{\,\mathrm{GeV}}
\newcommand{\nn}{\nonumber}

\newcommand{\mi}{{\mu}}
\newcommand{\as}{\alpha_s}

\newcommand{\msbar}{\overline{\textrm{MS}}}

\newcommand{\alphaew}{\alpha_{\rm em}}
\newcommand{\ntllp}{\mbox{N${}^3$LL${}^\prime$}\xspace}
\newcommand{\ntll}{\mbox{N${}^3$LL}\xspace}
\newcommand{\muj}{\mu_J}
\newcommand{\muh}{\mu_H}
\newcommand{\mus}{\mu_S}
\newcommand{\pci}{1\xspace}
\newcommand{\pcii}{2\xspace}
\newcommand{\pciii}{3\xspace}

  \newcount\hour \newcount\minute
  \hour=\time \divide \hour by 60
  \minute=\time
  \newcommand{\mydate}{\ \today \ - \number\hour :\ifnum \minute<10 0\fi 
\number\minute}

\allowdisplaybreaks[4]

\begin{document}

\preprint{\vbox{
%\hbox{arXiv:1006.xxxx}
\hbox{MIT--CTP 4101}\hbox{MPP--2010--7}
}}

\title{\boldmath 
Thrust at N${}^3$LL with Power Corrections and a Precision Global Fit 
for $\alpha_s(m_Z)$
\vspace{0.1cm}
}

\author{Riccardo Abbate}
\affiliation{Center for Theoretical Physics, Massachusetts Institute of
Technology, Cambridge, MA 02139\vspace{0.cm}}

\author{Michael Fickinger}
\affiliation{Department of Physics, University of Arizona, Tucson, AZ 85721 
}

\author{Andr\'e H. Hoang}
\affiliation{Max-Planck-Institut f\"ur Physik (Werner-Heisenberg-Institut) 
F\"ohringer Ring 6, 80805 M\"unchen, Germany\vspace{0.3cm}}

\author{Vicent Mateu} 
\affiliation{Max-Planck-Institut f\"ur Physik
(Werner-Heisenberg-Institut) F\"ohringer Ring 6, 80805 M\"unchen, 
Germany\vspace{0.3cm}}

\author{Iain W.\ Stewart\vspace{0.5cm}}
\affiliation{Center for Theoretical Physics, Massachusetts Institute of
Technology, Cambridge, MA 02139\vspace{0.cm}}

\begin{abstract}
  
  We give a factorization formula for the $e^+e^-$ thrust distribution ${\rm
    d}\sigma/{\rm d}\tau$ with $\tau=1-T$ based on soft-collinear effective
  theory.  The result is applicable for all $\tau$, i.e. in the peak, tail, and
  far-tail regions.  The formula includes ${\cal O}(\alpha_s^3)$ fixed-order QCD
  results, resummation of singular partonic $\alpha_s^j\ln^k(\tau)/\tau$ terms
  with \ntll accuracy, hadronization effects from fitting a universal
  nonperturbative soft function defined in field theory, bottom quark mass
  effects, QED corrections, and the dominant top mass dependent terms from the
  axial anomaly.  We do not rely on Monte Carlo generators to determine
  nonperturbative effects since they are not compatible with higher order
  perturbative analyses.  Instead our treatment is based on fitting
  nonperturbative matrix elements in field theory, which are moments $\Omega_i$
  of a nonperturbative soft function.  We present a global analysis of all
  available thrust data measured at center-of-mass energies $Q=35$ to $207$~GeV
  in the tail region, where a two parameter fit to $\alpha_s(m_Z)$ and the first
  moment $\Omega_1$ suffices.  We use a short distance scheme to define
  $\Omega_1$, called the R-gap scheme, thus ensuring that the perturbative ${\rm
    d}\sigma/{\rm d}\tau$ does not suffer from an ${\cal O}(\Lambda_{\rm QCD})$
  renormalon ambiguity.  We find $\alpha_s(m_Z)=0.1135 \pm (0.0002)_{\rm expt}
  \pm (0.0005)_{\rm hadr} \pm (0.0009)_{\rm pert}$, with $\chi^2/{\rm
    dof}=0.91$, where the displayed $1$-sigma errors are the total experimental
  error, the hadronization uncertainty, and the perturbative theory uncertainty,
  respectively. The hadronization uncertainty in $\alpha_s$ is significantly
  decreased compared to earlier analyses by our two parameter fit, which
  determines $\Omega_1=0.323\,{\rm GeV}$ with 16\% uncertainty.

\end{abstract}

\maketitle

\section{Introduction}
\label{sec:intro}

A traditional method for testing the theory of strong interactions (QCD) at
high-precision is the analysis of jet cross sections at $e^{+}\, e^{-}$
colliders.  Event shape distributions play a special role as they have been
extensively measured with small experimental uncertainties at LEP and earlier
$e^{+}\, e^{-}$ colliders, and are theoretically clean and accessible to
high-order perturbative computations. They have been frequently used to make
precise determinations of the strong coupling $\alpha_{s}$, see e.g.\
Ref.~\cite{Kluth:2006bw} for a review. One of the most frequently studied event
shape variables is thrust~\cite{Farhi:1977sg},
\begin{align}
\label{Tdef}
T & \, = \,\mbox{max}_{\hat {\bf t}}\frac{\sum_i|\hat {\bf t}\cdot\vec{p}_i|}
{\sum_i|\vec{p}_i|}
\,,
\end{align}
where the sum $i$ is over all final-state hadrons with momenta $\vec{p}_i$.  The
unit vector ${\hat {\bf t}}$ that maximizes the right-hand side (RHS) of
Eq.~(\ref{Tdef}) defines the thrust axis.  We will use the more convenient
variable $\tau = 1-T$.  For the production of a pair of massless quarks at tree
level ${\rm d}\sigma/{\rm d}\tau \propto \delta(\tau)$, so the measured
distribution for $\tau>0$ involves gluon radiation and is sensitive to the value
of $\alpha_s$.  The thrust value of an event measures how much it resembles two
jets. For $\tau$ values close to zero the event has two narrow, pencil-like,
back-to-back jets, carrying about half the center-of-mass (c.m.)  energy into
each of the two hemispheres defined by the plane orthogonal to ${\hat {\bf t}}$.
For $\tau$ close to the multijet endpoint $1/2$, the event has an isotropic
multi-particle final state containing a large number of low-energy jets.

On the theoretical side, for $\tau< 1/3$ the dynamics is governed by three
different scales. The \emph{hard scale} $\muh\simeq Q$ is set by the $e^+e^-$
c.m.~energy $Q$.  The \emph{jet scale}, $\muj\simeq Q\sqrt{\tau}$ is the typical
momentum transverse to ${\hat {\bf t}}$ of the particles within each of the two
hemispheres, or the jet invariant mass scale if all energetic particles in a
hemisphere are grouped into a jet. There is also uniform soft radiation with
energy $\mus\simeq Q\tau$, called the \emph{soft scale}.  The physical
description of the thrust distribution can be divided into three regions,
\begin{align} \label{eq:regions}
  &  \text{peak region:} & & \tau  \sim  2\Lambda_{\rm QCD}/Q \nn\,, \\
  &  \text{tail region:} & & 2\Lambda_{\rm QCD}/Q  \ll \tau \lesssim 1/3 \,, \\
  & \text{far-tail region:} & & 1/3 \lesssim \tau \le 1/2 \,. \nn
\end{align}
In the {\em peak region} the hard, jet, and soft scales are $Q$,
$\sqrt{Q\Lambda_{\rm QCD}}$, and $\Lambda_{\rm QCD}$, and the distribution shows
a strongly peaked maximum.  Theoretically, since $\tau\ll 1$ one needs to sum
large (double) logarithms, $(\alpha_s^j\ln^k\!\tau)/\tau$, and account for the
fact that $\mus\simeq \Lambda_{\rm QCD}$, so ${\rm d}\sigma/{\rm d}\tau$ is
affected at leading order by a nonperturbative distribution. We call this
distribution the nonperturbative soft function.  The \emph{tail region} is
populated predominantly by broader dijets and 3-jet events.  Here the three
scales are still well separated and one still needs to sum logarithms, but now
$\mus\gg \Lambda_{\mathrm{QCD}}$, so soft radiation can be described by
perturbation theory and a series of power correction parameters $\Omega_i$.
Finally, the \emph{far-tail} region is populated by multijet events.  Here the
distinction of the three scales becomes meaningless, and accurate predictions
can be made with fixed-order perturbation theory supplemented with power
corrections.  The transition to this region must be handled carefully since
including a summation of $(\alpha_s^j\ln^k\!\tau)/\tau$ terms in this region
spoils the cancellations that take place at fixed order multijet thresholds, and
hence would induce uncertainties that are significantly larger than those of the
fixed-order results.

Recently two very important achievements were made improving the theoretical
description of event shape distributions in $e^+e^-$ annihilation.  First, in
the work of Gehrmann et al.~in
Refs.~\cite{GehrmannDeRidder:2007bj,GehrmannDeRidder:2007hr} and Weinzierl in
Refs.~\cite{Weinzierl:2008iv,Weinzierl:2009ms} the full set of ${\cal
  O}(\alpha_s^3)$ contributions to the 2-, 3- and 4-jet final states were
determined.  These results were made available in the program package
EERAD3~\cite{GehrmannDeRidder:2007bj}. Second, soft-collinear effective theory
(SCET)~\cite{Bauer:2000ew, Bauer:2000yr, Bauer:2001ct, Bauer:2001yt,
  Bauer:2002nz} provides a systematic framework to treat nonperturbative
corrections~\cite{Bauer:2002ie,Bauer:2003di} and to factorize and compute hard,
collinear and soft contributions for jet production to all orders in
$\alpha_s$~\cite{Fleming:2007qr,Schwartz:2007ib,Hoang:2007vb}, building on
earlier all orders QCD factorization
results~\cite{Korchemsky:1998ev,Korchemsky:1999kt,Korchemsky:2000kp}.  The SCET
framework allows for the summation of large logarithms at higher orders, as
demonstrated by the analytic calculation of the thrust distribution at \ntll
order by Becher and Schwartz in Ref.~\cite{Becher:2008cf}.\footnote{The
  calculation of Ref.~\cite{Becher:2008cf} also revealed a numerical problem at
  small $\tau$ in the initial fixed-order results of
  Refs.~\cite{GehrmannDeRidder:2007bj,GehrmannDeRidder:2007hr}.} In contrast,
the classic exponentiation techniques of Ref.~\cite{Catani:1992ua} for event
shapes have so far only been carried out to NLL order. Also, the anomalous
dimensions in SCET relevant for thrust are valid over perturbative momentum
scales, and there are no Landau pole ambiguities in the resummation at any
order.  In addition, as we will discuss in the body of our paper, SCET provides
a rigorous framework for including perturbative and nonperturbative
contributions, which can be used to connect power corrections in factorization
theorems to those in an operator expansion for thrust moments. Moreover it
provides a simple method to simultaneously treat the peak, tail, and far-tail
regions.

Several determinations of $\alpha_s$ in the tail region have been carried out
incorporating the fixed-order ${\cal O}(\alpha_s^3)$ results, which we have
collected in Tab.~\ref{tab:aseventshapes}.
\begin{table}[t] 
\begin{tabular}{cccl}
  &\ sum logs\ \ & \ power corrections\ \ & \hspace{0.5cm} $\alpha_s(m_Z)$\\
  \hline
  Ref.~\cite{Dissertori:2007xa} & no & Monte Carlo (MC)
   & $0.1240 \pm 0.0034^{*}$ \\
  Ref.~\cite{Becher:2008cf} & \ntll & uncertainty from MC\
   & $0.1172 \pm 0.0021^{*}$ \\
  Ref.~\cite{Davison:2008vx} & NLL & effective coupling 
   & $0.1164 \pm 0.0028^{\#}$ \\[-5pt]
   && model & \\[-2pt]
  Ref.~\cite{Bethke:2008hf} & NLL & Monte Carlo 
   & $0.1172\pm 0.0051^{**}$ \\ 
  Ref.~\cite{Dissertori:2009ik} & NLL & Monte Carlo 
   & $0.1224 \pm 0.0039^{*}$ \\
\end{tabular}
\caption{Recent thrust analyses which use the ${\cal O}(\alpha_s^3)$ fixed-order
  results. The theoretical component of the errors were determined as indicated,
  by either: ${}^*$ the error band method, ${}^{**}$ variation of the
  renormalization scale $\mu$,
  or ${}^\#$ by a simultaneous fit to 
  $\alpha_s(m_Z)$ and $\alpha_0$ (see text for more details). The analyses of
  Refs.~\cite{Becher:2008cf,Davison:2008vx} used thrust data only, while
  Refs.\cite{Dissertori:2007xa,Bethke:2008hf,Dissertori:2009ik} employed six 
  different event shapes.  
\label{tab:aseventshapes}}
\end{table}
They differ on which event shape data has been used for the fits, on the
accuracy of the partonic resummation of logarithms in the theory formula, the
approach for nonperturbative hadronization effects, and how the theory errors
are estimated.  It is instructive to compare the analyses by Dissertori~et al.\
\cite{Dissertori:2007xa,Dissertori:2009ik} and by Becher and
Schwartz~\cite{Becher:2008cf}, which both used the error band
method~\cite{Jones:2003yv} to determine theoretical uncertainties.  The improved
convergence and reduced theoretical uncertainty for $\alpha_s(m_Z)$ obtained by
Becher and Schwartz indicates that the summation of logarithms beyond NLL order
level is important.  Both the analyses by Dissertori~et al.\ and Becher and
Schwartz are limited by the fact that they used Monte Carlo (MC) generators to
estimate the size of nonperturbative corrections.

The use of $e^+e^-$ MC generators to estimate power corrections is problematic
since the partonic contributions are based on LL parton showers with at most
one-loop matrix elements, complemented by hadronization models below the shower
cutoff that are not derived from QCD.  The parameters of these models have been
tuned to LEP data, and thus unavoidably encode both nonperturbative effects as
well as higher order perturbative corrections. Hence, one must worry about
double counting, and this makes MC generators unreliable for estimating
nonperturbative corrections in higher order LEP analyses.  Moreover, purely
perturbative results for event shapes in the $\msbar$ scheme such as those in
Refs.~\cite{Dissertori:2007xa,Becher:2008cf,Dissertori:2009ik,Weinzierl:2009ms}
suffer from infrared effects known as infrared renormalons (see
Ref.~\cite{Beneke:1998ui} for a review of the early literature). These infrared
effects arise because fluctuations from large angle soft radiation down to
arbitrarily small momenta are included in the $\msbar$ perturbative series and
can cause unphysical large corrections already in low-order perturbative QCD
results.  On the other hand, the hard shower cutoff protects the parton level MC
from infrared renormalons. Hence one cannot rigorously combine MC hadronization
effects with strict perturbative $\msbar$ results.  From the two points raised
above, we conclude that the $\alpha_s(m_Z)$ results obtained in
Refs.~\cite{Dissertori:2007xa,Becher:2008cf,Dissertori:2009ik} contain a
systematic theoretical error from nonperturbative effects that can be quite
sizeable. We emphasize that this criticism also applies in part to the numerous
earlier event shape analyses which estimated nonperturbative corrections using MC
generators, see Ref.~\cite{Kluth:2006bw} for a review.

The presence of $\Lambda_{\rm QCD}/(Q\tau)$ power corrections in ${\rm
  d}\sigma/{\rm d}\tau$ have been discussed in earlier
literature~\cite{Manohar:1994kq,Webber:1994cp,Dokshitzer:1995zt,Akhoury:1995sp,Nason:1995hd,Korchemsky:1994is,MovillaFernandez:2001ed},
where it has been argued that the leading effect is a shift in the thrust
distribution, $\tau\to \tau-2\Lambda/Q$ with $\Lambda\sim\Lambda_{\rm QCD}$.
The analyses with the ${\cal O}(\alpha_s^3)$ results, that discuss
nonperturbative effects in the thrust tail region without relying on MC
generators are: Ref.~\cite{Becher:2008cf} which examined a $1/Q$ power
correction in a simple soft function model, but due to the large induced
uncertainty on $\alpha_s(m_Z)$ does not use it for their final error analysis, and
Ref.~\cite{Davison:2008vx} which uses the effective coupling model.

\begin{figure}[t]
      \vspace{0pt}
      \includegraphics[width=1\linewidth]{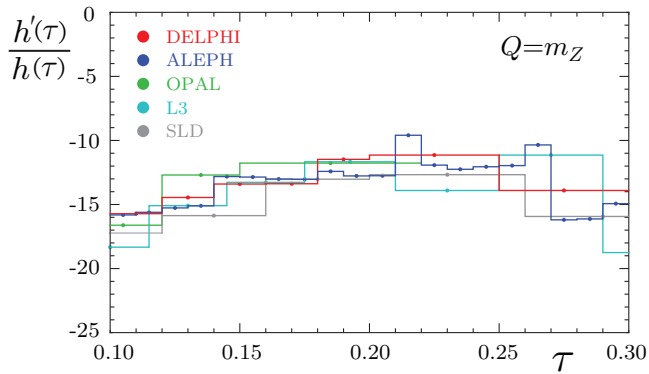}
      \caption{Plot of $h'(\tau)/h(\tau)$, the slope of
        $\ln[(1/\tau)\df\sigma/\df\tau]$, computed from experimental data at
        $Q=m_Z$. The derivative is computed using the central difference with
        neighboring experimental bins. }
      \label{fig:fracalphas}
\end{figure}

For the most accurate data at $Q=m_Z$ the change to the extracted
$\alpha_s(m_Z)$ from including the leading power correction can be quite
significant, at the 10\% level.  We can derive this estimate by a simple
calculation. We first write the cross section with a shift due to the power
correction, $(1/\sigma) \df\sigma/\df\tau = h(\tau-2\Lambda/Q)$.  Assuming $h$
is proportional to $\alpha_s$, and expanding for $\Lambda/Q \ll 1$, one can
easily derive that the change in the $\alpha_s$ value extracted from data due to
the existence of the power correction is
\begin{align} \label{eq:fracalphas}
  \frac{\delta\alpha_s}{\alpha_s} \simeq \frac{2\Lambda}{Q}\:
  \frac{h'(\tau)}{h(\tau)}\,.
\end{align}
The expression in \eq{fracalphas} gives a scaling estimate for the fractional
change in $\alpha_s$ from an analysis with the power correction compared to one
without, using data at $\tau$. To the extent that the assumptions stated above
are realized, the slope factor $h'(\tau)/h(\tau)$ should be constant.  In
Fig.~\ref{fig:fracalphas} we show the slope factor computed from
experimental data at $Q=m_Z$. The figure shows that the slope factor is indeed
only weakly depending on $\tau$ in the tail region and we get $h'(\tau)/h(\tau)
\simeq -14 \pm 4$.  The remaining visible variation in $\tau$ is related to
subleading nonperturbative and higher power $\alpha_s$ effects that are not
accounted for in our simple scaling estimate. For a QCD power correction of
natural size, $\Lambda=0.3\,{\rm GeV}$, \eq{fracalphas} gives
$\delta\alpha_s/\alpha_s \simeq -(9 \pm 3)\%$ for $Q=m_Z$. The magnitude of this
effect makes it important to treat power corrections as accurately as possible
in a fit to thrust data. We will show in later sections of this work that the
relative downward shift in the fitted $\alpha_s(m_Z)$ due to nonperturbative
effects is indeed at the level of the scaling estimate of~$-10\%$.

In the NLL/${\cal O}(\alpha_s^3)$ analysis by Davison and
Webber~\cite{Davison:2008vx} the nonperturbative effects are incorporated
through a power correction $\alpha_0$ which is fit together with $\alpha_s$ to
the experimental data.  The power correction is formulated from the low-scale
effective coupling model of Ref.~\cite{Dokshitzer:1995qm}, which modifies
$\alpha_s(\mu)$ below $\mu=\mu_I=2$~GeV, and defines $\alpha_0$ as the average
value of the coupling between $\mu=0$ and $\mu_I$.  It is important that the
effective coupling model correctly predicts the $Q$ dependence of the leading
nonperturbative power correction in
factorization~\cite{Dokshitzer:1995qm,Korchemsky:1999kt}. This model also
induces a subtraction of perturbative contributions below the momentum cutoff
$\mu_I$ (based on the running coupling approximation) and thus removes infrared
renormalon ambiguities.\footnote{Another thrust analysis where infrared
  renormalon contributions have been removed from the partonic contributions is
  by Gardi and Rathsman in Refs.~\cite{Gardi:2001ny,Gardi:2002bg}, which used a
  principal value prescription for the inverse Borel transformation of the
  thrust distribution.  Their analysis was prior to the new ${\cal
    O}(\alpha_s^3)$ fixed-order computations, and hence was not included in
  Table~\ref{tab:aseventshapes}.}  However the model is not based on
factorization, and hence this treatment of nonperturbative corrections is not
systematically improvable. It is therefore not easy to judge the corresponding
uncertainty.  Another problem of the effective coupling model is that its
subtractions involve large logs, $\ln(\mu_I/Q)$, which are not resummed. This
effects the $Q$ dependence in the interplay between perturbative and
nonperturbative effects.

In this paper we extend the event shape formalism to resolve the theoretical
difficulties mentioned above. Our results are formulated in the SCET framework,
and hence are rigorous predictions of QCD.  The formula we derive has a
\ntll order summation of logarithms for the partonic singular
$\alpha_s^j\ln^k(\tau)/\tau$ terms, and ${\cal O}(\alpha_s^3)$ fixed-order
contributions for the partonic nonsingular terms.  Our theoretical improvements
beyond earlier work include:
\begin{itemize}
\item A factorization formula that can be simultaneously applied to data in the
  peak and the tail regions of the thrust distribution and for multiple
  c.m.~energies $Q$, as well as being consistent with the multijet thresholds in
  the far-tail region.
\item In the factorization formula a nonperturbative soft function defined from
  field theory is implemented using the method of Ref.~\cite{Hoang:2007vb} to
  incorporate hadronization effects. To achieve independence of a particular
  analytic ansatz in the peak region, the nonperturbative part of the soft
  function uses a linear combination of orthogonal basis functions that converge
  quickly for confined functions~\cite{Ligeti:2008ac}.
\item In the tail region the leading power correction to ${\rm d}\sigma/{\rm
    d}\tau$ is determined by a nonperturbative parameter $\Omega_1$ that appears
  through a factorization theorem for the singular distribution.  $\Omega_1$ is
  a field theory matrix element of an operator, and is also related to the first
  moment of the nonperturbative soft function. In the tail region the effects of
  $\Omega_1$ hadronization corrections are included for the nonsingular
  corrections that are kinematically subleading in the dijet limit, based on
  theoretical consistency with the far-tail region.
\item Defining the matrix element $\bar\Omega_1$ in $\msbar$, the perturbative
  cross section suffer from an ${\cal O}(\Lambda_{\rm QCD})$ renormalon. In our
  analysis this renormalon is removed by using an R-gap scheme for the
  definition of $\Omega_1$~\cite{Hoang:2007vb}. This scheme choice induces
  subtractions on the leading power $\msbar$ cross section which simultaneously
  remove the renormalon there.  Large logarithms in the subtractions are summed
  to all orders in $\alpha_s$ using R-evolution equations given in
  Refs.~\cite{Hoang:2008yj,Hoang:2009yr}.
\item Finite bottom quark mass corrections are included using a factorization
  theorem for event shapes involving massive quarks, derived in
  Refs.~\cite{Fleming:2007qr,Fleming:2007xt}.
\item QED corrections at NNLL order are incorporated, counting $\alphaew\sim
\alpha_s^2$. This includes QED Sudakov effects, final state radiation, and
QED/QCD renormalization group interference.
\item The 3-loop finite term $h_3$ of the quark form factor in
  $\msbar$ is extracted using the results of Ref.~\cite{Baikov:2009bg}, 
  and is included in our analysis.
\item The most important corrections from the axial anomaly are included. The
  anomaly modifies the axial-vector current contributions at ${\cal
    O}(\alpha_s^2)$ by terms involving the top quark mass.
\end{itemize}
Electroweak effects from virtual $W$ and $Z$ loops mostly effect the
normalization of the cross section and so their dominant contribution drops out
of $(1/\sigma) {\rm d}\sigma/{\rm d}\tau$~\cite{Denner:2009gx,Denner:2010ia}. These
corrections are not included in our analysis.

For the numerical analyses carried out in this work we have created within our
collaboration two completely independent codes. One code within
Mathematica~\cite{mathematica} implements the theoretical expressions exactly as
given in this paper, and one code is based on theoretical formulae in Fourier
space and realized as a fast Fortran code suitable for parallelized runs on
computer clusters. These two codes agree for the thrust distribution at the
level of $10^{-6}$.

While the resulting theoretical code can be used for all values of $\tau$, in
this paper we focus our numerical analysis on a global fit of $e^+e^-$ thrust
data in the tail region, for c.m.\ energies $Q$ between $35$ and $207$~GeV, to
determine $\alpha_s(m_Z)$.\footnote{Throughout this paper we use the $\msbar$
  scheme for $\alpha_s$ with five light flavors.}  Our global fit exhibits
consistency across all available data sets, and reduces the overall experimental
uncertainty.  For a single $Q$ we find a strong correlation between the effect
of $\alpha_s(m_Z)$ and $\Omega_1$ on the cross section. This degeneracy is
broken by fitting data at multiple $Q$s. The hadronization uncertainty is
significantly decreased by our simultaneous global fit to $\alpha_s(m_Z)$ and
$\Omega_1$.  To estimate the perturbative uncertainty in the fit we use a random
scan in a 12-dimensional theory parameter space.  This space includes 6
parameters for $\mu$-variation, 3 parameters for theory uncertainties related to
the finite statistics of the numerical fixed-order results, one parameter for
the unknown 4-loop cusp anomalous dimension, and two parameters for unknown
constants in the perturbative 3-loop jet and 3-loop soft functions.  The scan
yields a more conservative theory error than the error band
method~\cite{Jones:2003yv}.  Despite this we are able to achieve smaller
perturbative uncertainties than earlier analyses due to our removal of the ${\cal
  O}(\lqcd)$ renormalon and the inclusion of $h_3$. We also analyze in detail
the dependence of the fit results on the range in $\tau$ used in the fit.

The outline of the paper is as follows. In Sec.~\ref{sec:bsg} we introduce
the theoretical framework and discuss the various theoretical ingredients in our
final ${\rm d}\sigma/{\rm d}\tau$ formula. In Sec.~\ref{sec:profile} we
present the profile functions which allow us to simultaneously treat multiple
$\tau$ regions, and discuss the 6 parameters used for $\mu$-variation in the
analysis of the perturbative uncertainty. In Sec.~\ref{sec:model} we review
the parametrization of the nonperturbative function. In
Sec.~\ref{sec:convergence} we discuss the normalization of our distributions
and compare results at different orders in perturbation theory for: fixed-order
results, adding the log resummation, adding the nonperturbative corrections, and
adding the renormalon subtractions.  In Sec.~\ref{sec:expedata} we discuss
the experimental data and the fit procedure. Our results for $\alpha_s(m_Z)$ and
the soft function moment $\Omega_1$ from the global fit are presented in
Sec.~\ref{sec:fit}, including a discussion of the theory errors.  In
Sec.~\ref{sec:fartailpeak} we use our tail fit results to make predictions in
the far-tail and peak regions, and compare with data.  Cross checks on our code
are discussed in Sec.~\ref{sec:comparison}, including using it to reproduce
the earlier lower precision fits of Dissertori et al.~\cite{Dissertori:2007xa}
and Becher and Schwartz~\cite{Becher:2008cf}.  Section~\ref{sec:conc} contains
our conclusions and outlook, including prospects for future improvements based
on the universality of the parameter $\Omega_1$.  The analytic theoretical
expressions that went into our analysis for massless quarks and QCD effects are
presented in condensed form in Appendix~\ref{app:appendix}. In
Appendix~\ref{app:SoftOPE} we use the operator product expansion for the soft
function in the tail region, discussing uniqueness and deriving an all order
relation for the Wilson coefficient of $\Omega_1$.  In
Appendix~\ref{app:MomentOPE} we use an OPE for the first moment of the thrust
distribution to show that it involves the same $\Omega_1$ at lowest order.
Readers most interested in our numerical results can skip directly to
Sections~\ref{sec:expedata} and~\ref{sec:fit}.

\section{Formalism}
\label{sec:bsg}

\subsection{Overview}

The factorization formula we use for the fits to the experimental thrust data is
\begin{align} \label{eq:masterformula}
& \dfrac{\mathrm{d}\sigma}{\mathrm{d}\tau}  
= \! \int\!\! \mathrm{d}k\left(
\dfrac{\mathrm{d}\hat{\sigma}_{\rm s}}{\mathrm{d}\tau}+
\dfrac{\mathrm{d}\hat{\sigma}_{\rm ns}}{\mathrm{d}\tau}+
\dfrac{\Delta\mathrm{d} \hat{\sigma}_{b}}{\mathrm{d}\tau}\!\right)
\!\!\bigg(\!\tau- \frac{k}{Q}
\bigg) 
 \nn\\
& \quad \times S_{\tau}^{\mathrm{mod}}\Big(k\!-\!2\bar\Delta(R,\mu_S)\Big)
 + {\cal O}\Big(\sigma_0\, \alpha_s \frac{\Lambda_{\rm QCD}}{Q}\Big)
 \,.
\end{align}
Here $\mathrm{d}\hat{\sigma}_{\rm s}/\mathrm{d}\tau$ contains the singular
partonic QCD corrections $\alpha_s^j\,[\ln^k(\tau)/\tau]_+$ and
$\alpha_s^j\,\delta(\tau)$ with the standard plus-functions as defined in
Eq.~(\ref{eq:cLna_def}). It also contains the singular
partonic QED corrections depending on $\alphaew$ which are discussed in
\subsec{QED}.  This $\mathrm{d}\hat{\sigma}_{\rm s}/\mathrm{d}\tau$ term
accounts for matrix element corrections and the resummation of $\ln\tau$ terms
within the SCET formalism up to \ntll order, which we discuss in
\subsec{singular}.  Our definition of \ntll, \ntllp, and other orders is
discussed in detail in Sec.~\ref{subsec:orders} (see also
Tab.~\ref{tab:orders}).

The term $\mathrm{d}\hat{\sigma}_{\rm ns}/\mathrm{d}\tau$, which we call the
nonsingular partonic distribution, contains the thrust distribution in strict
fixed-order expansion with the singular terms $\propto
\alpha_s^j\ln^k(\tau)/\tau$ subtracted to avoid double counting. The most
singular terms in $\mathrm{d}\hat{\sigma}_{\rm ns}/\mathrm{d}\tau$ scale as
$\ln^k\tau$ for $\tau\to 0$.\footnote{For ${\rm d}\hat\sigma_{\rm ns}/{\rm
    d}\tau$ the resummation of $\ln\tau$ terms is currently unknown. These terms
  could be determined with subleading factorization theorems in SCET.} Our
implementation of nonsingular terms is discussed in detail in
Sec.~\ref{subsec:NSdist}.

Finally, $\Delta {\rm d}\hat\sigma_b/{\rm d}\tau$ contains corrections to the
singular and nonsingular cross sections due to the finite mass of the bottom
quark.  The $b$-mass corrections are implemented as a difference of the massive
and massless cross sections computed at NNLL order as discussed in
\subsec{bottom}.

The function $S_{\tau}^{\rm mod}$ that is convoluted with these partonic cross
sections in \eq{masterformula} describes the nonperturbative effects from soft
gluons including large angle soft
radiation~\cite{Korchemsky:1999kt,Berger:2003pk}. The definition of $S_\tau^{\rm
  mod}$ also depends on the hemisphere prescription inherent to the thrust
variable. This is a hadronic function that enters in a universal way for both
massless and massive cross sections, and is independent of the value of $Q$.
The universality of $S_{\tau}^{\rm mod}$ in \eq{masterformula} follows from the
leading power thrust factorization
theorem~\cite{Korchemsky:1999kt,Fleming:2007qr,Schwartz:2007ib}, and the thrust
factorization theorem for massive quarks in
Refs.~\cite{Fleming:2007qr,Fleming:2007xt}. Our treatment of the convolution of
$S_{\tau}^{\rm mod}$ with $\mathrm{d}\hat{\sigma}_{\rm ns}/\mathrm{d}\tau$
yields a consistent treatment of multijet thresholds and the leading power
correction to the operator expansion for the first moment of thrust. Details of
our implementation of power corrections and nonperturbative corrections are
discussed in Sec.~\ref{subsec:NSfac} and Sec~\ref{sec:model}. The function
$S_{\tau}^{\rm mod}$ is normalized to unity and can be determined from
experimental data.  Its form depends on a gap parameter $\bar \Delta$ and
additional moment parameters $\Omega_i$ which are discussed below.

The factorization formula given in \eq{masterformula} can be applied
simultaneously in the peak, tail, and the far-tail regions of \eq{regions},
i.e.\ for all $\tau$ values. In the peak region ${\rm d}\hat\sigma_{\rm ns}/{\rm
  d}\tau$ is significantly smaller than ${\rm d}\hat\sigma_{\rm s}/{\rm d}\tau$,
and the full analytic form of the soft nonperturbative function $S_{\tau}^{\rm
  mod}(k)$ is relevant to determine the $\tau$-distribution since $\mus\simeq
\Lambda_{\rm QCD}$. Because $\muh\gg \muj\gg \mus$, the summation of logarithms
of $\tau$ is also crucial to achieve an accurate theoretical description.

For much of the tail region the summation of $\ln\tau$ terms remains important,
although this is no longer the case when we reach $\tau\simeq 1/3$. Likewise,
the dominance of the singular partonic contributions remains as long as $\tau <
1/3$, but the nonsingular terms become more important for increasing $\tau$ (see
Fig.~\ref{fig:sigcomponents} below).  Near $\tau\simeq 1/3$ the nonsingular
terms become equal in size to the singular terms with opposite sign. Since
$\mus\gg \Lambda_{\rm QCD}$ in the tail region the effects of $S_{\tau}^{\rm
  mod}$ can be parameterized in terms of the moments
\begin{align} \label{eq:omega1}
\Omega_i \,& = \int\!{\rm d}k\, 
\bigg(\frac{k}{2}\bigg)^i\, S_{\tau}^{\rm mod}(k-2\bar\Delta)\, ,
\end{align}
where $\Omega_0=1$ since $S_{\tau}^{\rm mod}$ is normalized. Their importance is
determined by $\Omega_i/(Q\tau)^i$ as discussed in Sec.~\ref{subsec:NSfac},
so the first moment $\Omega_1$ parameterizes the dominant power correction and
higher moments provide increasingly smaller corrections. The first moment is
defined by
\begin{align}
\label{eq:O1def}
  \Omega_1 \equiv \bar\Delta 
  + \frac{1}{2N_c} \big\langle 0 \big| {\rm tr}\ \overline Y_{\bar
    n}^T(0) Y_n(0)\, i\widehat\partial\: Y_n^\dagger(0) \overline Y_{\bar
    n}^*(0) \big| 0 \big\rangle \,,
\end{align}
where $Y_n^\dagger(0) = {\rm P} \exp\,( ig \int_0^\infty ds\, n\cdot A(ns) )$,
$\overline Y_{\bar n}^\dagger$ is similar but in the $\overline 3$
representation, and we trace over color. Here 
\begin{align} \label{eq:ptau}
 i\widehat\partial\equiv
\theta(i\bar n\cdot\partial\!-\!in\cdot\partial)\, in\cdot\partial + \theta(i
n\cdot\partial\!-\!i\bar n\cdot\partial)\, i\bar n\cdot\partial,
\end{align}
is a derivative operator\footnote{Note that $i\widehat\partial$ is defined in
  the c.m. frame of the colliding $e^+e^-$. One may also write
  $i\hat\partial=\int\!{\rm d}\eta\: e^{-|\eta|} \hat{\cal E}_T(\eta)$ where
  $\hat{\cal E}_T(\eta)$ measures the sum of absolute transverse momenta at a
  given rapidity $\eta$ with respect to the thrust axis $\hat
  t$~\cite{Korchemsky:1999kt,Belitsky:2001ij}.} involving light-like vectors
$n=(1,\hat {\bf t})$ and $\bar n=(1,-\hat {\bf t})$.  $\Omega_1$ is the field
theory analog of the parameter $\alpha_0$ employed in the low-scale effective
coupling approach to power corrections.  Since the renormalon subtractions
depend on a cutoff scale $R$ and the renormalization scale $\mus$, all moments
$\Omega_i(R,\mu_S)$ as well as $\bar\Delta(R,\mu_S)$ are scale and scheme
dependent quantities.  The scheme we use to define $\Omega_1(R,\mu_S)$ is
described in \subsec{gap}.  In our fit to experimental data we use the R-gap
scheme, and extract the first moment at a reference scale
$R_\Delta=\mu_\Delta=2$~GeV, i.e. we use $\bar\Delta(R_\Delta,\mu_\Delta)$ and
hence $\Omega_1=\Omega_1(R_\Delta,\mu_\Delta)$. In the factorization theorem the
gap appears evaluated at $\bar\Delta(R,\mu_S)$ and the scales ($R,\mu_S$) are
connected to the reference scales ($R_\Delta,\mu_\Delta$) using renormalization
group equations.

Finally, in the far-tail region $\tau \simeq 0.3$ the singular and the
nonsingular partonic contributions $\mathrm{d}\hat{\sigma}_{\rm
  s}/\mathrm{d}\tau$ and $\mathrm{d}\hat{\sigma}_{\rm ns}/\mathrm{d}\tau$ become
nearly equal with opposite signs, exhibiting a strong cancellation.  This is due
to the strong suppression of the fixed-order distribution in the three- and
four-jet endpoint regions at $\tau \gtrsim 1/3$ in fixed-order perturbation
theory.  In this region the summation of logarithms of $\tau$ must be switched
off to avoid messing up this cancellation.  Here our \eq{masterformula} reduces
to the pure fixed-order partonic thrust distribution supplemented with power
corrections coming from the convolution with the soft function.  All three
regions are smoothly joined together in \eq{masterformula}.  The proper
summation (or non-summation) of logarithms is achieved through $\tau$-dependent
renormalization scales, $\muj(\tau)$, $\mus(\tau)$, and $R(\tau)$ which we call
{\it profile functions}. They are discussed in detail in \sec{profile}.

In the following subsections various ingredients of the factorization formula of
Eq.~(\ref{eq:masterformula}) are presented in more detail. Compact results for
the corresponding analytic expressions for massless quarks in QCD are given in
\app{appendix}. In Secs.~\ref{subsec:bottom} and \ref{subsec:QED} we describe
how finite bottom mass and QED corrections are included in our analysis. The
full formulae for these corrections will be presented in a future publication.

\subsection{Order Counting} 
\label{subsec:orders}

\begin{table*}[t!]
\hspace{-1.4cm}
a) Perturbative Corrections \hspace{5.2cm}
b) Nonperturbative Corrections with $\Omega_k\sim \Lambda_{\rm QCD}^k$\\[3pt]
\begin{tabular}{c|ccccccc}
&cusp&non-cusp&matching& $\beta[\alpha_s]$ &nonsingular
 & $\gamma_{\Delta}^{\mu,R}$ & $\delta$ \\
\hline
LL&1&-&tree&1&- & - & - \\
{\rm NLL}&2&1&tree&2&- & 1 & -\\
NNLL&3&2&1&3&1 & 2 & 1\\
\ntll &$4^{\rm pade}$&3&2&4&2 & 3 & 2\\
\hline
NLL${}^\prime$&2&1&1&2&1 & 1 & 1\\
NNLL${}^\prime$&3&2&2&3&2 &2 & 2\\
\ntllp &$4^{\rm pade}$&3&3&4&3 & 3 & 3\\
\end{tabular}
\hspace{0.5cm}
\begin{tabular}{c|cc}
&peak (any $k$)  & tail and far-tail ($k=0,1,2$)\\
\hline \\[-12pt]
 \!\!\! {\large \mbox{$\frac{\df\hat\sigma_s}{\df\tau}$ }}\!\!
 & $\alpha_s^i $ \!\!\! {\large \mbox{$\frac{\ln^j\!\!\tau}{\tau}
 \frac{\Omega_k}{(Q\tau)^k}$ }}
 & $\alpha_s^i $ \!\!\!  {\large \mbox{$\frac{\ln^j\!\!\tau}{\tau}
\frac{\Omega_k}{(Q\tau)^k}
$ }}
\\[5pt]
 \!\!\! {\large \mbox{$\frac{\df\hat\sigma_{n_s}}{\df\tau}$ }}\!\!
 & $\alpha_s^i\, f_{ik}(\tau)$\!\!\!\! {\large \mbox{ 
 $\frac{\Omega_k}{(Q\tau)^k}$
 }}
 & $\alpha_s^i\, f_{ik}(\tau)$\!\!\!\! {\large \mbox{ 
$\frac{\Omega_k}{(Q\tau)^k}$
}}
\\[5pt]
 \!\!\! {\large \mbox{$\frac{\df\hat\sigma_{b}}{\df\tau}$ }} \!\!
 &
 $ \alpha_s^i\, g_{ik}\big(\tau,\!\frac{m_b}{Q}\big)$\!\!\!\!
 {\large \mbox{
 $\frac{\Omega_k}{(Q\tau)^k}$
 }} 
 & $\alpha_s^i\, g_{ik}\big(\tau,\!\frac{m_b}{Q}\big)$\!\!\!\! 
 {\large \mbox{
$\frac{\Omega_k}{(Q\tau)^k}$
 }}\!\!\!\!\!
\\[10pt]
p.c.\phantom{x}
  & $\alpha_s$\!\!\!\! {\large \mbox{ $\frac{\Lambda_{\rm QCD}}{Q}$ }} 
  & $\alpha_s$\!\!\!\! {\large \mbox{ $\frac{\Lambda_{\rm QCD}}{Q}$ }}
  \\
\end{tabular}
\caption{Perturbative and nonperturbative corrections included in our analysis. 
  a) (left panel) Loop orders $j$ for perturbative corrections of ${\cal
    O}(\alpha_s^j)$. Here cusp, non-cusp, and $\gamma_\Delta^{\mu,R}$ refer 
  to anomalous dimensions, while matching, nonsingular, and the gap subtraction 
  $\delta$ refer to fixed-order series.  For
  convenience in our numerical analysis we use the four-loop
  beta function for the $\alpha_s$ running in all orders displayed. 
  b) (right panel) Nonperturbative corrections
  included in ${\rm d}\sigma/{\rm d}\tau$ with implicit sums over $i$ and 
  $k$. All powers $\Omega_k/(Q\tau)^k$ can be included in the peak region with the function 
  $S_{\tau}^{\rm mod}$, while only a fixed set of power correction parameters  
  are included in the tail and far-tail regions. The row labeled p.c. shows
  the scaling of the the first power correction that is not entirely determined
  by the earlier rows and hence yield corrections to \eq{masterformula}.
}
\label{tab:orders}
\end{table*}

In the classic order counting used for fits to event shape distributions it is
common to separately quote orders for the summation of logarithms and the
fixed-order matching contributions.  For fixed-order contributions the ${\cal
  O}(\alpha_s)$ contributions are called LO, the ${\cal O}(\alpha_s^2)$
contributions are called NLO, etc.  This counting is motivated from the fact
that at tree level the fixed-order thrust distribution vanishes for $\tau>0$.
For the summation one refers to LL (leading-log) summation if the one-loop cusp
anomalous dimension is used to sum the double Sudakov logs, and NLL
(next-to-leading-log) if the two-loop cusp and the one-loop non-cusp anomalous
dimension terms are also included.

In our analysis the summation orders (LL, NLL, ...) match the classical
language.  For the fixed-order contributions we account for the tree level
$\delta(\tau)$ in LL and NLL, and we include ${\cal O}(\alpha_s)$ corrections at
NLL$^\prime$ and NNLL, etc, as shown in Tab.~\ref{tab:orders}a. In SCET the
summation can be carried out at both NNLL and \ntll~\cite{Becher:2008cf}.  The
corresponding loop orders for the anomalous dimensions are also shown in
Tab.~\ref{tab:orders}a.  Within SCET the summation of logarithms is achieved by
renormalization group evolution and the fixed-order corrections enter as series
evaluated at each of the transition scales $\muh$, $\muj$, and $\mus$ which we
refer to as matching or matrix element corrections.  The logs in the singular
thrust cross section exponentiate to all orders if we use $y$, the
Fourier-transformed variable to $\tau$.  The orders we consider correspond to
summing the terms
\begin{align}
  \ln\Big[ \frac{{\rm d}\tilde\sigma_s}{{\rm d}y}\Big] & \sim \bigg[L \sum_{k=1}^\infty
  (\alpha_s L)^k \bigg]_{\rm LL} + \bigg[ \sum_{k=1}^\infty (\alpha_s L)^k
  \bigg]_{\rm NLL}
  \\
  &\!\! + \bigg[ \alpha_s \sum_{k=0}^\infty (\alpha_s L)^k \bigg]_{\rm NNLL} +
  \bigg[ \alpha_s^2 \sum_{k=0}^\infty (\alpha_s L)^k \bigg]_{\text{N${}^3$LL}}
  \nn \,,
\end{align}
where $L=\ln(iy)$, and the series in the exponent makes clear the structure of
the large logs that are summed at each order.

The nonsingular counting in Tab.~\ref{tab:orders}a for the fixed-order series in
${\rm d}\hat\sigma_{\rm ns}/{\rm d}\tau$ must be the same as for the matching
and matrix element corrections to ensure that we exactly reproduce the
fixed-order cross section when the resummed result is expanded.  Since the
relative importance of the log resummation and the nonsingular terms varies
depending on the $\tau$-region, we also consider an alternative ``primed''
counting scheme. In the primed counting all series for fixed-order quantities
are included to one higher order in $\alpha_s$. In this counting scheme the
${\cal O}(\alpha_s^3)$ fixed-order results occur in \ntllp, which is the order
we use for our final analysis.

Also shown in Tab.~\ref{tab:orders}a are columns for the fixed-order gap
subtractions $\delta= \delta(R,\mu)$, and the gap anomalous dimensions
$\gamma_{\Delta}^{\mu,R}$. These terms are required to remove the leading ${\cal
  O}(\Lambda_{\rm QCD})$ renormalon from the perturbative corrections, while
still maintaining the same level of log resummation for terms in the cross
section. The resummation of these large logarithms is missing in the recent
analysis of Ref.~\cite{Davison:2008vx} and is discussed further in \subsec{gap}.

A crucial aspect of our analysis is the inclusion of power corrections in a
rigorous manner through field theoretic techniques. In the effective theory
there are several types of power corrections which arise from the possible
ratios of the scales $\muh$, $\muj$, $\mus$, and $\lqcd$:
\begin{align} \label{eq:pctype}
 &1)\ \ \frac{\Lambda_{\rm QCD}}{\mus}=\frac{\Lambda_{\rm QCD}}{Q\tau} \,,\nn\\
 &2)\ \ \frac{\mus^2}{\muj^2}=\tau \,, 
  \nn\\
 &3)\ \ \frac{\Lambda_{\rm QCD}}{\muh}=\frac{\Lambda_{\rm QCD}}{Q} \,.
\end{align}
Any $\Lambda_{\rm QCD}/\mu_J$ power correction can be taken as a cross-term
between types 1) and 2) for the purpose of enumeration.  The type \pci power
corrections are enhanced by the presence of the soft scale and are encoded by
the moments $\Omega_k\sim \Lambda_{\rm QCD}^k$ of the soft function. Type \pcii
are kinematic power corrections that occur because of the expansion about small
$\tau$, and can be computed with perturbation theory.  The importance of these
first two types depends on the region considered in \eq{regions}, with all terms
in type \pcii becoming leading order for the far-tail region.  Type \pciii are
non-enhanced power correction that are of the same size in any region. There are
also cross-terms between the three types.

In our analysis we keep all power corrections of types \pci and \pcii, and the
dominant terms of type \pciii. Our treatment of the nonsingular cross section
also includes cross-terms between \pci and \pcii in a manner that is discussed in
\subsec{NSfac}.  For the different thrust regions we display the relevant terms
kept in our analysis in Tab.~\ref{tab:orders}b. The nonsingular cross section
corrections fully account for the power corrections of type \pcii. The factor
$[\Lambda_{\rm QCD}/(Q\tau)]^k$ in the peak region denotes the fact that we sum
over all type \pci power corrections from the leading soft function. In the tail
and multijet regions we only consider the first three orders: k=0 (partonic
result), k=1 (power correction involving $\Omega_1$) and , k=2 (power correction
involving $\Omega_2$). Here $k=2$ terms are used in our error analysis for our
simultaneous fit to $\alpha_s(m_Z)$ and $\Omega_1$. The leading power correction
that is not fully captured in all regions is of type \pciii, and are of ${\cal
  O}(\alpha_s\Lambda_{\rm QCD}/Q)$. Since our analysis is dominated by $Q=m_Z$
or larger, parametrically this gives an uncertainty of
\begin{align} \label{eq:Lam1}
 \Big[ \frac{\delta\alpha_s}{\alpha_s}\Big]_{{\rm p.c.}} \sim \frac{\Lambda_{\rm
     QCD}}{Q} \simeq 0.3\%
\end{align}
in our final fit (taking $\Lambda_{\rm QCD}=0.3\,{\rm GeV}$ to obtain the number
here).  This estimate has been validated by running our fits in the presence of an
additional $\alpha_s\Lambda_{\rm QCD}/Q$ power correction.\footnote{To perform
  this test we include an $\alpha_s(\mu_{\rm ns}) \Lambda_1/Q$ correction in the
  normalized thrust cross section, vary $\Lambda_1=\pm 1.0\,{\rm GeV}$, and
  perform our default fit to $\alpha_s(m_Z)$ and $\Omega_1$ as described in
  Sec.~\ref{sec:expedata}. This variation causes only a $\pm 0.1\%$ change to
  these fit parameters, which is smaller than the estimate in \eq{Lam1}.}

\subsection{Singular Partonic Distribution}
\label{subsec:singular}

The singular partonic thrust distribution $\df\hat\sigma_s/\df\tau$ contains the
most singular terms $\propto \alpha_s^j\ln^k(\tau)/\tau$ and
$\alpha_s^j\delta(\tau)$ that arise from perturbation theory. Using SCET one can
derive a factorization theorem for these terms which allows for the resummation
of the logarithmic terms to all orders in perturbation theory. In massless QCD the
factorization formula for the perturbative corrections involving $\alpha_s$
reads
\begin{align}
\label{eq:singular}
 & \dfrac{\mathrm{d}\hat{\sigma}_{\rm s}^{ \mbox{\tiny QCD}}}{\mathrm{d}\tau}(\tau)
= Q \sum_I \sigma_{0}^{I}\, H^I_{Q}(Q,\mu_H)\, U_{H}(Q,\mu_H,\mu)\!\!
\int\!\mathrm{d}s\,\mathrm{d}s^\prime\, \nonumber \\
& \ \times J_{\tau}(s^\prime,\mu_J)\,
 U_{J}^{\tau}(s-s^\prime,\mu,\mu_J)\, \int\! \df k'\: U_S^\tau(k',\mu,\mu_S) \nn \\
& \ \times e^{-2\frac{\delta(R,\mu_s)}{Q}\frac{\partial}{\partial \tau}} \ 
S_{\tau}^{\rm part}\Big(\!Q\tau-\frac{s}{Q}-k',\mu_S\!\Big). 
\end{align}
Here $\sigma_0^I$ is the total partonic $e^+e^-$ cross section for quark pair
production at tree level from a current of type $I =\{uv,dv,bv,ua,da,ba\}$ as
explained below. Large logs are summed by the renormalization group factors
$U_{H}$ between the hard scale and $\mu$, $U_{J}^{\tau}$ between the jet scale
and $\mu$, and $U_S^{\tau}$ between the soft scale and $\mu$. The choice of
$\mu$ is arbitrary and the dependence on $\mu$ cancels out exactly when working
at any particular order in the resummed expansion. Short distance virtual
corrections are contained in the hard function $H_Q^I$. The term $J_{\tau}$ is
the thrust jet function. The term $S^{\rm part}_\tau$ is the partonic soft
function and the $\delta(R,\mu_S)$-dependent exponential implements the
perturbative renormalon subtractions.  There are four renormalization scales
governing the factorization formula, the hard scale $\muh\sim Q$, the jet scale
$\muj$, the soft scale $\mus$, and the renormalon subtraction scale $R$. We have
$R\sim \mus$ to properly sum logarithms related to the renormalon subtractions,
and there is also a renormalization group evolution in $R$.  The typical values
for $\muj$, $\mus$, and $R$ depend on $\tau$ as discussed in
Sec.~\ref{sec:profile}.

The total tree level partonic $e^+e^-$ cross section
$\sigma_0^I=\sigma_0^I(Q,m_Z,\Gamma_Z)$ depends on the c.m.\ energy $Q$, the
$Z$-mass, and $Z$-width, and has six types of components, $\sigma_0^{uv}$,
$\sigma_0^{ua}$, $\sigma_0^{dv}$, $\sigma_0^{da}$, $\sigma_0^{bv}$,
$\sigma_0^{ba}$, where the first index denotes flavor, $u={\rm up}+{\rm charm}$,
$d={\rm down}+{\rm strange}$, and $b={\rm bottom}$, and the other index denotes
production through the vector ($v$) and axial-vector ($a$) currents. For QCD
corrections we have the hard functions $H_Q^v\equiv H_Q^{uv}=H_Q^{dv}=H_Q^{bv}$,
$H_Q^{ua}$, $H_Q^{da}$, and $H_Q^{ba}$, where the vector current terms do not
depend on the flavor of the quark.  For massless quark production the
axial-vector hard functions differ from the vector due to flavor singlet
contributions.  All six $\sigma_0^I$'s and $H_Q^I$'s are relevant for the
implementation of the $b$-mass and QED corrections.  Since we use data taken for
energies close to the Z pole we adopt $i/(q^2-m_Z^2+i\, Q^2\Gamma_z/m_Z)$ as the
Z-boson propagator which is the form of the width term used for thrust data
analyses.  The modifications of \eq{singular} required to include QED effects
are discussed in \subsec{QED}.

The hard factor $H_Q$ contains the hard QCD effects that arise from the matching
of the two-jet current in SCET to full QCD. For $\muh=Q$ we have $H_Q^v(Q,Q) = 1
+ \sum_{j=1}^3 h_j [\alpha_s(Q)/4\pi]^j$, and the full hard function with
$\ln(\muh/Q)$ dependence is given in \eq{Hardnumeric}. For the flavor nonsinglet
contributions where the final-state quarks are directly produced by the current
one can obtain the matching coefficient from the on-shell quark vector current
form factor, which is known to ${\cal O}(\alpha_s^3)$
\cite{Matsuura:1987wt,Matsuura:1988sm,Gehrmann:2005pd,Moch:2005id,Lee:2010cg,Baikov:2009bg}.
Converting the bare result in Ref.~\cite{Lee:2010cg} 
(see also Refs.~\cite{Baikov:2009bg,Heinrich:2009be}) to the $\msbar$ scheme
and subtracting $1/\epsilon_{\rm IR}^k$ divergences present in SCET graphs, the
three-loop non-singlet constant, which is one of the new ingredients in our
analysis, is
\begin{align} \label{eq:h3}
 &h_3=
C_F^3 \bigg[\!
-460 \zeta (3)
-\frac{140 \pi ^2 \zeta (3)}{3}
+32 \zeta (3)^2
+1328\zeta (5)
\nn\\ &
-\frac{5599}{6}
+\frac{4339 \pi ^2}{36}
-\frac{346 \pi ^4}{15}
+\frac{27403 \pi ^6}{17010}\bigg]
\nn\\ &
+C_A C_F^2 \bigg[
-\frac{52564 \zeta(3)}{27}
+\frac{1690 \pi ^2 \zeta (3)}{9}
+\frac{592 \zeta (3)^2}{3}
\nn\\ &
-\frac{5512 \zeta (5)}{9}
+\frac{824281}{324}
-\frac{406507 \pi ^2}{972}
+\frac{92237 \pi^4}{2430}
\nn\\ &
-\frac{1478 \pi ^6}{1701}\bigg]
+C_A^2 C_F \bigg[
\frac{505087 \zeta (3)}{243}
-\frac{1168 \pi ^2 \zeta (3)}{9}
\nn\\ &
-\frac{2272 \zeta (3)^2}{9}
-\frac{868 \zeta(5)}{9}
-\frac{51082685}{26244}
+\frac{596513 \pi ^2}{2187}
\nn\\ &
-\frac{4303 \pi^4}{4860}
+\frac{4784 \pi ^6}{25515}\bigg]
+C_F^2 n_f \bigg[
\frac{26080 \zeta (3)}{81}
-\frac{148 \pi ^2 \zeta (3)}{9}
\nn\\ &
-\frac{832 \zeta (5)}{9}
-\frac{56963}{486}
+\frac{13705 \pi ^2}{243}
-\frac{1463 \pi ^4}{243}\bigg]
\nn\\ &
+C_A C_F n_f \bigg[
-\frac{8576 \zeta (3)}{27}
+\frac{148 \pi ^2 \zeta (3)}{9}
-\frac{8 \zeta(5)}{3}
\nn\\ &
+\frac{3400342}{6561}
-\frac{201749 \pi ^2}{2187}
-\frac{35 \pi ^4}{243}\bigg]
+C_F n_f^2 \bigg[
-\frac{832 \zeta (3)}{243}
\nn\\ &
-\frac{190931}{6561}
+\frac{1612 \pi ^2}{243}
+\frac{86 \pi ^4}{1215}\bigg]
\nn\\ &
= 20060.0840 -2473.4051 n_f + 52.2009 n_f^2
\,.
\end{align}
For $n_f=5$ we have $h_3=8998.080$, which is the value used for our
analysis.\footnote{
The analytic expression for $h_3$ in Eq.~(\ref{eq:h3}) is consistent with
Eq.~(7.3) given in Ref.~\cite{Gehrmann:2010ue}.} 

\begin{figure}[t]
      \vspace{0pt}
      \includegraphics[width=0.7\linewidth]{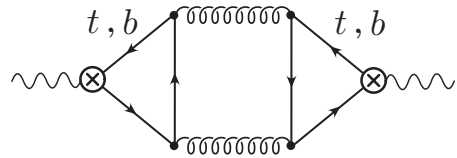}
      \caption{Two-loop singlet correction to the axial current. Its cuts
        contribute to the hard coefficient and nonsingular terms.}
      \label{fig:Haxial}
\end{figure}

The axial-vector hard functions $H_Q^{ua}$ and $H_Q^{da}$ are equal to $H_Q^v$
up to additional singlet corrections that enter at ${\cal O}(\alpha_s^{2})$ and
${\cal O}(\alpha_s^{3})$. The fact that the SCET hard functions have these
singlet corrections was discussed in Ref.~\cite{Stewart:2009yx}. At ${\cal
  O}(\alpha_s^{2})$ only the axial-vector current gets a singlet correction. It
arises from the axial-vector anomaly, from suitable cuts of the graph shown in
\fig{Haxial} where each axial current is connected to a triangle.
Summing over the light quarks $u,d,s,c$ gives a vanishing contribution from this
graph, but it does not vanish for heavy quarks due to the large bottom-top mass
splitting~\cite{Kniehl:1989qu}. Since for the $Q$s we consider top-pairs are
never produced, the required terms can be obtained in the limit $m_b/m_t\to 0$.
For the axial current the hard correction arises from the $b\bar b$ cut and
gives $H_Q^{ua}= H_Q^{da}=H_Q^{v}$, and $H_Q^{ba}=H_Q^{v} + H_Q^{\rm singlet}$,
where
\begin{align} \label{eq:Hsinglet}
 H_Q^{\mathrm{singlet}}(Q,r_t,\muh) 
 =\dfrac{1}{3}\left(\dfrac{\alpha_{s}(\muh)}{\pi}\right)^{2}\, I_{2}(r_t)\,.
\end{align}
Here $r_t=Q^2/(4 m_t^2)$ and the function $I_2(r_t)$ from
Ref.~\cite{Kniehl:1989qu} is given in \eq{I2fn}.  Throughout our analysis we use
$m_t=172\,$GeV.  $H_Q^{\rm singlet}$ is a percent level correction to the cross
section at the Z peak and hence is non-negligible at the level of precision of
our analysis. (The uncertainty in the top mass is numerically irrelevant.) At
${\cal O}(\alpha_s^3)$ the singlet corrections for vector currents are known
\cite{Baikov:2009bg}, but they are numerically tiny.  We therefore neglect the
${\cal O}(\alpha_s^3)$ vector current singlet corrections together with the
unknown ${\cal O}(\alpha_s^3)$ singlet corrections for the axial-vector current.
Likewise we do not account for ${\cal O}(\alpha_s^3)$ singlet corrections
to the nonsingular distributions discussed in Sec.~\ref{subsec:NSdist}.

The full anomalous dimension of $H_Q^{I}$ is known at three-loops, ${\cal
  O}(\alpha_s^3)$~\cite{vanNeerven:1985xr,Matsuura:1988sm,Moch:2005id}. It
contains the cusp anomalous dimension, responsible for the resummation of the
Sudakov double logarithms, and the non-cusp anomalous dimension.  To determine
the corresponding hard renormalization group factor $U_H$ at the orders
N${}^3$LL${}^\prime$ and N${}^3$LL we need the ${\cal O}(\alpha_s^4)$ cusp
anomalous dimension $\Gamma^{\rm cusp}_3$ which is still unknown and thus
represents a source of theory error in our analysis. We estimate the size of
$\Gamma^{\rm cusp}_3$ from the order [1/1] Pad\'e approximant in $\alpha_s$
built from the known lower order coefficients, which is within 13\% of the two
other possible Pad\'e approximants, [0/2] and [0/1]. For our theory error
analysis we assign 200\% uncertainty to this estimate and hence scan over values
in the range $\Gamma^{\rm cusp}_3=1553.06\pm 3016.12$.

The thrust jet function $J_\tau$ is the convolution of the two hemisphere jet
functions that describe collinear radiation in the $\hat {\rm t}$ and $-\hat
{\rm t}$ directions,
\begin{align} \label{eq:Jtau}
 J_{\tau}(s,\mu) &= \int{\rm d}s^\prime\, J(s^\prime,\mu)\, J(s-s^\prime,\mu)
  \nn\\
  &= \frac{1}{\mu^2} \sum_{n=-1}^\infty J_n[\alpha_s(\mu)] {\cal L}_n(s/\mu^2) 
 \,.
\end{align}
Here the coefficients $J_n$ are multiplied by the functions
\begin{align} \label{eq:Ln}
 & {\cal L}_{-1}(x) =\delta(x)\,,
 & {\cal L}_{n}(x) = \Big[ \frac{\ln^n x}{x}\Big]_+ \,,
\end{align}
where $n\ge 0$. Here ${\cal L}_{n\ge 0}(x)$ are the standard plus-functions, see
\eq{cLna_def}. At ${\cal O}(\alpha_s^3)$ only $J_{-1}(\alpha_s)$ through
$J_5(\alpha_s)$ are nonzero. The results are summarized in \eq{Jncoeff}.
In SCET the inclusive jet function is defined as
\begin{align}
& J(Q r^+,\mu) = \nonumber \\
& \frac{-1}{4\pi N_c Q } \, \textrm{Im} \bigg[ i\! \int \!\! d^4 x \, 
  e^{i r\cdot x} \,
\langle 0|{\rm T}\{ \bar\chi_{n}(0)\slash\!\!\!\bar{n}  \chi_n(x)\}|0 \rangle
  \bigg],
\end{align}
where the $\chi_n$ are quark fields multiplied by collinear Wilson lines.  The
hemisphere jet function has been computed at ${\cal
  O}(\alpha_s)$~\cite{Lunghi:2002ju,Bauer:2003pi} and ${\cal
  O}(\alpha_s^2)$~\cite{Becher:2006qw}. Its anomalous dimension is known at
three loops, and can be obtained from Ref.~\cite{Moch:2004pa}.  At the order
N${}^3$LL${}^\prime$ we need the ${\cal O}(\alpha_s^3)$ corrections to the jet
function. From the anomalous dimension we know the logarithmic terms, $J_{0}$ to
$J_{-5}$ in \eq{Jtau}, at three loops. In the non-logarithmic term $J_{-1}$ at
${\cal O}(\alpha_s^3)$ there is an unknown coefficient $j_3$ (which we define as
the constant non-logarithmic 3-loop coefficient in the position space hemisphere
jet function). We
estimate a range for $j_3$ from the largest value obtained from the three Pad\'e
approximations for the position space hemisphere jet function that one can
construct from the available results. This gives $j_3=0\pm 3000$ for the range
of variation in our theory error analysis.  We note that for the ${\cal
  O}(\alpha_s^3)$ coefficient $h_3$ the corresponding Pad\'e estimate $h_3=0\pm
10000$ covers the exact value given in \eq{h3}.

The renormalization group factors of the thrust jet function $U^\tau_J$ and
thrust soft function $U^\tau_S$ sum up large logs involving the jet and the soft
scales. The required cusp and non-cusp anomalous dimensions are fully known at
three-loops, but again there is dependence on the four-loop cusp anomalous
dimension $\Gamma^{\rm cusp}_3$.  This dependence is included when we scan
this parameter as described above in our description of the hard evolution.

The hadronic thrust soft function $S_\tau$ describes soft radiation between the
two jets.  It is defined by
\begin{align}
\label{eq:Sdef}
 S_{\tau}(k,\mu) 
&=
  \frac{1}{N_c} 
  \big\langle 0 \big|{\rm tr}\:
   \overline Y_{\!\bar n}^T Y_n\, \delta(k-i\widehat\partial) 
   Y_n^\dagger \overline Y_{\!\bar n}^* 
    \big| 0 \big\rangle  
 \,,
\end{align} 
where $Y_n=Y_n(0)$ and $\overline Y_{\!\bar n}=\overline Y_{\!\bar n}(0)$ are
defined below \eq{O1def}. The soft function factorizes into a partonic
perturbative part $S_{\tau}^{\rm part}$ and a nonperturbative part $S_\tau^{\rm
  mod}$, $S_\tau = S_{\tau}^{\rm part}\otimes S_{\tau}^{\rm mod}$, as discussed
in detail in Sec.~\ref{subsec:NSfac}. This factorization has already been used above
in \eqs{masterformula}{singular}.

At the partonic level the soft function is
\begin{align} \label{eq:Stau}
 S_{\tau}^{\rm part}(k,\mu) = \frac{1}{\mu} \sum_{n=-1}^\infty
 S_n[\alpha_s(\mu)] {\cal L}_n(k/\mu) \,, 
\end{align}
where $S_{-1}$ to $S_{5}$ are the only nonzero coefficients at ${\cal
  O}(\alpha_s^3)$, and ${\cal L}_n(x)$ is defined in \eq{Ln}.  Results for these
$S_k(\alpha_s)$ are summarized in \eq{Sncoeff}.  $S_\tau^{\rm pert}$ was
calculated at ${\cal O}(\alpha_s)$ in
Ref.~\cite{Schwartz:2007ib,Fleming:2007qr}.  At ${\cal O}(\alpha_s^2)$ the
non-logarithmic correction was determined in
Refs.~\cite{Becher:2008cf,Hoang:2008fs} using numerical output from
EVENT2~\cite{Catani:1996jh,Catani:1996vz}. The numerical constant that appears
in the non-logarithmic ${\cal O}(\alpha_s^2)$ term $S_{-1}$ is referred to as
$s_2$ (which is defined as the constant 2-loop coefficient in the logarithm of
the position space soft function). We use $s_2=-39.1\pm
2.5$~\cite{Hoang:2008fs}, and this uncertainty is taken into account in our
theory error analysis.\footnote{Note that in Ref.~\cite{Hoang:2008fs} our $s_2$
  was called $s_1$.} The anomalous dimension of the soft function is a linear
combination of the anomalous dimensions of the hard and jet functions which can
be obtained from the consistency conditions~\cite{Fleming:2007xt,Becher:2008cf}.
As for the jet function we need the ${\cal O}(\alpha_s^3)$ corrections to
$S_\tau^{\rm part}$.  From its anomalous dimension we know the logarithmic terms
at three loops, namely $S_{0}$ to $S_{5}$ in \eq{Stau}. The only unknown is the
${\cal O}(\alpha_s^3)$ non-logarithmic correction in $S_{-1}$, referred to as
$s_3$ (which is defined as the constant non-logarithmic term in the logarithm of
the position space hemisphere soft function).  Just like for the constant $j_3$
we estimate a value for $s_3$ from the largest value obtained from the three
possible Pad\'e approximations to the position space soft function that one can
construct from the available results.  This yields the range $s_3 =0 \pm 500$,
which we scan over in our theory error analysis.

As already mentioned, in Ref.~\cite{Becher:2008cf} an analytic expression for
the resummed singular thrust distribution was presented. Their derivation relies
on the Laplace transform of the jet and soft functions. In our analysis we have
derived the resummed cross section using two independent procedures, performing
all convolutions either in momentum space (as presented in \app{appendix}), or
in Fourier space.  These two approaches have been implemented in two independent
codes and we have checked that they give exactly the same results. We note that
the Fourier transform method is equivalent to the Laplace procedure used by
Becher and Schwartz in Ref.~\cite{Becher:2008cf} through a contour deformation,
and we find agreement with their quoted $\ntll$ formula including matrix
elements and anomalous dimensions. Furthermore, we also agree with their result
for the fixed-order singular terms up to ${\cal O}(\alpha_s^3)$.
 
In summary, the singular terms in the thrust factorization theorem are known at
\ntll order, up to the unknown constant $\Gamma^{\rm cusp}_3$.  The effect
of the cusp anomalous dimension at 4-loops is much smaller than one might
estimate, so for numerical purposes the cross section is known at this order.
The constants $s_3$ and $j_3$ only enter for our \ntllp order. For the singular
terms they predominantly affect the peak region with spread into the tail region
only due to RG evolution. Thus in the tail region the numerically dominant
\ntllp terms are all known. The uncertainties from $\Gamma^{\rm cusp}_3$,
$s_3$, and $j_3$ are discussed more explicitly in \sec{fit}

\subsection{$\Omega_1$ and Nonperturbative Corrections}
\label{subsec:NSfac}

In this section we discuss nonperturbative corrections to the thrust
distribution included in our analysis, as shown in Tab.~\ref{tab:orders}b (right
panel). We focus in particular on those associated to the first moment 
parameter~$\Omega_1$.  Our analysis includes the operator product expansion
(OPE) for the 
soft function in the tail region, and combining perturbative and nonperturbative
information to smoothly connect the peak and tail analyses. We also discuss our
treatment of nonperturbative corrections in the far-tail region, and for the
nonsingular terms in the cross section.

In the tail region where $k\sim Q\tau \gg \Lambda_{\rm QCD}$ we can perform an
operator product expansion of the soft function in \eq{Sdef}.  At tree level this
gives~\cite{Lee:2006fn,Lee:2006nr}
\begin{align} \label{eq:Sopetree}
  S_{\tau}(k,\mu) = \delta(k) - \delta'(k)\: 2 \bar\Omega_1 + \ldots \,.
\end{align}
where the nonperturbative matrix element $\bar\Omega_1$ is defined in the
$\overline {\rm MS}$ scheme as
\begin{align} \label{eq:O1bar}
 \bar\Omega_1 = \frac{1}{2N_c} \big\langle 0 \big| {\rm tr}\ \overline Y_{\bar
    n}^T(0) Y_n(0)\, i\widehat\partial\, Y_n^\dagger(0) \overline Y_{\bar
    n}^*(0) \big| 0 \big\rangle \,.
\end{align}
Dimensional analysis indicates that $\bar\Omega_1\sim \Lambda_{\rm QCD}$.  When
the OPE is performed beyond tree level we must add perturbative corrections at a
scale $\mu\simeq k$ to \eq{Sopetree}. The first operator in the OPE is the
identity, and its Wilson coefficient is the partonic soft function. Thus
$\delta(k) \to S_{\tau}^{\rm part}(k,\mu)$ when the matching of the leading
power operator is performed at any fixed order in perturbation theory. Here we
derive the analog for the Wilson coefficient of the $\bar\Omega_1$ matrix
element and prove that
\begin{align} \label{eq:Sope}
  S_{\tau}(k,\mu) = S_{\tau}^{\rm part}(k) -  \frac{{\rm d}S_{\tau}^{\rm part}(k)}{{\rm d}k}
    \: 2\bar\Omega_1 + \ldots \,.
\end{align} 
This result implies that the leading perturbative corrections that multiply the
power correction are determined by the partonic soft function to all orders in
perturbation theory. The proof of \eq{Sope} is given in \app{SoftOPE}. The
uniqueness of the leading power correction $\bar\Omega_1$ to all orders in the
perturbative matching can be derived following Ref.~\cite{Lee:2006fn}, and we
carry out an all orders matching computation to demonstrate that the Wilson
coefficient is $\df S_\tau^{\rm part}(k)/\df k$. At first order in
$\bar\Omega_1/k \ll 1$ \eq{Sope} shows that the perturbative corrections in the
OPE are consistent with a simple shift to $S_\tau(k-2\bar\Omega_1,\mu)$. This
type of shift was first observed in the effective coupling
model~\cite{Dokshitzer:1995qm}.

To smoothly connect the peak and tail regions we use a factorized soft
function~\cite{Hoang:2007vb,Korchemsky:2000kp,Ligeti:2008ac}
\begin{align}\label{eq:Sfact}
 S_{\tau}(k,\mu) = \int\!\! \df k'\ S_\tau^{\rm part}(k-k',\mu)\ S_{\tau}^{\rm
   mod}(k') \,,
\end{align}
where $S_\tau^{\rm part}$ is a fixed-order perturbative $\msbar$ expression for
the partonic soft function, and $S_{\tau}^{\rm mod}$ contains the
nonperturbative ingredients. In the tail region this expression can be expanded
for $k' \ll k$ and reduces to precisely the OPE in \eq{Sope} with the
identification 
\begin{align}\label{eq:O1barS1}
 2\bar\Omega_1 = \int {\rm d}k' \: k'\, S_{\tau}^{\rm mod}(k') \,,
\end{align}
and normalization condition $\int {\rm d}k'\: S_{\tau}^{\rm mod}(k')
=1$~\cite{Hoang:2007vb}. All moments of $S_{\tau}^{\rm mod}(k')$ exist so it has
an exponential tail, whereas the tail for $S_{\tau}^{\rm part}(k)$ is a power
law. In the peak region the full nonperturbative function $S_{\tau}^{\rm
  mod}(k)$ becomes relevant, and \eq{Sfact} provides a nonperturbative function
whose $\mu$ dependence satisfies the $\overline {\rm MS}$ renormalization group
equation for the soft function. In position space the convolution in \eq{Sfact}
is a simple product, making it obvious that \eq{Sfact} provides a completely
general parametrization of the nonperturbative corrections. The complete basis
of functions used to parameterize $S_{\tau}^{\rm mod}(k)$ in the peak region is
discussed in \sec{model}.

The expression in \eq{Sfact} also encodes higher order power corrections of type
\pci from \eq{pctype} through the moments $2^i\, \bar \Omega_i=\int {\rm d}k
\:\, k^i\, S_{\tau}^{\rm mod}(k)$, which for tree level matching in the OPE can
be identified as the matrix elements $\langle 0| {\rm tr}\, \overline Y_{\bar
  n}^T(0) Y_n(0)\, (i\widehat\partial)^i\, Y_n^\dagger(0) \overline Y_{\bar
  n}^*(0) | 0 \rangle/N_c$.  For $i\ge 2$ perturbative $\alpha_s$ corrections to
the soft function OPE would have to be treated in a manner similar to
\app{SoftOPE} to determine the proper Wilson coefficients, and whether
additional operators beyond the powers $(i\widehat\partial)^i$ start
contributing. The treatment of perturbative corrections to these higher order
nonperturbative corrections is beyond the level required for our analysis.

Using \eq{Sfact} the hadronic version of the singular factorization theorem
which involves $S_\tau$ immediately yields \eq{singular} and the first term in
\eq{masterformula}. The conversion of $S_{\tau}^{\rm part}(k)$ and
$\bar\Omega_1$ from $\overline {\rm MS}$ to a renormalon-free scheme is discussed
in \subsec{gap}.

Next we turn to the effect of $\Lambda_{\rm QCD}$ power corrections on the
nonsingular terms in the cross section in \eq{masterformula}. The form of these
power corrections can be constrained by factorization theorems for subleading
power corrections when $\tau \ll 1$, and by carrying out an OPE analysis for
power corrections to the moments of the thrust distribution. In the following we
consider both of these.

Based on the similarity of the analysis of power corrections to thrust with
those in $B\to X_s\gamma$~\cite{Lee:2004ja,Paz:2009ut}, the factorization
theorems for the nonsingular corrections involves subleading hard functions, jet
functions and soft functions. They have the generic structure
$H_Q^{(a,b)}(Q,\tau,x_i) \otimes J_\tau^{(a)}(s_j,x_i) \otimes
S_\tau^{(b)}(Q\tau,s_j/Q)$, where the $x_i$ and $s_j$ are various convolution
variables. Here $S_\tau^{(b)}$ includes the leading order soft function in
\eq{Sdef} as well as power suppressed soft functions. Neglecting nonperturbative
corrections the nonsingular cross section yields terms we refer to as kinematic
power corrections of type \pcii in \eq{pctype}. If we do not wish to sum large
logs in the nonsingular partonic terms, they can be treated in fixed-order
perturbation theory and determined from the full fixed-order computations. In
the tail region these $\tau$-suppressed terms grow and become much more
important than the $\Lambda_{\rm QCD}/Q$ power corrections of type \pciii from
subleading soft functions. In the transition to the far-tail region, near
$\tau=1/3$, they become just as important as the leading perturbative singular
terms.  In this region there are large cancellations between the singular and
nonsingular terms (shown below in \fig{sigcomponents}), and one must be careful
with the treatment of the nonsingular terms not to spoil this.

We require the nonsingular cross section terms to yield perturbative corrections
at leading power in $\lqcd$ that are consistent with the fixed-order results and
with multijet thresholds. Our treatment of power corrections in the nonsingular
terms is done in a manner consistent with these goals and with the OPE for the
first moment of the thrust distribution. To achieve this we use
\begin{align} \label{eq:ns1}
&
  \int\!\! {\rm d}k'\:
  \frac{\df\hat\sigma_{\rm ns}}{\df\tau}\Big(\tau-\frac{k'}{Q},
  \frac{\mu_{\rm ns}}{Q}\Big)
  S_\tau^{\rm  mod}(k') ,
\end{align}
where $\df\hat\sigma_{\rm ns}/\df\tau$ is the partonic nonsingular cross section
in fixed-order perturbation theory, whose determination we discuss in
\subsec{NSdist}. \eq{ns1} is independent of the renormalization scale
$\mu_{\rm ns}$ order by order in its series expansion in $\alpha_s(\mu_{\rm
  ns})$. The convolution with the same $S_{\tau}^{\rm mod}(k')$ as the singular
terms allows the perturbative corrections in $\df\hat\sigma_{s}/\df\tau +
\df\hat\sigma_{\rm ns}/\df\tau$ to smoothly recombine into the fixed-order
result in the far-tail region as required by the multijet thresholds. \eq{ns1}
yields the second term in \eq{masterformula}.  We will treat the conversion of
$\bar\Omega_1$ and $S_{\tau}^{\rm mod}$ to a renormalon-free scheme in the same
manner as for the singular cross section, which again for consistency requires a
perturbative subtraction for the partonic $\df\hat\sigma_{\rm ns}/\df\tau$
that we treat in Sec.~\ref{subsec:gap}.

Note that \eq{ns1} neglects the fact that not all of the $\tau$ dependence in
$\df\hat\sigma_{\rm ns}/\df\tau$ must necessarily be convoluted with
$S_\tau^{\rm mod}$. This causes a deviation which is $\sim \alpha_s \tau
\Lambda_{\rm QCD}/(Q\tau)$ and hence is at the same level as other power
corrections that we neglect. The largest uncertainty from our treatment of
$\df\hat\sigma_{\rm ns}/{\df\tau}$ arises from the fact that we do not sum
$\ln\tau$ terms, which would require anomalous dimensions for the subleading
soft and hard functions for these nonsingular terms.  These logs are most
important in the peak region, and less relevant in the tail region. The size of
missing higher order nonsingular terms such as log enhanced terms will be
estimated by varying the scale $\mu_{\rm ns}$.

Our setup is also consistent with the OPE for the first moment
of the thrust distribution. \eq{masterformula} yields
\begin{align} \label{eq:thrustmoment1}
  \int\!\! \df\tau\: \tau \frac{\df\sigma}{\df\tau} 
  = \int\!\!\df\tau\: \tau \Big[ \frac{\df\hat\sigma_s}{\df\hat\tau} 
  +  \frac{\df\hat\sigma_{\rm ns}}{\df\hat\tau} \Big] 
  + \sum_I \sigma_0^I \: \frac{2\bar\Omega_1}{Q} + \ldots\,,
\end{align}
where the ellipses denote ${\cal O}(\alpha_s\Lambda_{\rm QCD}/Q)$ and ${\cal
  O}(\Lambda_{\rm QCD}^2/Q^2)$ power corrections. In \app{MomentOPE} we
demonstrate that a direct OPE computation for the thrust moment also gives the
same result, and in particular involves precisely the same matrix element
$\bar\Omega_1$ at this order. The theoretical expression in \eq{masterformula}
simultaneously includes the proper matrix elements that encode power corrections
in the peak region, tail region, and for moments of the thrust distribution.
This implies a similar level of precision for the multijet region. Although
\eq{masterformula} does not encode all $\alpha_s\Lambda_{\rm QCD}/Q$
corrections, it turns out that the ones it does encode, involving $\Omega_1$,
numerically give an accurate description of the multijet cross section. (This is
visible in \fig{fartailfit} and will be discussed further in
Sec.~\ref{sec:fit}.) This agreement provides additional support for our
treatment of nonperturbative corrections in the nonsingular cross section in
\eq{ns1}.

\subsection{Nonsingular Distribution}
\label{subsec:NSdist}

The nonsingular partonic thrust distribution ${\rm d}\hat\sigma_{\rm ns}/{\rm
  d}\tau$ accounts for contributions in the thrust distribution that are
kinematically power suppressed. We write
\begin{align} \label{eq:signs}
  \frac{\df\hat\sigma_{\rm ns}}{\df\tau} (\tau) 
  &= \sum_I \sigma_0^I\ e^{-2\frac{\delta(R,\mu_s)}{Q}
\frac{\partial}{\partial \tau}} f^I\Big(\tau,\frac{\mu_{\rm ns}}{Q}\Big) \,,
\end{align} 
with the same superscript $I$ notation for different currents as in
\eq{singular}.  The presence of the $\delta(R,\mu_{S})$-dependent exponent
arises because $S^{\rm mod}_{\tau}$ depends on $\Omega_1$ and we use the same
renormalon-free definition for $\Omega_1$ as for the singular terms. In our
numerical evaluation we integrate by parts so that the $\partial/\partial\tau$
derivative acts on $S_{\tau}^{\rm mod}$ in \eq{masterformula}.  This exponent is
discussed in detail in \subsec{gap}.

In this section we discuss our determination of the functions $f^I$ in pure QCD
with massless quarks, while the generalization to include $m_b$ effects is
discussed in \subsec{bottom} and to include QED effects in \subsec{QED}. For
pure QCD there is one function $f^v_{\rm qcd}=f^{uv}_{\rm qcd}=f^{dv}_{\rm
  qcd}=f^{bv}_{\rm qcd}$ for the vector current, and functions $f^{ua}_{\rm
  qcd}=f^{da}_{\rm qcd}$, and $f^{ba}_{\rm qcd}$ for the axial-vector currents.
In general $f^I$ is the partonic fixed-order distribution where the singular
terms which are already contained in ${\rm d}\hat\sigma_{\rm s}/{\rm d}\tau$ are
subtracted to avoid double counting.  Setting the renormalization scale
$\mu_{ns}=Q$ they have the form
\begin{align} \label{eq:nonsingular}
f^v_{\rm qcd}(\tau,1)  &= 
\frac{\alpha_s}{2\pi}\,
f_1(\tau)+
\frac{\alpha_s^2}{(2\pi)^2}\,
f_2(\tau)+
\frac{\alpha_s^3}{(2\pi)^3}\,
f_3(\tau)+\ldots,
 \nn\\
f^{ua}_{\rm qcd}(\tau,1) &= f^{da}_{\rm qcd}(\tau,1) = f^v_{\rm qcd}(\tau,1) \,,
\nn\\
f^{ba}_{\rm qcd}(\tau,1) &= f^v_{\rm qcd}(\tau,1) + \frac{\alpha_s^2}{(2\pi)^2} 
 f_{\rm singlet}\big(\tau,r_t) \,,
\end{align}
where here $\alpha_s=\alpha_s(Q)$ and $r_t=Q^2/(4 m_t^2)$.  The required results
\begin{figure}[t!]
      \vspace{0pt}
      \includegraphics[width=0.995\linewidth]{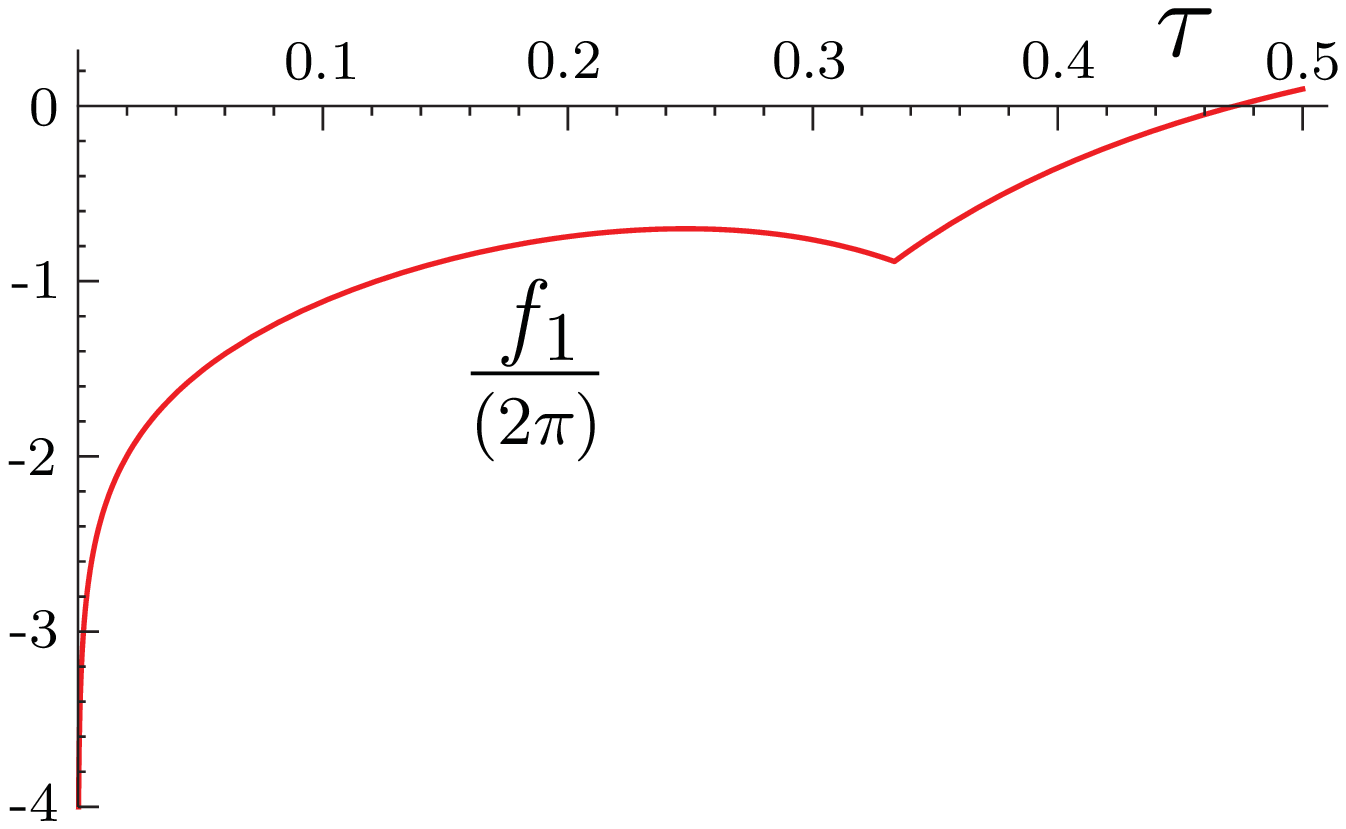}
\caption{${\cal O}(\alpha_s)$ nonsingular thrust distribution.}
\label{fig:nonsingalpha1}
\end{figure}
\begin{figure}[t!]
      \vspace{0pt}
      \includegraphics[width=0.995\linewidth]{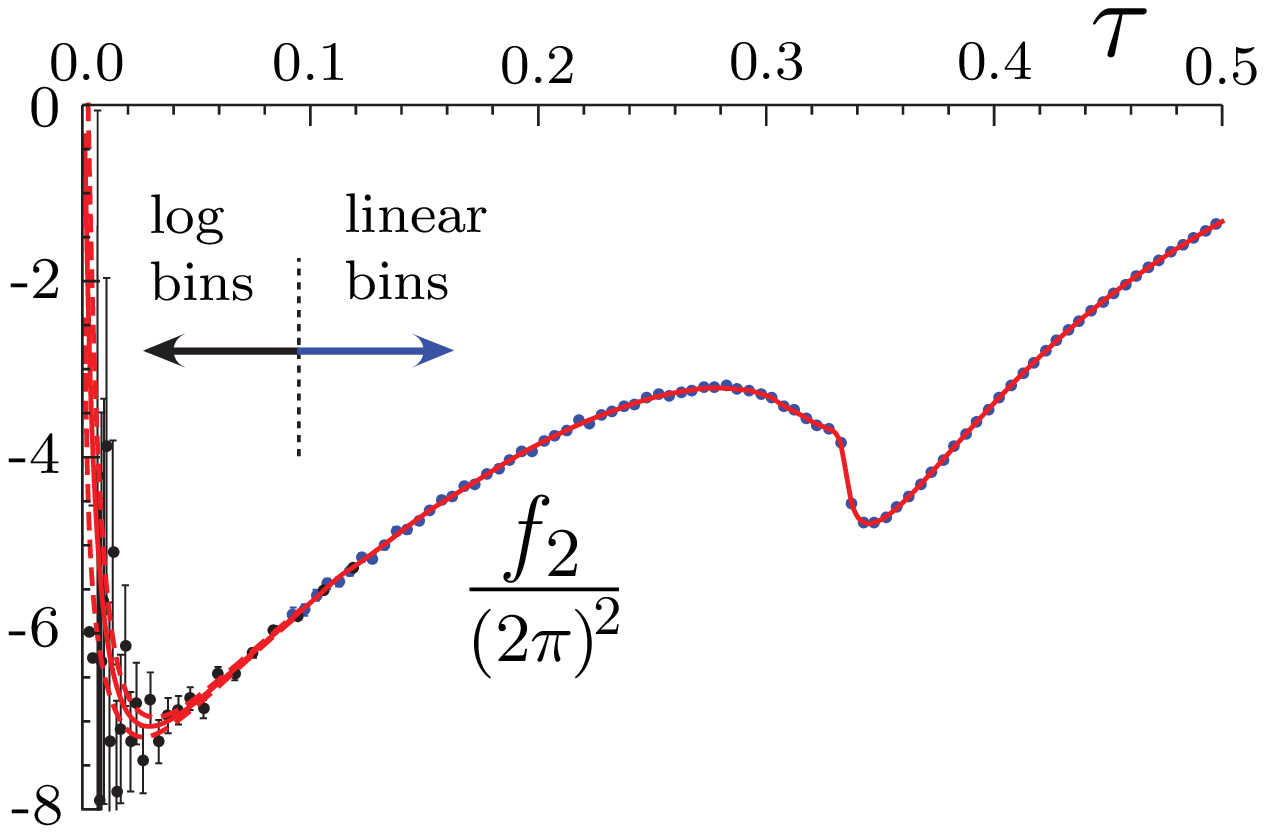}
\caption{${\cal O}(\alpha_s^2)$ nonsingular thrust distribution.}
\label{fig:nonsingalpha2}
\end{figure}
\begin{figure}[t!]
      \vspace{0pt}
      \includegraphics[width=0.995\linewidth]{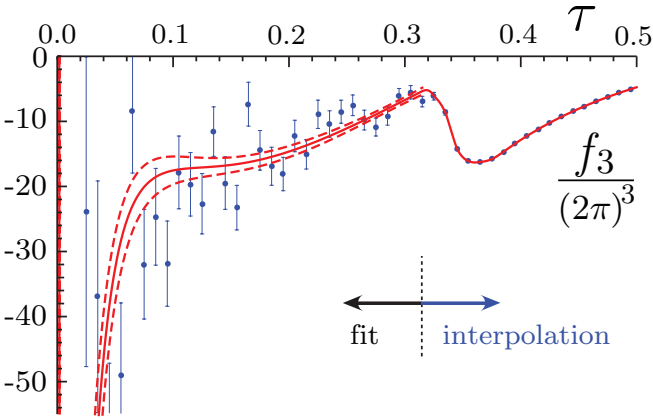}
      \caption{${\cal O}(\alpha_s^3)$ nonsingular thrust distribution. For
        simplicity we only show the data binned with $0.01$ bin size.}
      \label{fig:nonsingalpha3}
\end{figure}
for $f^I(\tau,\mu_{\rm ns}/Q)$ can be obtained by shifting $\alpha_s(Q)$ to
$\alpha_s(\mu_{\rm ns})$ using the fixed-order relation between these couplings
at ${\cal O}(\alpha_s^2)$.

The full ${\cal O}(\alpha_s)$ partonic thrust distribution has been known
analytically for a long time~\cite{Ellis:1980wv}. For the one-loop nonsingular
distribution it gives
\begin{align}
&f_{1}(\tau)\,=\,\dfrac{4}{3\,\tau\,(\tau-1)}\left[\left(-\,6\,\tau^{2}+6\,\tau-4\right)
\log\left(\frac{1}{\tau}-2\right)\right.\nonumber\\
&+9\,\tau^{3}-3\,\tau^{2}-9\,\tau+3\bigg]\theta\left(\dfrac{1}{3}-\tau\right)
+\frac{4}{3\,\tau }\,[\,3 +4 \log (\tau )\,] \,.
\end{align}
This result is plotted in \fig{nonsingalpha1}. The kink at $\tau=1/3$ appears
because the full one-loop distribution vanishes at this value with a nonzero
slope, and there is an 
exact cancellation between the fixed-order singular and nonsingular one-loop
expressions. For $\tau> 1/3$ the one-loop nonsingular distribution is precisely
the negative of the one-loop fixed-order singular distribution.

The ${\cal O}(\alpha_s^2)$ and ${\cal O}(\alpha_s^3)$ QCD distributions are
available in numeric form from the Fortran programs
EVENT2~\cite{Catani:1996jh,Catani:1996vz} and
EERAD3~\cite{GehrmannDeRidder:2007bj} (see also
Ref.~\cite{GehrmannDeRidder:2007hr,Weinzierl:2008iv,Weinzierl:2009ms}),
respectively. These programs are used to derive results for our $f_2(\tau)$ and
$f_3(\tau)$ nonsingular distributions in a manner discussed below. At ${\cal
  O}(\alpha_s^2)$ there is also the singlet correction $f_{\rm singlet}(\tau,r)$
for the axial-vector contribution arising from the large bottom-top mass
splitting. The three-parton quark-antiquark-gluon cut from \fig{Haxial}
contributes to the nonsingular distribution, and we have included this
contribution analytically~\cite{Hagiwara:1990dx}. The formula for $f_{\rm
  singlet}(\tau,r)$ is given in \eq{three-partons-cut}. There is also a
contribution from the four-parton cut. Its contribution to $f_{\rm
  singlet}(\tau,r)$ is unknown, but it is tiny for the total cross
section~\cite{Kniehl:1989qu} and can therefore be safely neglected.

At ${\cal O}(\alpha_s^2)$ we use linear binned EVENT2 results for $\tau>0.095$
and log-binning results for $\tau<0.095$ each obtained from runs with $10^{10}$
events and infrared cutoff $y_0=10^{-8}$. For $\tau>0.095$ (using a $0.005$ bin
size) the resulting statistical uncertainties in the nonsingular distribution
are always below the percent level and negligible and we can use an
interpolation of numerical tables for $f_2(\tau)$.  For $\tau < 0.095$ the
singular terms dominate the distribution which leads to large cancellations and
an enhancement of the statistical uncertainties. Here we use the ansatz
$f_2(\tau)= \sum_{i=0}^3 a_i\ln^i\tau +\tau\sum_{i=2}^3b_i\ln^i\tau$ and fit the
coefficients $a_i$ and $b_i$ to the EVENT2 output, including the constraint that
the integral over the full distribution reproduces the known ${\cal
  O}(\alpha_s^2)$ coefficient for the total cross section. The result has the
form $f_2(\tau)+\epsilon_2\, \delta_2(\tau)$, where $f_2$ represents the best
fit and $\delta_2$ is the $1$-sigma error function with all correlations
included. The term $\epsilon_2$ is a parameter which we vary during our
$\alpha_s$-$\Omega_1$ fit procedure to account for the error.  Here $f_2$ and
$\delta_2$ also depend on the coefficient $s_2$ in the partonic soft function
$S_\tau$ which is known only numerically. In \fig{nonsingalpha2} we plot the
EVENT2 data we used, along with our $f_2(\tau)$ with $s_2 = -39.1$. The dashed
curves show the result for $\epsilon_2=\pm 1$, with the region inbetween
corresponding to the $1$-sigma error band.

For the determination of $f_3$ at ${\cal O}(\alpha_s^3)$ we implement a similar
approach as for $f_2$, using results from EERAD3~\cite{GehrmannDeRidder:2007bj}
computed with $6\times 10^7$ events for the three leading color structures and
$10^7$ events for the three subleading ones, using an infrared cutoff
$y_0=10^{-5}$. We employ linearly binned results with $0.01$ bin size for
$\tau>0.315$ (keeping the statistical error below the percent level) and with
$0.005$ bin size for $\tau<0.315$.  For the fit for $\tau<0.315$ our ansatz
function has the form $f_3(\tau)= \sum_{i=1}^5 c_i\ln^i\tau$ and the result has
the form $f_3(\tau)+\epsilon_3 \,\delta_3(\tau)$, with $f_3$ being the best fit
and $\delta_3$ the $1$-sigma error function. The constant $\epsilon_3$ is the
analog of $\epsilon_2$ and is varied in the error analysis.  We note that $f_3$
and $\delta_3$ depend on the constant $s_2$ and on the constants $s_3$ and $j_3$
that account for the unknown non-logarithmic terms in the ${\cal O}(\alpha_s^3)$
soft and jet functions. This dependence is included in our error analysis. In
\fig{nonsingalpha3} we plot the EERAD3 data with bin size $0.01$, along with our
$f_3(\tau)$ with $s_2 = -39.1$, $h_3=8998.08$, $j_3=s_3=0$.  The dashed curves
show the result for $\epsilon_3=\pm 1$, with the region inbetween corresponding
to the $1$-sigma error band.

In our analysis we use the values $-1,0,1$ for $\epsilon_2$ and $\epsilon_3$ to
account for the numerical uncertainties of our fit functions in the small $\tau$
region.  The nonsingular partonic distribution depends on one common
renormalization scale $\mu_{\rm ns}$ which is varied in our theory error
analysis as given in Sec.~\ref{sec:profile}.

\subsection{Gap Formalism}
\label{subsec:gap}

The partonic soft function $S^{\rm part}_\tau(k)$ computed perturbatively in
$\msbar$ has an ${\cal O}(\Lambda_{\rm QCD})$ renormalon ambiguity. The same
renormalon is present in the partonic $\msbar$ thrust cross section with or
without resummation. This is associated with the fact that the partonic
threshold at $k=0$ in $S_{\tau}^{\rm part}(k)$ is not the same as the physical
hadronic threshold for the distribution of soft radiation that occurs in
$S_\tau(k)$. One can see this explicitly in the large-$\beta_0$ approximation,
where it is associated to a pole at $u=1/2$ in the Borel
transform~\cite{Hoang:2007vb}
\begin{align}
\label{borelS}
B\Big[S^{\rm part}_{\tau}(k,\mu)\Big]\Big(u\approx\frac{1}{2}\Big)
\,=\,\mu
\frac{16 C_F e^{-5/6}}{\pi \beta_0\,(u-\frac{1}{2})}\,
\frac{\partial}{\partial k}S^{\rm part}_{\tau}(k,\mu)
\,.
\end{align}
This result shows that $S^{\rm part}_\tau(k)$ in the $\msbar$ scheme suffers
from the renormalon ambiguity for all $k\ge 0$.  The $\msbar$ matrix element
$\bar \Omega_1$ defined in \eq{O1bar} also has an ${\cal O}(\Lambda_{\rm QCD})$
renormalon ambiguity. Together, the renormalon in this power correction and in
the perturbative series for $S^{\rm part}_\tau(k)$ combine to give a soft
function $S_\tau(k)$ that is free from this ${\cal O}(\Lambda_{\rm QCD})$
renormalon.  If left unsubtracted this renormalon ambiguity leads to numerical
instabilities in perturbative results for the thrust distribution and in the
large order dependence for the determination of the soft nonperturbative
function $S_\tau^{\rm mod}$. In this section we resolve this problem by
switching to a new scheme for $\Omega_1$. This scheme change induces
subtractions on ${\rm d}\sigma^{\rm part}/{\rm d}\tau$ that render it free of
this renormalon. We start by reviewing results from Ref.~\cite{Hoang:2007vb}.

Consider a class of soft nonperturbative functions with a gap parameter
$\Delta$, which only have support for $k\ge \Delta$, so $S_\tau^{\rm mod}(k)\to
S_\tau^{\rm mod}(k-2\Delta)$. Here the $\msbar$ moment relation in
\eq{O1barS1} becomes
\begin{align} \label{eq:O1barS1wD}
 2\Delta + \int\!\! {\rm d}k\: k\: S_{\tau}^{\rm mod}(k) = 2 \bar\Omega_1 \,,
\end{align}
where $\Delta$ accounts for the complete renormalon ambiguity contained in
$\bar\Omega_1$. 
We can now obtain a renormalon-free definition for $\Omega_1$ by splitting $\Delta$
into a nonperturbative component $\bar\Delta(R,\mu_S)$ that is free of the
${\cal O}(\Lambda_{\rm QCD})$ renormalon, and a suitably defined perturbative
series $\delta(R,\mu_S)$ that has the same renormalon ambiguity as
$\bar\Omega_1$. The parameter $\Delta$ is scheme and renormalization group
invariant, while $\bar\Delta$ and $\delta$ individually depend on the
subtraction scale $R$ and in general also on the soft scale $\mus$.  Writing
\begin{align}
\Delta  = \bar\Delta(R,\mus) + \delta(R,\mus)\,,
\end{align} 
the factorization of perturbative and nonperturbative components in \eq{Sfact}
becomes
\begin{align}\label{eq:SfactNEW}
 & S_{\tau}(k,\mus) 
  = \!\! 
  \int\! \df k'\ S_\tau^{\rm part}(k\!-\!k'\!-\!2\delta,\mus)\
  S_{\tau}^{\rm mod}(k'\!-\!2\bar\Delta) 
   \nn\\
 &= \int\! \df k'\: \Big[ e^{-2\delta\frac{\partial}{\partial k}} 
  S_\tau^{\rm part}(k\!-\!k',\mus)\Big]
  S_{\tau}^{\rm mod}(k'\!-\!2\bar\Delta) \,.
\end{align}
Here the exponential operator induces perturbative subtractions (in powers of
$\alpha_s(\mus)$) on the $\msbar$ series in $S_{\tau}^{\rm part}(k)$ that render
it free of the renormalon. This exponential modifies perturbative results for
the cross section in the manner we have shown earlier in
\eqs{singular}{signs}. The convolution of the nonsingular cross-section with
$S_{\tau}^{\rm mod}$ in \eq{ns1} now becomes
\begin{align} \label{eq:ns2}
&
  \int\!\! {\rm d}k'\:
  \frac{\df\hat\sigma_{\rm ns}}{\df\tau}\Big(\tau-\frac{k'}{Q},
  \frac{\mu_{\rm ns}}{Q}\Big)
  S_\tau^{\rm  mod}(k'-2\bar\Delta) \,.
\end{align}
Furthermore, with \eq{SfactNEW} the result in \eq{O1barS1wD} becomes
\begin{align} \label{eq:O1S1wD}
  2\bar\Delta(R,\mu_S) + \int\!\! {\rm d}k\: k\:
  S_{\tau}^{\rm mod}(k) = 2 \Omega_1(R,\mu_S) \,,
\end{align}
where here $\Omega_1(R,\mus)$ is renormalon-free. Combining
\eqs{O1S1wD}{O1barS1wD} we see that the scheme conversion formula from $\msbar$
to the new scheme is
\begin{align}
  \Omega_1(R,\mus) = \bar\Omega_1 - \delta(R,\mus) \,.
\end{align}
Thus, the precise scheme for $\Omega_1(R,\mus)$ is specified by the choice of
the subtraction series $\delta(R,\mus)$. Note that in general the gap parameter
$\bar\Delta$ is an additional nonperturbative parameter that can be determined
together with other parameters in the function $S_\tau^{\rm mod}$ from
fits to experimental data. However, in the tail region the power corrections are
dominated by a single parameter, $\Omega_1(R,\mu_S)$, which encodes the
dependence on $\bar\Delta$.

In Ref.~\cite{Hoang:2008fs} a convenient scheme for $\delta(R,\mus)$ was derived
(based on a scheme proposed in Ref.~\cite{Jain:2008gb}) where
\begin{align}
  \label{eq:gapdef}
\delta(R,\mu) = \frac{R}{2}\,e^{\gamma_E}\,
{\textstyle \frac{d}{d\ln(i x)}} \!
\left.\Big[\ln S_{\tau}(x,\mu)\,\Big]\right|_{x=(i R
e^{\gamma_E})^{-1}}.
\end{align}   
Here $S_\tau(x,\mu)$ is the position space partonic soft function, and the fact
that we write this result for $S_\tau$ rather than for the hemisphere soft
function explains the extra factor of $1/2$ relative to the formula in
Ref.~\cite{Hoang:2008fs}. The cutoff parameter $R$, having mass dimension $1$,
is a scale associated with the removal of the infrared renormalon.  To achieve
the proper cancellation of the renormalon in \eq{SfactNEW} one has to expand
$\delta(R,\mus)$ together with $S_\tau^{\rm part}(k,\mus)$ order by order in
$\alpha_s(\mus)$. The perturbative series for the subtraction is
\begin{align} \label{eq:deltaseries}
  \delta(R,\mus) = 
e^{\gamma_E} R \sum_{i=1}^\infty \alpha_s^i(\mus)\, \delta_i(R,\mus) \,,
\end{align}
where the $\delta_{i\ge 2}$ depend on both the adjoint Casmir $C_A=3$ and the
number of light flavors in combinations that are unrelated to the QCD beta
function. For five light flavors the one, two, and three-loop coefficients
are~\cite{Hoang:2008fs}
\begin{align} \label{eq:d123}
\delta_1(R,\mus) &= -0.848826 L_R \,, \nn \\
\delta_2(R,\mus) &= -0.156279 - 0.46663 L_R
 - 0.517864 L_R^2 \,, \nn \\
\delta_3(R,\mus)&= 
0.0756831 + 0.01545386\, s_2
- 0.622467 L_R
  \notag\\&\quad - 0.777219 L_R^2 - 0.421261 L_R^3 \,,
\end{align}
with $L_R=\ln(\mus/R)$.  We will refer to the scheme defined by \eq{gapdef} as
the R-gap scheme for $\Omega_1$. 

\begin{figure}[t!]
      \vspace{0pt}
      \includegraphics[width=\linewidth]{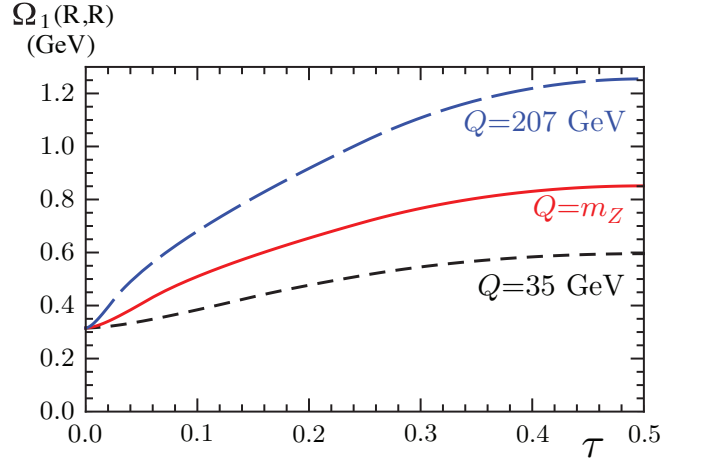}
      \caption{The running of $\Omega_1(R,R)$ with $R=R(\tau)$, plotted as a
        function of $\tau$ for $Q=35, 91.2, 207 \,{\rm GeV}$.}
      \label{fig:RunningO1} 
\end{figure}

From the power counting $\bar\Omega_1\sim \Lambda_{\rm QCD}$ one expects that a
cutoff $R\sim 1\,{\rm GeV}$ should be used, such that $\Omega_1\sim \Lambda_{\rm
  QCD}$ and perturbation theory in $\alpha_s(R)$ remains applicable. We refer to
this as the power counting criterion for $R$. Since in the tail region $\mus\sim
Q\tau\gg 1$~GeV the factors of $L_R$ in \eq{d123} are then large logs. To avoid
large logarithms in the subtractions $\delta_i(R,\mus)$ it is essential to
choose $R\sim\mus$, so that the subtraction scale $R$ is dependent on $\tau$
much like the soft scale $\mus$. We refer to this as the large-log criterion for
$R$.  To resolve the conflict between these two criteria, and sum the large logs
while keeping $\bar\Delta(R,\mus\sim R)$ renormalon-free, we make use of
R-evolution~\cite{Hoang:2008yj,Hoang:2009yr}. Formulas for the gap case were
given in Ref.~\cite{Hoang:2008fs} and are reviewed here. In this scheme
$\bar\Delta(R,\mu)$ satisfies an  R-RGE and $\mu$-RGE
\begin{align}\label{eq:RRGE}
  R \frac{d}{dR} \bar\Delta(R,R)
  &= - R \sum_{n=0}^\infty \gamma_n^R
  \Big(\frac{\alpha_s(R)}{4\pi}\Big)^{n+1}\,,
  \nn\\
  \mu \frac{d}{d\mu} \bar\Delta(R,\mu)
  &= 2 R e^{\gamma_E} \sum_{n=0}^\infty \Gamma^{\rm cusp}_n
  \Big(\frac{\alpha_s(\mu)}{4\pi}\Big)^{n+1} \,,
\end{align}
so that $\gamma_\Delta^\mu=-2 e^{\gamma_E} \Gamma^{\rm cusp}[\alpha_s]$.  For
five flavors the anomalous dimension coefficients up to three loops are
\begin{align}\label{eq:gammaR}
  \gamma_0^R &=0 \,,
  \qquad \gamma_1^R  = -43.954260 \,, \nn
  \\
  \gamma_2^R & = 1615.42228 +54.6195541\, s_2
\,,
\end{align}
while the coefficients $\Gamma_n^{\rm cusp}$ are given in \eq{betaCusp}.
The solution of \eq{RRGE} at N$^k$LL is
\begin{align} \label{eq:DeltaSoln}
 \bar\Delta(R,\mu) &= \bar\Delta(R_\Delta,\mu_\Delta)
  + R e^{\gamma_E} \omega[\Gamma^{\rm cusp},\mu,R]
  \nn\\
 & + R_\Delta e^{\gamma_E} \omega[\Gamma^{\rm cusp},R_\Delta,\mu_\Delta] 
  \nn\\
 & + \Lambda_{\rm QCD}^{(k)}
   \, D^{(k)}\big[\alpha_s(R),\alpha_s(R_\Delta)\big],
\end{align}
where the resummed $\omega[\Gamma^{\rm cusp},\mu,\mu_0]$ is given in \eq{w} and
the resummed $D^{(k)}[\alpha_s(R),\alpha_s(R_\Delta)]$ is given in \eq{Dk}. Both
the gap subtraction and R-evolution equations at ${\cal O}(\alpha_s^3)$ depend
on the constant $s_2$ which we vary within its errors in our theory error scan.
In our analysis, when quoting numerical results, we always use the parameter
$\bar\Delta(R_\Delta,\mu_\Delta)$ at the reference scales
$R_\Delta=\mu_\Delta=2$~GeV to satisfy the power counting criterion for $R$. We
then use \eq{DeltaSoln} to run up to the scale $R\sim \mus$ in order to satisfy
the large-log criterion. The precise $R$ value is a function of $\tau$, 
$R=R(\tau)$, and given in 
\sec{profile} with our discussion of the profile functions. The RGE solution for
$\bar\Delta(R,\mu_S)$ in \eq{DeltaSoln} yields a similar solution for a
running $\Omega_1(R,\mu_S)$ using \eq{O1S1wD}. In Fig.~\ref{fig:RunningO1} we
show the result for the running $\Omega_1(R,R)$ with the boundary value
$\Omega_1(R_\Delta,\mu_\Delta)=0.323\,{\rm GeV}$. The anomalous dimension and
$R(\tau)$ profile function cause an increase in the size of the power correction
for increasing $\tau$ and for increasing $Q$.

Note that our R-gap subtraction scheme differs from the subtractions in the
low-scale effective coupling model of Ref.~\cite{Dokshitzer:1995qm}, which is
not based on the factorization of the soft large angle radiation but on the
assumption that the ${\cal O}(\Lambda_{\rm QCD})$ renormalon ambiguity is
related entirely to the low-energy behavior of the strong coupling
$\alpha_s$. In the 
effective coupling model the subtractions involve logarithms, $\ln(\mu/\mu_I)$,
where $\mu$ is the usual renormalization scale of perturbation theory and
$\mu_I$ is the low-momentum subtraction scale, which is set to $\mu_I=2\,{\rm
  GeV}$. The scale $\mu_I$ plays a role very similar to the scale $R$ in the
R-gap scheme. These logarithms are 
the analogs of $L_R$ in \eq{d123} and, since $\mu\propto Q$
these logarithms also become large. In the effective
coupling model an appropriate resummation formalism for large logs in the
subtractions remains an open question.

\begin{figure}[t!]
      \vspace{0pt}
      \includegraphics[width=\linewidth]{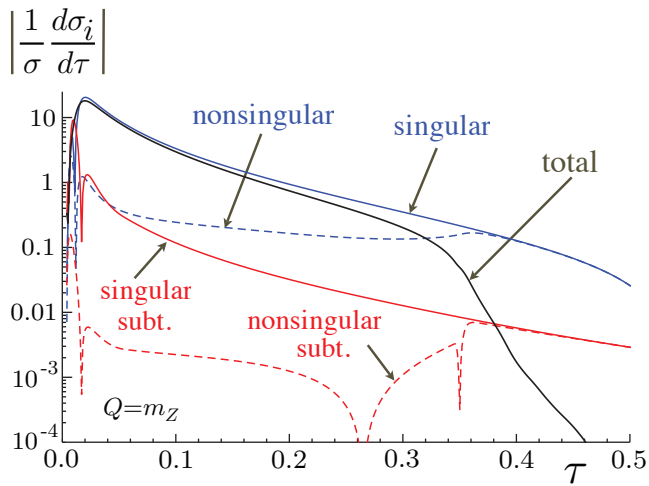}
      \caption{Components of the pure QCD cross section. Here $\Omega_1 = 0.35~{\rm
          GeV}$ and $\alpha_s(m_Z)=0.114$.}
      \label{fig:sigcomponents}
\end{figure}

In Fig.~\ref{fig:sigcomponents} we plot the absolute value of four components of
our cross section for our complete QCD result at $\ntllp$ order in the R-gap
scheme at $Q=m_Z$.  The cross section components include the singular terms
(solid blue), nonsingular terms (dashed blue), and separately the contributions
from terms that involve the subtraction coefficients $\delta_i$, for both
singular subtractions (solid red) and nonsingular subtractions (dashed red). The
sum of these four components gives the total cross section (solid black line).
The subtraction components are a small part of the cross section in the tail
region, but have an impact at the level of precision obtained in our
computation. In the peak region at very small $\tau$ the solid red singular
subtraction grows to be the same size as the solid blue singular term, and is
responsible for yielding a smooth positive definite total cross section. In both
the peak and tail regions the singular cross section dominates over the
nonsingular cross section. But as we approach the threshold $\tau\sim 1/3$ for
the far-tail region they appear with opposite signs and largely cancel. This is
clear from the figure where individually the singular and nonsingular lines are
larger than the total cross section in this region.  The same cancellation
occurs for the singular subtraction and nonsingular subtraction terms.

\subsection{Bottom Mass Effects}
\label{subsec:bottom}

In this work we implement bottom mass effects using the SCET factorization
framework for massive quarks~\cite{Fleming:2007qr,Fleming:2007xt}. We
include $m_b$-dependence in the kinematics, which starts at tree level, and in the
${\cal O}(\alpha_s)$ corrections in the partonic singular and nonsingular
distributions. We also account for the resummation of large logs and for
hadronization effects in the $m_b$-dependent terms. The mass dependent
factorization theorem implies that the renormalization group summation of
logarithms is identical to the one for massless quarks, and that all power
corrections of type 1 from \eq{pctype} are described by the nonperturbative soft
function $S_\tau^{\rm mod}$ already defined for the massless
case~\cite{Fleming:2007qr,Fleming:2007xt}.  We have already indicated this
with the convolution $\Delta \df\hat\sigma_b/\df\tau \otimes S_{\tau}^{\rm mod}$
shown in \eq{masterformula}. Since for the numerical analysis in this work we
fit to data in the tail region, where $Q\tau>6$~GeV, and since the massive quark
thrust factorization theorem implies for the soft scale $\mus\sim
Q\tau>6$~GeV, we do not have to account for any flavor threshold in the
renormalization group evolution and can always use $n_f=5$. The mass dependent
factorization theorem further implies that the only nontrivial $m_b$-dependence
in the singular distribution arises in the thrust jet function. Thus the jet
scale $\muj\sim Q\sqrt{\tau}\gg m_b$ for the region of our fit and we use the
$\overline{\rm MS}$ bottom mass $\overline m_b(\muj)$ to parameterize the $m_b$
corrections with $\bar m_b(\bar m_b)=4.2\,{\rm GeV}$ as our input value.  Using
the $\msbar$ mass rather than the pole mass avoids the appearance of large
higher order effects related to the ${\cal O}(\Lambda_{\rm QCD})$ pole mass
renormalon.

We implement the partonic bottom mass corrections as an additive term to the
massless partonic $\ntllp$ cross section. These corrections come from the
production of bottom quarks by the virtual $\gamma$ or $Z$,
\begin{align}
   \frac{\Delta \df\hat\sigma_b}{\df\tau}=
   \frac{\df\hat\sigma_b}{\df\tau}
   - \frac{\df\hat\sigma_b^{\bar m_b=0}}{\df\tau} \,,
\end{align}
where both $\df\hat\sigma_b/\df\tau$ and $\df\hat\sigma_b^{\bar m_b=0}/\df\tau$
are computed at NNLL. Because the effect of $\bar m_b\ne 0$ in
$\Delta\df\hat\sigma_b/\df\tau$ is expected to be a percent level correction to
the tail cross section, we anticipate that the NNLL level of precision suffices.
(This is also justified a posteriori by the relatively small effect of the $m_b$
corrections on our fit results.)

An important aspect in the discussion of the finite quark mass effects is in
which way hadron and heavy quark masses need to be accounted for in the
definition of thrust in Eq.~(\ref{Tdef}). In the experimental analyses Monte
Carlo generators are used to convert the actual measurements to the momentum
variables needed to compute $\tau$, and this conversion depends on hadron
masses. Since the final state stable hadrons are light, these effects are
related to nonperturbative physics.  Theoretically they are therefore implicitly
encoded within our fit of the nonperturbative corrections. In the partonic
theoretical computation light hadron masses are neglected in the computation of
the $\tau$ distribution, and it is consistent to set $\sum_i|\vec{p}_i|=Q$ in
the denominator of Eq.~(\ref{Tdef}).

To understand how the heavy quark masses affect the definition of thrust in
Eq.~(\ref{Tdef}) we recall that the partonic computation relies on the inclusive
nature of the measurements and that, experimentally, only light and long-lived
hadrons reach the detectors and are accounted for in the $\vec p_i$ momenta that
enter in computing $\tau$. Thus for heavy hadrons containing bottom (or charm)
quarks, it is their light and long-lived hadronic decay products that enter the
particle sum $\sum_i$. Due to energy conservation it is therefore necessary to
set $\sum_i|\vec{p}_i|=Q$ in the denominator of the thrust definition of
Eq.~(\ref{Tdef}) for the leading power partonic computations involving heavy
quarks. On the other hand, due to three-momentum conservation, it is consistent
to use the heavy quark three-momentum in the numerator of Eq.~(\ref{Tdef}) for
the partonic computations. This makes the partonic thrust computations involving
heavy quarks simple because we do not need to explicitly account for the heavy
quark decay in the calculations. Together with the relation
$\sum_i|\vec{p}_i|=Q$ in the denominator of Eq.~(\ref{Tdef}) this induces a
shift of the observed thrust distribution for $b$ quarks to larger $\tau$
values. Comparing to the massless quark situation, the small-$\tau$ endpoint is
moved from $0$ to
\begin{align} \label{eq:taubmin}
\tau_b^{\rm min} = 1-\sqrt{1-4 \bar m_b^2/Q^2}\,,
\end{align}
where here $\bar m_b =\bar m_b(\mu_J)$.  At tree level this shifts $\delta(\tau)
\to \delta (\tau-\tau_b^{\rm min})$.  For the fixed-order result at ${\cal
  O}(\alpha_s)$ the three-jet endpoint is moved from $1/3$ to $\tau_b^{3 {\rm
    jet}}=5/3-4/3\sqrt{1-3 \bar m_b^2/Q^2}$.  At leading order in $\bar
m_b^2/Q^2\ll 1$ we have~$\tau_b^{\rm min}=2 \bar m_b^2/Q^2+{\cal O}(\bar
m_b^4/Q^4)$ and $\tau_b^{3{\rm jet}}=1/3+2\bar m_b^2/Q^2+{\cal O}(\bar
m_b^4/Q^4)$, so the shift is the same for both endpoints.  Numerically, for
$\bar m_b=4.2\,$GeV and $Q=(35,\,91.2,\,207)\,$GeV, $\tau$ is shifted by $(0.029,
0.004, 0.0008)$.  This shift is also observed experimentally in flavor tagged
thrust analyses~\cite{Achard:2004sv,Buskulic:1995wb,Abbiendi:1999fs}.

In the following we outline the method used to compute the partonic
$\df\hat\sigma_b/\df\tau$. Like for the massless case the distribution is
divided into singular and nonsingular parts
\begin{align}
  \frac{\df\hat\sigma_b}{\df\tau} 
    &= \frac{\df\hat\sigma_b^{{\rm s}}}{\df\tau}
   + \frac{\df\hat\sigma_b^{{\rm ns}}}{\df\tau} \,.
\end{align}
The implementation of the bottom mass effects into the singular distribution
${\rm d}\hat\sigma_b^{\rm s}/{\rm d}\tau$ follows the NLL$^\prime$ analysis in
Ref.~\cite{Fleming:2007xt}, except that the evolution in the present work is
incorporated fully at NNLL order and that the exact partonic threshold at $\tau
= \tau_b^{\rm min}$ is accounted for,
\begin{align} \label{eq:bsing}
   \frac{\df\hat\sigma_b^{{\rm s}}}{\df\tau}
 &= Q\: \sigma_0^b\Big(\frac{\bar m_b}{Q}\Big)\,
 H^v_{Q}(Q,\mu_H)\, U_{H}(Q,\mu_H,\mu)\!\!
\int\!\mathrm{d}s\,\mathrm{d}s^\prime\, \nonumber \\
& \times J_{\tau b}(s^\prime,\bar m_b,\mu_J)\,
 U_{J}^{\tau}(s-s^\prime,\mu,\mu_J)\, \int\! \df k\: U_S^\tau(k,\mu,\mu_S) \nn \\
&  \times e^{-2\frac{\delta(R,\mu_s)}{Q}\frac{\partial}{\partial \tau}} \ 
S_{\tau}^{\rm part}\Big(\!Q\tau-Q\tau_b^{\rm min}-\frac{s}{Q}-k,\mu_S\!\Big)
\nn\\
 & + (\overline{\rm MS} \text{-pole mass scheme change terms}) \,,
\end{align}
where $\sigma_0^b(x)= \sigma_0^{bv} \sqrt{1-4x^2}(1+2x^2)+\sigma_0^{ba}
(1-4x^2)^{3/2}$. Perturbative bottom mass effects in the soft function start at
two loops, so at ${\cal O}(\alpha_s)$ $S_{\tau}^{\rm part}$ remains unchanged.
Since we have $\bar m_b/Q\ll 1$, only the thrust jet function for bottom quark
production, $J_{\tau b}(s,\bar m_b,\mu)$~\cite{Boos:2005qx}, receives
modifications from the finite $m_b$.  These modifications lead to a shift of the
partonic threshold of the thrust jet function from invariant mass $p^2=0$ to
$p^2=\bar m_b^2$. In $J_{\tau b}(s,\bar m_b,\mu)$ the variable $s=p^2-\bar
m_b^2$, and the presence of the mass leads to $\tau_b^{\rm min}$ in \eq{bsing}.
It also gives a more complicated form for ${\cal O}(\alpha_s)$ corrections in
$J_{\tau b}$ involving regular functions of $m_b^2/s$ in addition to singular
terms $\propto \delta(s)$ and $[\ln^k(s/\mu^2)/(s/\mu^2)]_+$ familiar from the
massless quark jet function.  More details and explicit formulae can be found in
Refs.~\cite{Fleming:2007qr,Fleming:2007xt}.

The bottom quark mass effects in the nonsingular partonic distribution 
${\rm d}\hat\sigma_b^{\rm ns}/{\rm d}\tau$ are more
complicated since finite mass effects at ${\cal O}(\alpha_s)$ differ for vector
and axial-vector current induced jet production,
\begin{align}
  \frac{\df\hat\sigma_b^{{\rm ns}}}{\df\tau} &=
  e^{-2\frac{\delta(R,\mu_s)}{Q}\frac{\partial}{\partial \tau}} 
  \bigg[ \sigma_0^{bv} 
   f^{v}_b\Big(\tau,\frac{\bar m_b}{Q},\frac{\mu_{\rm ns}}{Q}\Big) \nn\\
 &  + \sigma_0^{ba} 
   f^{a}_b\Big(\tau,\frac{\bar m_b}{Q},\frac{\mu_{\rm ns}}{Q}\Big) 
  \bigg] \nn\\
& + (\overline {\rm MS}\text{-pole mass scheme change terms}) \,.
\end{align}
In our analysis we implement analytic expressions for the nonsingular functions
$f_b^v$ and $f_b^a$.  The full ${\cal O}(\alpha_s)$ distributions for $\tau>0$
can be obtained from integrating the known double differential $b\bar b$ energy
distribution for vector-induced and axial-vector-induced production,
respectively, see e.g.\ Refs.~\cite{Ioffe:1978dc,Nilles:1980ic}.\footnote{
Results for bottom mass corrections at ${\cal O}(\alpha_s^2)$ were determined in
Refs.~\cite{Brandenburg:1997pu,Nason:1997nw,Rodrigo:1999qg}, but are not used in
our analysis due to the small effect the bottom mass corrections have in our fits. 
} 
The corresponding ${\cal O}(\alpha_s)$ coefficient of the $\delta (\tau-\tau_b^{\rm
  min})$ term is obtained using the one-loop correction to the total $b\bar b$
cross section as a constraint.  To determine the nonsingular distributions
$f_b^{v,a}$ we proceed much like for the massless case and subtract the singular
contributions expanded to ${\cal O}(\alpha_s)$ from the full ${\cal
  O}(\alpha_s)$ distribution. Further details and explicit formulas for
$f_b^{v,a}$ will be given in a future publication.

\subsection{QED Corrections}
\label{subsec:QED}

For the electroweak corrections to the thrust distribution we can distinguish
purely weak contributions and QED effects. The dominant effects to jet
production from the purely weak interactions are given by virtual one-loop
corrections to the hard Wilson coefficient $H_Q$. Since the contribution of the
singular thrust distribution ${\rm d}\hat\sigma_{\rm s}/{\rm d}\tau$ dominates
in the $\tau$ ranges we use for our fits as well as in the total cross section
$\sigma_{\rm tot}=\int {\rm d}\tau\, {\rm d}\sigma/{\rm d}\tau$ (see
Fig.~\ref{fig:sigcomponents}), the purely weak corrections largely drop out when
the distribution is normalized to the total cross section. This is consistent
with the explicit computations carried out in
Refs.~\cite{Denner:2009gx,Denner:2010ia}, where purely weak corrections were
found to be tiny. In our analysis we therefore neglect purely weak effects.

For QED corrections the situation is more complicated because, apart from
virtual effects which again largely cancel in the normalized distribution, one
also has corrections due to initial state and final state radiation. In
addition, one has to account for the fact that the treatment of QED effects in
the thrust measurements depends on the experiment. In general, using Monte Carlo
simulations, all experimental data were corrected to eliminate the effects from
initial state radiation. However, they differ concerning the treatment of final
state photon corrections, which were either eliminated or included in the
corrected data sets. In Sec.~\ref{sec:expedata} we review information on the
approach followed by the various experimental collaborations. Since many
experiments did not remove final state radiation, we have configured a version of
our code that adds final state photons and QED Sudakov effects, and does so on
an experiment by experiment basis. A parametric estimate of the potential impact
of these QED effects on the measurement of $\alpha_s(m_Z)$ is $\sim
-0.244\,\alpha_{\rm em}/(C_F \alpha_s)\sim -1\%$, where $0.244$ is the average of
the square 
of the electromagnetic charges for the five lightest flavors. 

We implement the leading set of QED corrections to all components that go into
the main factorization formula of Eq.~(\ref{eq:masterformula}) in the massless
quark limit counting $\alpha_{\rm em}\sim \alpha_s^2$ to make a correspondence
with Tab.~\ref{tab:orders} and remembering to include cross terms such as terms
of ${\cal O}(\alpha_{\rm em}\alpha_s)$.  Exceptions where QED corrections are
not included are the gap subtraction $\delta(R,\mus)$ and the R-evolution
equation for the gap parameter $\bar\Delta$. This is because QED effects do not
lead to ${\cal O}(\Lambda_{\rm QCD})$ infrared renormalon ambiguities. Most of
the required QED results can be obtained in a straightforward manner from
modifications of the known QCD corrections.

Our implementation of QED effects is briefly described as follows: For the
evolution of the strong coupling we included the ${\cal O}(\alpha_s^2\alpha_{\rm
  ew})$ corrections to the QCD beta function.  There are also effects from the
evolution of the QED coupling $\alpha_{\rm em}(\mu)$ which we define in the
$\overline{\rm MS}$ scheme.  In the beta function for the QED coupling
$\alpha_{\rm em}$ we account for the dominant ${\cal O}(\alpha_{\rm em}^2)$
and the next-to-leading ${\cal O}(\alpha_{\rm em}^2\alpha_s)$ contributions.
For the full singular partonic distribution which includes both QCD and QED
effects we have
\begin{align}
\label{eq:qedsingular}
 & \dfrac{\mathrm{d}\hat{\sigma}_{\rm s}}{\mathrm{d}\tau}
= Q\sum_I \sigma_{0}^{I}\, H^I_{Q}(Q,\mu_H)\, U_{H}^I(Q,\mu_H,\mu)\!\!
\int\!\mathrm{d}s\,\mathrm{d}s^\prime\, \nonumber \\
& \ \times J_{\tau}^I(s^\prime,\mu_J)\,
 U_{J}^{\tau I}(s-s^\prime,\mu,\mu_J)\, \int\! \df k\: U_S^{\tau I}(k,\mu,\mu_S) \nn \\
& \ \times e^{-2\frac{\delta(R,\mu_s)}{Q}\frac{\partial}{\partial \tau}} \ 
S_{\tau}^{{\rm part} I}\Big(\!Q\,\tau-\frac{s}{Q}-k,\mu_S\!\Big), 
\end{align}
where all factors now depend on the index $I$ due to their dependence on the
electromagnetic charges $q^{I=uv,ua} = +2/3$ and $q^{I=dv,da,bv,ba}=-1/3$.  We
implement one-loop QED corrections in the hard factor $H_Q^I$, the jet function
$J_\tau^I$ and the soft functions $S_\tau^{{\rm part} I}$. In the
renormalization group evolution factors $U_H^I$, $U_J^{\tau I}$, $U_S^{\tau I}$
we account for the one-loop QED corrections to the cusp and the non-cusp
anomalous dimensions.  In the nonsingular partonic distribution ${\rm
  d}\hat\sigma_{\rm ns}/{\rm d}\tau$ the same approach is employed. Here the
${\cal O}(\alpha_{\rm em})$ contributions that are analogous to the ${\cal
  O}(\alpha_s)$ terms are included by writing the full functions $f^I$ to be
used in \eq{signs} as
\begin{align}
  f^{I}\Big(\tau,\frac{\mu_{\rm ns}}{Q}\Big) 
 = f^{I}_{\rm qcd}\Big(\tau,\frac{\mu_{\rm ns}}{Q}\Big)  
 + \frac{3\,\alpha(\mu)\, (q^I)^2}{8\pi} f_1(\tau) \,.
\end{align}
The 1\% parametric estimate and
the moderate size of the QED effects we observe from the results of our fits
justifies the neglect of higher order QED effects. A more precise treatment of
QED effects is also not warranted given the level of accuracy of the Monte Carlo
generators used to correct the experimental data. More details and explicit
formulae for the QED corrections discussed here will be given in a future
publication. 

\section{Profile functions}
\label{sec:profile}

The factorization formula for the singular partonic distribution
${\mathrm{d}\hat{\sigma}_{\rm s}}/{\mathrm{d}\tau}$ in Eq.~(\ref{eq:singular})
is governed by three renormalization scales, the hard scale $\muh$, the jet
scale $\muj$, and the soft scale $\mus$. To avoid large logarithms appearing in
the corrections to the hard coefficient $H_Q$, the jet function $J_\tau$ and the
soft function $S_\tau$, the corresponding scales must satisfy the following
theoretical constraints in the three $\tau$ regions:
\begin{align}
\label{eq:profile123}
  &  \text{1) peak:} & &
    \muh\sim Q \,,\ \ 
    \muj\sim \sqrt{\Lambda_{\rm QCD}Q}\,,\  
    & \mus &\gtrsim \Lambda_{\rm QCD}\,, 
    \nn\\
  &  \text{2) tail:} & & 
    \muh\sim Q \,,\ \
    \muj\sim Q\sqrt{\tau} \,,\
    &\mus &\sim Q\tau \,,
    \nn\\
  & \text{3) far-tail:} & & 
    \muh=\muj = \mus \sim Q \,.
\end{align}
In the peak region, where the full nonperturbative function $S_\tau^{\rm mod}$
is relevant we have $\muh\gg \muj\gg\mus\sim \Lambda_{\rm QCD}$.  In the tail
region, where the nonperturbative effects are described by a series of moments
of the soft function we have $\muh\gg \muj\gg\mus\gg \Lambda_{\rm QCD}$. To
achieve an accurate theoretical description, we resum logarithms of $\tau$ in
the peak and tail region where $\muh$, $\muj$, and $\mus$ are separated.
Finally, in the far-tail region the partonic contributions are described by
usual fixed-order perturbation theory, and a proper treatment of fixed order
multijet thresholds requires that the three $\mu$ parameters merge close
together in the far-tail region and become equal at $\tau=0.5$, with
$\muh=\muj=\mus\sim Q \gg \Lambda_{\rm QCD}$. Thus in the far-tail region
logarithms of $\tau$ are not summed. The merging of $\muh$, $\muj$, and $\mus$
in the far-tail region is of key importance for the cancellations between
singular and nonsingular cross sections shown in Fig.~\ref{fig:sigcomponents}.
To obtain a continuous factorization formula that is applicable in all three
regions we use $\tau$-dependent renormalization scales, which we call {\it
  profile functions}.  These are smooth functions of $\tau$ which satisfy the
theoretical constraints listed in \eq{profile123}.

In addition to the three renormalization scales of the singular partonic
distribution there are two more scales, $\mu_{\rm ns}$ and $R$. The
renormalization scale $\mu_{\rm ns}$ governs the perturbative series for the
function $f^I$ contained in the nonsingular partonic distribution
${\mathrm{d}\hat{\sigma}_{\rm ns}}/{\mathrm{d}\tau}$.  The subtraction scale $R$
arises when we implement the gap subtractions in the R-gap scheme for $\Omega_1$
that remove the ${\cal O}(\Lambda_{\rm QCD})$ renormalon contained in the
$\msbar$ soft function. This $R$ also corresponds to the endpoint of the
R-evolution for $\bar\Delta(R,\mu_S)$ given in \eq{DeltaSoln}. To avoid large
logarithms in the subtraction $\delta(R,\mus)$, the value of $R$ needs to be
chosen of order $\mus$ and is therefore also a function of $\tau$.

The factorization formula~(\ref{eq:masterformula}) is formally invariant under
${\cal O}(1)$ changes of the profile function scales, that is, changes that do
not modify the hierarchies. The residual dependence on the choice of profile
functions constitutes one part of the theoretical uncertainties and provides a
method to estimate higher order perturbative corrections. We adopt a set of six
parameters that can be varied in our theory error analysis which encode this
residual freedom while still satisfying the constraints in \eq{profile123}.

For the profile function at the hard scale, we adopt
\begin{align}
\muh=&\,e_H\,Q,
\end{align}  
where $e_H$ is a free parameter which we vary from $1/2$ to $2$ in our theory
error analysis.

For the soft profile function we use the form
\begin{align}
\mus(\tau)=\left\{\begin{array}{ll}\mu_0+\frac{b}{2 t_1} \tau^2,\hskip1.6cm
    & 0\leq \tau\leq t_1,\\
  b \, \tau+d,   & t_1\leq \tau\leq  t_2,\\
  \muh-\frac{b}{1-2 t_2}(\frac{1}{2}-\tau)^2, &
    t_2\leq \tau\leq \frac{1}{2}.\end{array} \right.  
\end{align}
Here, $t_1$ and $t_2$ represent the borders between the peak, tail and far-tail
regions. $\mu_0$ is the value of $\mu_S$ at $\tau=0$. Since the thrust value
where the peak region ends and the tail region begins is $Q$ dependent, we
define the Q-independent parameter $n_1=t_1\,(Q/1\,\mbox{GeV})$. To ensure that
$\mus(\tau)$ is a smooth function, the quadratic and linear forms are joined by
demanding continuity of the function and its first derivative at $\tau=t_1$ and
$\tau=t_2$, which fixes
$b=2\,\big(\muh-\mu_0\big)/\big(t_2-t_1+\frac{1}{2}\big)$ and
$d=\big[\mu_0(t_2+\frac{1}{2})-\muh \,t_1\big]/\big(t_2-t_1+\frac{1}{2}\big)$.  In
our theory error analysis we vary the free parameters $n_1$, $t_2$ and $\mu_0$.

The profile function for the jet scale is determined by the natural relation
between the hard, jet, and soft scales
\begin{align}
\muj(\tau)=\bigg(1+
e_J\Big(\frac{1}{2}-\tau\Big)^2\bigg)\,\sqrt{\muh\,\mus(\tau)}. 
\end{align}
The term involving the free ${\cal O}(1)$-parameter $e_J$ implements a
modification to this relation and vanishes in the multijet region where
$\tau=1/2$. We use a variation of $e_J$ to include the effect of such
modifications in our estimation of the theoretical uncertainties.

For the subtraction scale $R$ the choice $R=\mus(\tau)$ ensures that we avoid
large logarithms in the $\delta_i(R,\mu_S)$ subtractions for the soft function.
In the peak region, however, it is convenient to deviate from this choice so
that the ${\cal O}(\alpha_s)$ subtraction term $\delta_1(R,\mus)= -0.848826
\ln(\mus/R)$ is nonzero (see Eq.~(\ref{eq:d123})). We therefore use the form
\begin{align}
R(\tau)=\left\{\begin{array}{ll}R_0+\mu_1\,\tau+\mu_2 \tau^2,\hskip0.65cm
    & 0\leq \tau\leq t_1,\\
  \mus(\tau), & t_1\leq \tau\leq 0.5 \,.
\end{array}\right.
\end{align} 
Imposing continuity of $R(\tau)$ and its first derivative at $\tau=t_1$ requires
$\mu_1=(2 d-2 R_0+b\,t_1)/t_1$ and $\mu_2=(- d+R_0)/t_1^2$.  The only free
parameter is $R_0$ which sets the value of $R$ at $\tau=0$. We take $R_0=0.85
\mu_0$ to give the one loop subtraction $\delta_1(R,\mu_S)$ the appropriate sign
to cancel the renormalon in the peak region. Since our focus here is not the
peak region, we leave further discussion of the appropriate choice of $R_0$ to a
future publication.

In our theory error analysis we vary $\mu_{\rm ns}$ to account for our ignorance
on the resummation of logarithms of $\tau$ in the nonsingular corrections.  We
account for the possibilities
\begin{align}
\mu_{\rm ns}(\tau)=\left\{\begin{array}{ll}
    \muh,\hskip3.08cm  & n_s=1,\\
    \muj(\tau), & n_s=0, \\
    \frac{1}{2}(\muj(\tau)+\mus(\tau)),   & n_s=-1.
\end{array}\right. 
\end{align}
We do not include the choice $\mu_{\rm ns}=\mus$ since we find that the choice
of this small scale enhances the nonsingular contributions in an unnatural way.

\begin{figure}[t!]
\includegraphics[scale=1.5]{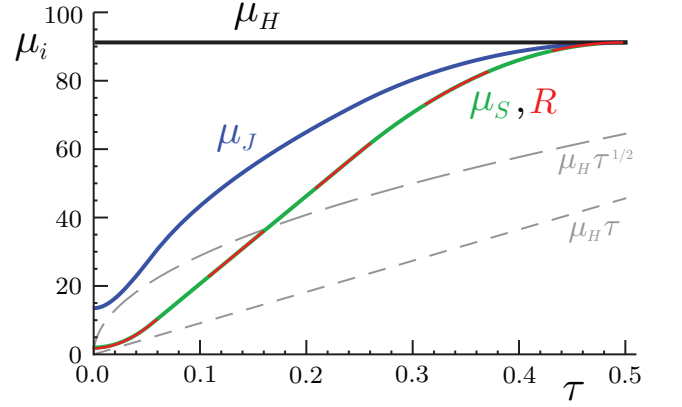}
\caption{Profile functions for the renormalization scales 
$\mu_J(\tau)$, $\mu_S(\tau)$, and subtraction scale $R(\tau)$ that appear in the
factorization theorem. Shown are results for the central parameter values at $Q=m_Z$. }
\label{fig:Profile}%
\end{figure}
In total, we have introduced six free parameters which we vary to account for
renormalization scale uncertainties. In our analysis we use the following
central values and variations: $\mu_0=2^{+0.5}_{-0.5}$~GeV, $n_1=5^{+3}_{-3}$,
$t_2=0.25^{+0.05}_{-0.05}$, $e_J=0^{+1}_{-1}$, $e_H=2^h$ with $h=0^{+1}_{-1}$
and $n_s=(-1,0,1)$.  In Fig.~\ref{fig:Profile} we show the form of the profile
functions for $Q=m_Z=91.2$~GeV and all profile parameters at their central
values.  The dashed lines represent the functions $Q\sqrt{\tau}$ and $Q\tau$
which were the central choices for $\muj(\tau)$ and $\mus(\tau)$ used in
Ref.~\cite{Becher:2008cf}, but which do not meet in the multijet region. In
order for our profile for $\mus(\tau)$ to join smoothly onto $\muh$ and
$\muj(\tau)$ it is necessary for $\mus(\tau)$ to have a slope $\sim 2 Q\tau$ in
the tail region. Since $\ln 2$ is not large our profiles sum the same
$\ln\tau$'s as with the choice in Ref.~\cite{Becher:2008cf}, but satisfy the
criteria necessary to treat the multijet thresholds.\footnote{ In
  Ref.~\cite{Davison:2008vx} where NLL resummation is achieved by
  exponentiation, the log resummation is turned off at a predefined threshold
  $\tau_{\rm max}$ with the log-R method~\cite{Catani:1992ua}. In this approach
  the transition to fixed order results in the multijet region differs from
  ours.}

\section{Nonperturbative Model Function}
\label{sec:model}

The soft nonperturbative function $S_\tau^{\rm mod}(k)$ parameterizes the
dominant nonperturbative hadronic effects in the thrust distribution. It
describes the hadronization contributions that arise from how soft hadrons that
are radiated in between the jets enter the thrust variable in Eq.~(\ref{Tdef}).
It is normalized, has the property $S_\tau(0)=0$, is positive definite and has
support for $k\ge 0$. To keep the representation of $S_\tau^{\rm mod}$ as much
as possible independent of a particular analytic parametrization we adopt the
approach of Ref.~\cite{Ligeti:2008ac} and write the soft nonperturbative
function as a linear combination of an infinite set of basis functions which can
in principle describe any function with the properties mentioned above. The
model function we use has the form
\begin{align} \label{eq:basis1}
S_\tau^{\rm mod}(k,\lambda,\{c_i\}) = 
 \frac{1}{\lambda}\, \bigg[  \sum_{n=0}^N
c_n \, f_n\bigg(\frac{k}{\lambda}\bigg)
\bigg]^2,
\end{align}
where the basis functions are~\cite{Ligeti:2008ac}
\begin{align} \label{eq:basis2}
f_n(z) & = 8 \sqrt{\frac{2 z^3 (2 n + 1)}{3}}\,\, e^{-2 z}\, P_n\Big(g(z)\Big),
\nonumber\\
g(z) & = \frac{2}{3}\Big( 3 - e^{-4 z}\,(3 + 12 z + 24 z^2 + 32 z^3)\Big) -1,
\end{align}
and $P_n$ are Legendre polynomials. For $\sum_i c_i^2=1$ the norm of $S^{\rm
  mod}_\tau(k)$ is unity, $\Omega_0=1$.  The choice of basis in
\eqs{basis1}{basis2} depends on specifying one dimensionful parameter $\lambda$
which is characteristic of the width of the soft function. With $N=\infty$ the
parameter $\lambda$ would be redundant, but in practice we truncate the sum in
\eq{basis1} at a finite $N$, and then $\lambda$ is effectively an additional
parameter of the model function.

In this work we fit to experimental thrust data in the tail region where the
predominant effects of the soft model function are described by its first moment
$\Omega_1(\lambda,\bar\Delta,\{c_i\})$.  As explained below, we use the second moment
$\Omega_2(\lambda,\bar\delta,\{c_i\})$ to validate our error analysis and
confirm the validity of neglecting this parameter in the fit. Since in the tail
region the exact form of the soft model function is not relevant, we take $N=2$
setting  $c_{n>2}=0$. Variations of the parameter $c_1$ are highly correlated
with variations of $\lambda$ and are hence not necessary for our purposes, so we
set $c_1=0$. For this case
\begin{align} \label{eq:Omega12}
  \Omega_1 &= \bar\Delta +
  \frac{\lambda}{2} \big[ c_0^2 + 0.201354 c_0 c_2 + 1.10031 c_2^2 \big]
  \,,\nn\\
  \Omega_2 &= \bar\Delta^2 + \bar\Delta \lambda  \big[ c_0^2 + 0.201354 c_0 c_2
  + 1.10031 c_2^2 \big] \nn\\
  & \ + \frac{\lambda^2}{4} \big[ 1.25 c_0^2 + 1.03621 c_0 c_2
  + 1.78859 c_2^2  \big]\,,
\end{align}
and the normalization condition $c_0^2 + c_2^2 =1$ can be used to eliminate
$c_0>0$.  Recall that in the soft model function in the factorization theorem we
must use $S_{\tau}^{\rm mod}\big(k-2\bar\Delta(R,\mu_S),\lambda,\{c_i\}\big)$
where $R=R(\tau)$ and $\mu_S=\mu_S(\tau)$ are determined by the profile
functions. When we quote numbers for parameters we use
$\bar\Delta=\bar\Delta(R_\Delta,\mu_\Delta)$ and hence $\Omega_{1,2} =
\Omega_{1,2}(R_\Delta,\mu_\Delta)$ with reference scales
$\mu_\Delta=R_\Delta=2\,{\rm GeV}$. The running between the scales $(R,\mu_S)$
and $(R_\Delta,\mu_\Delta)$ is determined by \eq{DeltaSoln}.

For our default fit in the tail region only the parameter $\Omega_1$ is
numerically relevant so without loss of generality we can take $c_0=1$, $c_2=0$,
and set $\bar\Delta(R_\Delta,\mu_\Delta)=0.05$~GeV.  In this case all higher
moments $\Omega_{n>1}$ are determined as a function of $\Omega_1$ and
$\bar\Delta$. For example we have $\Omega_2=(\bar\Delta^2-2\bar\Delta
\Omega_1+5\Omega_1^2)/4$ for the second moment.

In Sec.~\ref{sec:fit} we analyze the dependence of our fit results on
changes of $\Omega_2$.  Because $c_2$ has a rather strong correlation to
$\Omega_2$, we implement these $\Omega_2$ variations by using \eq{Omega12} and
setting $c_2$ to nonzero values. In this case we can hold $\Omega_1$ fixed by a
suitable choice of $\lambda$ for a given $c_2$. 

To obtain results from our code that do not include nonperturbative corrections
we can simply turn them off by setting $S_\tau^{\rm mod}(k)=\delta(k)$ and
$\bar\Delta=\delta =0$.

\section{Normalization and Convergence}
\label{sec:convergence}

The experimental data is normalized to the total number of events.  In our
prediction we therefore need to normalize the distribution to the total cross
section, i.e.\ we have to calculate $(1/\sigma) \df\sigma/\df\tau$. Since the
factorization formula in Eq.~(\ref{eq:masterformula}) is valid for all thrust
values we have the option to use either the integral of our $\df\sigma/\df\tau$
distribution for the norm, or the available fixed-order result for the total
hadronic cross section. 

The fixed-order total cross section is
\begin{align} \label{eq:sigtot}
&\sigma_{\rm tot}^{\rm FO}=\sum\sigma_0^I\,R^I\,,\quad
 R^{uv}=R^{dv}=R^{ua}=R^{da}=R_{{\rm Had}}\,,
 \nn\\
& R^{ba}=R_{{\rm Had}}+{R_A+}\dfrac{\alpha_s^2}{3\pi^2}\,I(r_t)\,,\quad
 R^{bv}=R_{{\rm Had}}{+R_V} .
\end{align}
Here $R_{\rm Had}$ is the pure QCD cross section for massless quarks, $R_{A,V}$
are mass corrections depending on $m_b/Q$, and $I(r_t)$ is the isosinglet
correction from the axial anomaly and large top-bottom mass
splitting~\cite{Kniehl:1989qu}.  Setting $\mu=Q$ the QCD cross
sections for massless quarks at three loops is
\begin{align} \label{eq:Rhad}
R_{{\rm Had}}\,&=\,1 + 0.3183099\,\alpha_s(Q) + 0.1427849\,\alpha_s^2(Q)
\nn\\
&\ -\,0.411757\,\alpha_s^3(Q)\,.
\end{align}
We refer to the review in Ref.~\cite{Chetyrkin:1996ia} for a discussion of the
fixed-order hadronic cross section. We note that the $\alpha_s$ series for the
fixed-order hadronic cross section exhibits an excellent and fast convergence.
At ${\cal O}(\alpha_s^3)$ the perturbative uncertainty is much below the
permille level and hence entirely negligible for the purpose of our analysis.

\begin{figure*}[t!]
\subfigure[]
{
\includegraphics[width=0.48\textwidth]{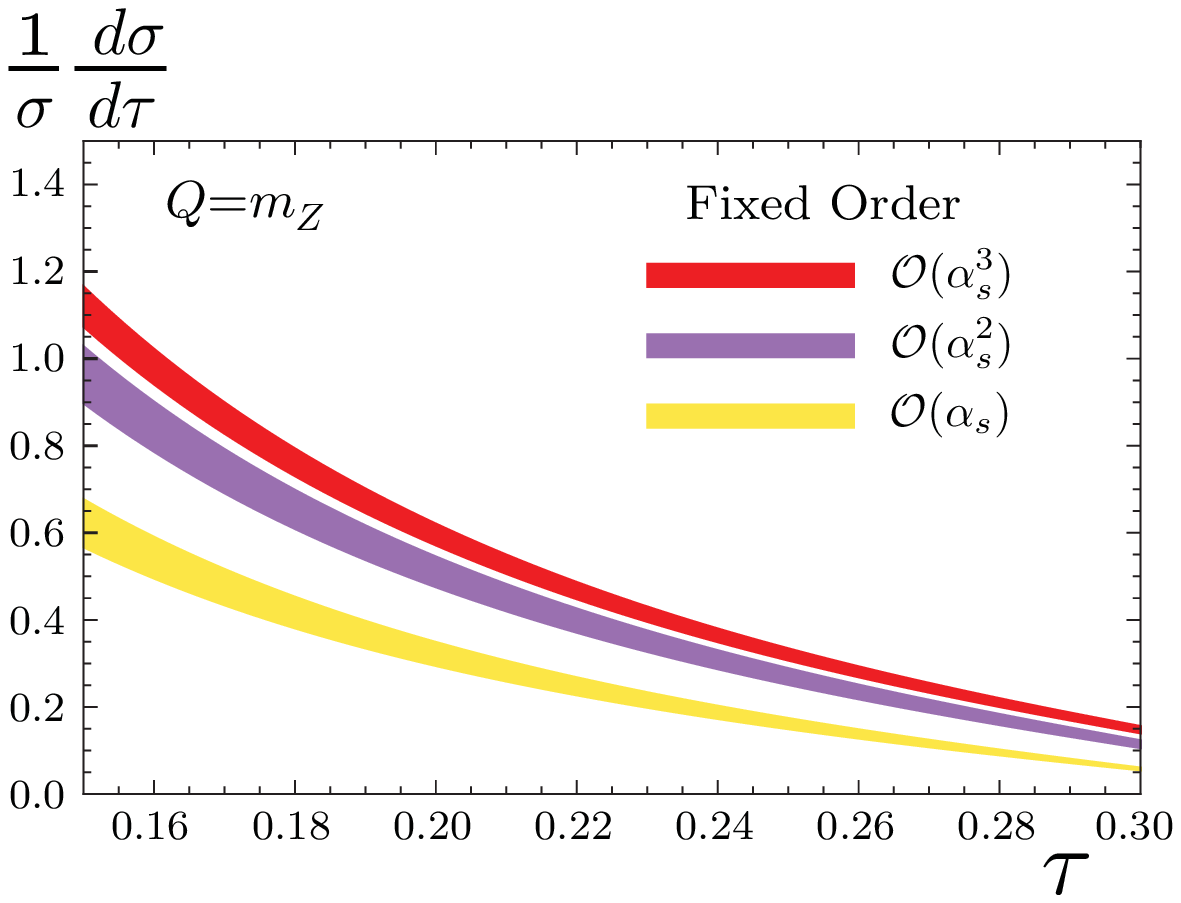}
\label{fig:tailbandFO}
}
\subfigure[]{
\includegraphics[width=0.48\textwidth]{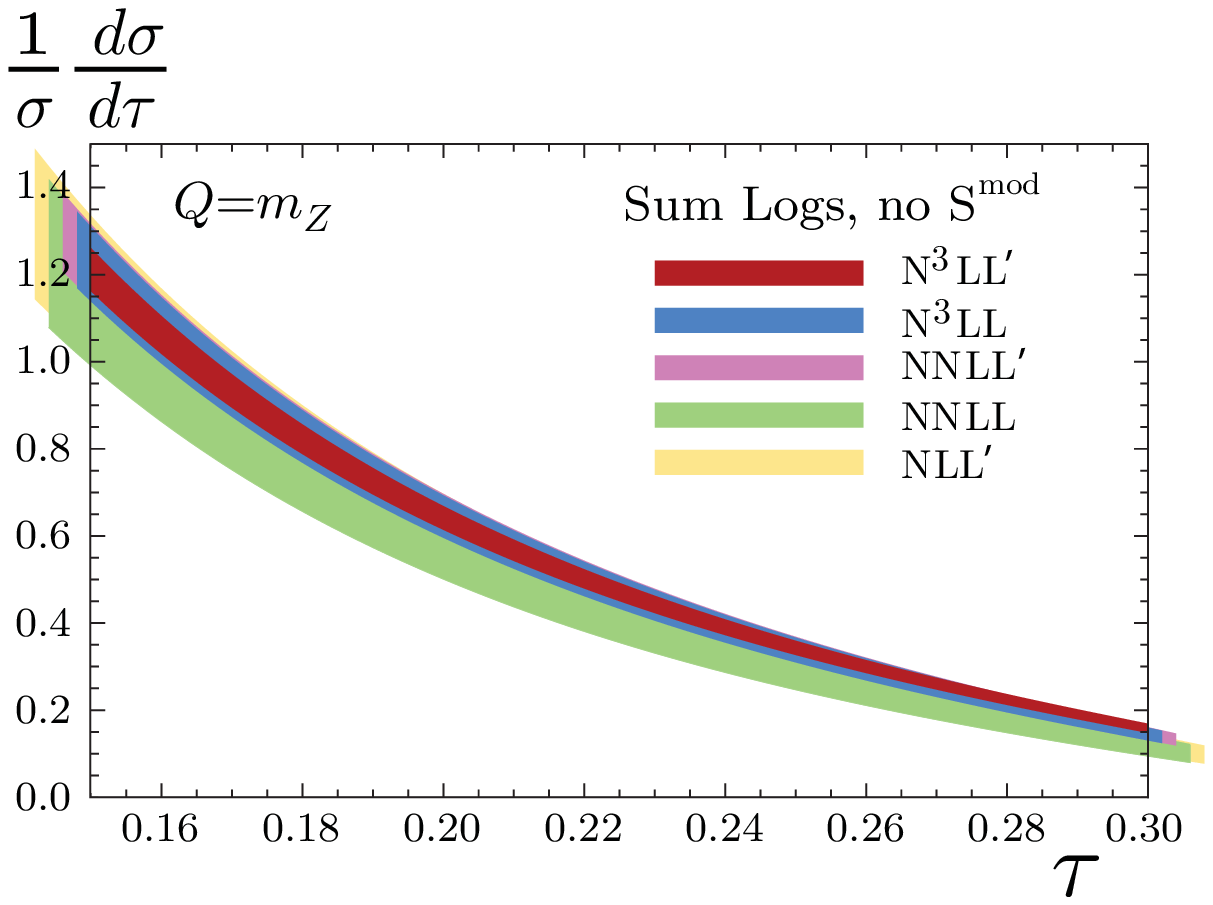}
\label{fig:tailbandnoF}
}
\subfigure[]{
\includegraphics[width=0.48\textwidth]{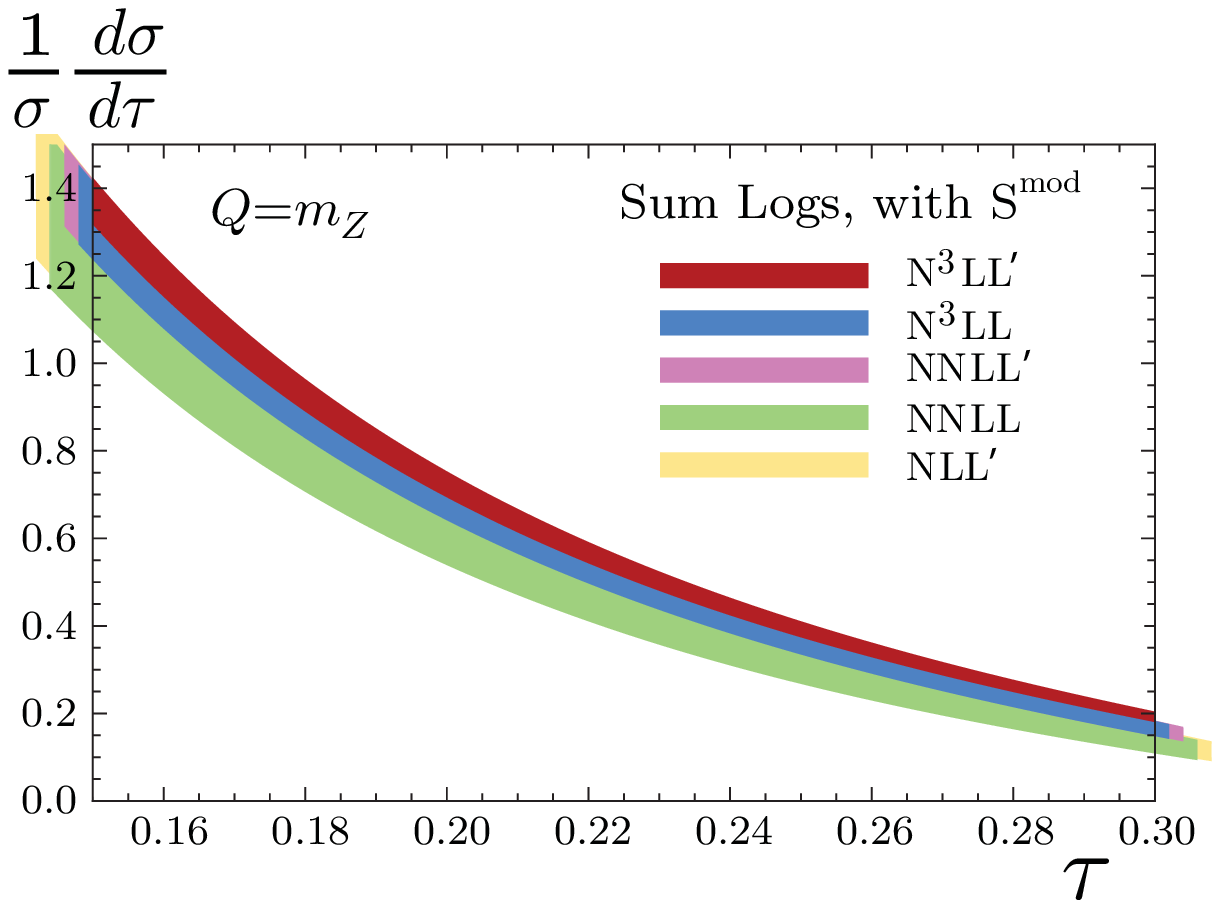}
\label{fig:tailbandnogap}
}
\subfigure[]{
\includegraphics[width=0.48\textwidth]{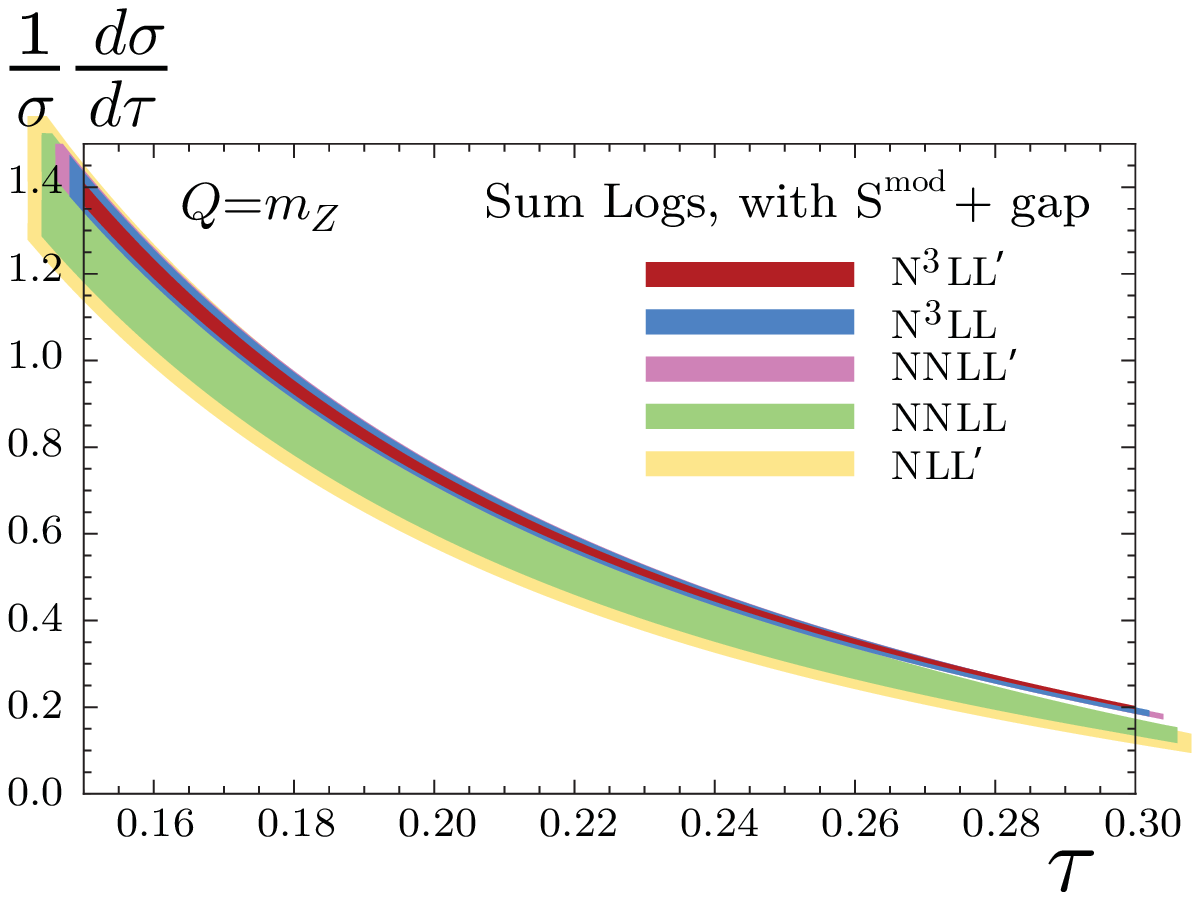}
\label{fig:tailbandwithgap}
}
\caption{Theory scan for errors in pure QCD with massless quarks. The panels are
  a) fixed-order, b) resummation with no nonperturbative function, c)
  resummation with a nonperturbative function using the $\msbar$ scheme for
  $\bar\Omega_1$ without renormalon subtraction, d) resummation with a
  nonperturbative function using the R-gap scheme for $\Omega_1$ with renormalon
  subtraction.}
\label{fig:4-plots}%
\end{figure*}

In the R-gap scheme in pure QCD, from a numerical analysis at $Q=m_Z$, we find
at N${}^3$LL${}^\prime$ order that the integrated norm of the thrust
distribution for the default setting of all theory parameters (see
Tab.~\ref{tab:theoryerr}) gives about $0.99\, \sigma_{\rm tot}^{\rm FO} $ at
${\cal O}(\alpha_s^3)$.  However we also find that the perturbative uncertainty
of the integrated norm (determined by the theory scan as described in
Sec.~\ref{sec:expedata}) is about $\pm 2.5\%$, which is substantially larger
than for the fixed-order cross section.  This larger uncertainty is due to the
perturbative errors of the thrust distribution in the peak region. At
N${}^3$LL${}^\prime$ order we therefore employ the fixed order cross section to
normalize the thrust distribution we use for the fits.

At the lower orders in the R-gap scheme ($\ntll$, NNLL$^\prime$, NNLL,
NLL$^\prime$) we find that the integrated norm for central theory parameters is
more appropriate since the order-by-order convergence to $\sigma_{\rm tot}^{\rm
  FO}$ is substantially slower than that of the rapid converging fixed-order QCD
result in \eq{Rhad}.  Again we find that the large perturbative uncertainties in
the peak region render the perturbative errors of the integrated norm larger
than those of the fixed-order norm. We therefore evaluate the integrated norms
at the lower orders with the theory parameters fixed at their default values
(see Tab.~\ref{tab:theoryerr}). This means that to estimate the theoretical
errors in our fits to experimental data at orders below $\ntllp$ in the R-gap
scheme, we vary the theory parameters only for the distribution and not for the
norm computation.  In the $\msbar$ scheme for $\bar\Omega_1$ we also adopt the
integrated norm at all orders. When we evaluate the thrust distribution with
log-resummation but without nonperturbative effects we use the same
normalization choices as for the R-gap scheme, which makes comparison to earlier
work in Sec.~\ref{sec:comparison} easier. For the situation where the
cross-section is evaluated at fixed-order, without resummation or
nonperturbative effects, we use the appropriate fixed order normalization at
each order.

As discussed in Sec.~\ref{sec:expedata}, to compare with the binned experimental
data we integrate our theoretical expression for the distribution
$(1/\sigma)(\df\sigma/\df\tau)$ over each bin $[\tau_1,\tau_2]$.  A potential
alternative is to use theoretical results for the cumulant
\begin{align}
  \Sigma(\tau) = \int_0^\tau \!\!\df\tau'\: \frac{1}{\sigma}\,
    \frac{\df\sigma}{\df\tau}(\tau') \,.
\end{align}
Here one sums large logs of $\tau$ rather than $\tau'$, and the SCET based
cumulant has $\tau$ dependent profiles, $\Sigma(\tau,\mu_i(\tau))$. The presence
of $\mu_i(\tau)$ implies that the derivative of the cumulant is not precisely
equal to the distribution,
\begin{align}
\label{dericumu1}
 \frac{\df}{\df\tau} \Sigma\big(\tau,\mu_i(\tau)\big) 
&=
\frac{1}{\sigma} \frac{\df\sigma}{\df\tau}\big(\tau, \mu_i(\tau)\big)
 \\
& + \frac{d\mu_i(\tau)}{d\tau} \frac{\partial}{\partial\mu_i} 
  \int_0^\tau d\tau' \frac{d\sigma}{d\tau'}\big(\tau',\mu_i(\tau)\big)
\,. \nn
\end{align} 
The difference coming from the second term in Eq.~(\ref{dericumu1}) can be
numerically important for certain observables. To test 
this we consider using for the cross-section integrated over the bin
$[\tau_1,\tau_2]$ the theoretical expression
\begin{align} \label{eq:sigdiff}
   \Sigma(\tau_2,\mu_i(\tilde\tau_2)) - \Sigma(\tau_1,\mu_i(\tilde \tau_1))\,,
\end{align}
and will examine several choices for $\tilde\tau_{1,2}$. 

One simple possibility is to use $\tilde \tau_1=\tau_1$ and
$\tilde\tau_2=\tau_2$, so that
$\Sigma(\tau_2,\mu_i(\tau_2))-\Sigma(\tau_1,\mu_i(\tau_1))$ is used.  In this
case there is a spurious contribution from outside the $[\tau_1,\tau_2]$ bin
associated to the second term in  Eq.~(\ref{dericumu1}), 
\begin{align}
 & \Sigma(\tau_1,\mu_i(\tau_2)) - \Sigma(\tau_1,\mu_i(\tau_1)) 
 \\
 & \simeq (\tau_2-\tau_1)
  \frac{d\mu_i(\tau_1)}{d\tau} \,\frac{\partial}{\partial\mu_i} 
  \int_0^{\tau_1} d\tau' \frac{d\sigma}{d\tau'}\big(\tau',\mu_i(\tau_1)\big)
  \,,\nn
\end{align}
where the $\simeq$ holds under the approximation that the derivative do not
change very much across the bin. With our default setup the deviation of this
simple choice for the cumulants from our integrated result for the distribution
is $2\%$ to $8\%$ for $\tau\in [0.1,0.3]$, bin-size $\tau_2-\tau_1=0.01$, and
$Q=91.2\,{\rm GeV}$.\footnote{For the profile functions used by Becher and
  Schwartz~\cite{Becher:2008cf}, discussed in section~\ref{sec:comparison}, this
  deviation has similar size but opposite sign.} In the far-tail region
$\tau_1\in [0.3,0.45]$, where the cross-section becomes small, the deviation
grows from $8\%$ to $1000\%$. These deviations are dominated by the spurious
contribution.  The size of the spurious contribution is not reduced by
increasing the bin-size to $\tau_2-\tau_1=0.05$, and is only mildly dependent on
$Q$.  Any choice in \eq{sigdiff} where $\tilde\tau_1\ne\tilde\tau_2$ leads to a
spurious contribution from $\tau\in [0,\tau_1]$.

If we instead use $\tilde \tau_1=\tilde \tau_2 = (\tau_1+\tau_2)/2$ then the
spurious contribution is identically zero. In this case the difference between
\eq{sigdiff} and our integrated thrust distribution is reduced to $0.5\%$ for
$\tau_1\in[0.1,0.3]$ and for $\tau_1\in [0.3,0.45]$ grows from $0.5\%$ to only
$20\%$.  Although dramatically reduced, the difference to the integrated
distribution in the far-tail region is
still quite sizeable. This discrepancy occurs because only for the distribution
$(1/\sigma)(\df\sigma/\df\tau)$ can the $\mu_i(\tau)$ profile functions be
constructed such that they satisfy exactly the criteria discussed in 
Sec.~\ref{sec:profile}.  Due to the above issues, and since the binned datasets
are intended as representations of the thrust distribution, we have determined
that our approach of integrating the thrust distribution is conceptually the
best.

In the rest of this section we discuss the perturbative behavior of the thrust
distribution in the tail region.  The values of the physical parameters used in
our numerical analysis are collected in Eq.~(\ref{eq:numerical_constants}). For
our lower order fits we always use the four-loop beta function in the running of
the strong coupling constant, as mentioned in the caption of
Tab.~\ref{tab:orders}.  Furthermore, we always consider five active flavors in
the running and do not implement bottom threshold corrections, since our lowest
scale in the profile functions (the soft scale $\mus$) is never smaller than
6~GeV in the tail where we perform our fit.

In Fig.~\ref{fig:4-plots} we display the normalized thrust distribution in the
tail thrust range $0.15<\tau<0.30$ at the different orders taking
$\alpha_s(m_Z)=0.114$ and $\Omega_1(R_\Delta,\mu_\Delta)=0.35~{\rm GeV}$ as
reference values, and neglecting $m_b$ and QED corrections.
We display the case $Q=m_Z$ where the experimental
measurements from LEP-I have the smallest statistical uncertainties. The
qualitative behavior of the results agrees with other c.m.\ energies. The
colored bands represent the theoretical errors of the predictions at the
respective orders, which have been determined by the scan method described in
Sec.~\ref{sec:expedata}.

In Fig.~\ref{fig:4-plots}a we show the ${\cal O}(\alpha_s)$ (light/yellow),
${\cal O}(\alpha_s^2)$ (medium/purple) and ${\cal O}(\alpha_s^3)$ (dark/red)
fixed-order thrust distributions without summation of large logarithms. The
common renormalization scale is chosen to be the hard scale $\muh$.  In the
fixed-order results the higher order corrections are quite large and our error
estimation obviously underestimates the theoretical uncertainty of the
fixed-order predictions. This panel including the error bands is very similar to
the analogous figures in Refs.~\cite{GehrmannDeRidder:2007hr}
and~\cite{Weinzierl:2009ms}. This emphasizes the importance of summing large
logarithms.

In Fig.~\ref{fig:4-plots}b the fully resummed thrust distributions at
NLL${}^\prime$ (yellow), NNLL (green), NNLL${}^\prime$ (purple), N${}^3$LL
(blue) and N${}^3$LL${}^\prime$ (red) order are shown, but without implementing
the soft nonperturbative function $S^{\rm mod}_\tau$ or the renormalon
subtractions related to the R-gap scheme. The yellow NLL${}^\prime$ error band
is mostly covered by the green NNLL order band, and similarly the purple
NNLL${}^\prime$ band is covered by the blue N${}^3$LL one.  Moreover the blue
N${}^3$LL band is within the purple NNLL band. Compared to the fixed-order
results, the improvement coming from the systematic summation of large
logarithms is obvious.  In particular we see that our way of estimating
theoretical uncertainties is appropriate once the logarithms are properly
summed. At N${}^3$LL and at N${}^3$LL${}^\prime$ order the relative
uncertainties of these resummed thrust distributions in the tail region $\tau \in
[0.1,0.3]$ are about $\pm\, 7.8$\% and $\pm\, 4.6$\%, respectively.

The results shown in Fig.~\ref{fig:4-plots}c are very similar to panel~b but now
include also the soft nonperturbative function $S^{\rm mod}_\tau$ without
renormalon subtractions, where $\bar\Omega_1$ is defined in the $\msbar$ scheme.
In the tail region the soft nonperturbative function leads to a horizontal shift
of the distribution towards larger thrust values by an amount $\delta\tau\propto
2\bar\Omega_1/Q$. This is clearly visible by comparing the values at $\tau=0.15$
where the curves intersect the $y$-axis.  Concerning the uncertainty bands and
the behavior of predictions at the different orders the results are very similar
to those in panel~b.

Finally, in Fig.~\ref{fig:4-plots}d we show the results with summation of large
logarithms including the soft model function with renormalon subtractions, where
$\Omega_1$ is defined in the R-gap scheme.  In the R-gap scheme the convergence
of perturbation theory is improved, and correspondingly the size of the
uncertainties from the same variation of the theory parameters is decreased. The
decrease of the uncertainties is clearly visible comparing the blue N${}^3$LL
and the red N${}^3$LL${}^\prime$ uncertainty bands with panel~c.  The relative
uncertainties of the thrust distribution at N${}^3$LL and at
N${}^3$LL${}^\prime$ order in the tail region $\tau\in [0.1,0.3]$ are now about
$\pm\, 3.4$\% and $\pm\, 1.7$\%, respectively. This improvement illustrates the
numerical impact of the ${\cal O}(\Lambda_{\rm QCD})$ renormalon contained in
the partonic soft function and shows the importance of eliminating the ${\cal
  O}(\Lambda_{\rm QCD})$ renormalon.

\section{Experimental data and fit procedure}
\label{sec:expedata}

Experimental data for thrust are available for various c.m.\ energies $Q$
between $14$ and $207$~GeV. In our analysis we fit the factorization
formula~(\ref{eq:masterformula}) in the tail region to extract $\alpha_s$ and
$\Omega_1$. As our default data set we use the thrust range $6/Q\le\tau\le
0.33$, and we only employ data from $Q\ge 35$~GeV. The lower boundary $6/Q$
removes data in the peak where higher order moments become important, while the
upper boundary of $0.33$ removes data in the far-tail region where the
$\alpha_s\Lambda_{\rm QCD}/Q$ power corrections become more important. We take
$Q\ge 35\,{\rm GeV}$ since a more sophisticated treatment of $b$ quark effects
is required at lower energies.  The data we use are from TASSO with $Q=\{35,
44\}$~GeV~\cite{Braunschweig:1990yd}, AMY with $Q=55.2$~GeV~\cite{Li:1989sn},
JADE with $Q=\{35, 44\}$~GeV~\cite{MovillaFernandez:1997fr}, SLC with
$Q=91.2$~GeV~\cite{Abe:1994mf}, L3 with $Q=\{41.4$, $55.3$, $65.4$, $75.7$,
$82.3$, $85.1$, $91.2$, $130.1$, $136.1$, $161.3$, $172.3$, $182.8$, $188.6$,
$194.4$, $200.0$, $206.2\}$~GeV~\cite{Achard:2004sv,Adeva:1992gv}, DELPHI with
$Q=\{45$, $66$, $76$, $89.5$, $91.2$, $93$, $133$, $161$, $172$, $183$, $189$,
$192$, $196$, $200$, $202$, $205$,
$207\}$~GeV~\cite{Abdallah:2003xz,Abreu:2000ck,Wicke:1999zz,Abreu:1999rc}, OPAL
with $Q=\{91$, $133$, $161$, $172$, $177$, $183$, $189$,
$197\}$~GeV~\cite{Abbiendi:2004qz,Ackerstaff:1997kk,Abbiendi:1999sx} and ALEPH
with $Q=\{91.2$, $133$, $161$, $172$, $183$, $189$, $200$,
$206\}$~GeV~\cite{Heister:2003aj}. (For TASSO and AMY we have separated
statistical and systematic errors using information from the experimental
papers.) All data is given in binned form, and we therefore integrate
Eq.~(\ref{eq:masterformula}) over the same set of bins to obtain appropriate
theory results for the fit to the experimental numbers. For the case that either
$\tau=6/Q$ or $\tau=0.33$ are located within an experimental bin, that bin is
excluded from the data set if more than half of it lies outside the chosen
interval. For the $Q>m_Z$ data we removed five bins with downward fluctuations
that were incompatible at the $> 10$-sigma level with the cross section implied
by neighboring data points and other experimental data in the same region. The
list of these bins is: L3 ($136.1$~GeV): $[0.25,0.275]$, DELPHI ($161$~GeV):
$[0.32,0.40]$, DELPHI ($183$~GeV): $[0.08,0.09]$, DELPHI ($196$~GeV):
$[0.16,0.18]$, ALEPH ($200$~GeV): $[0.16,0.20]$.\footnote{Four out of these bins
  lie in our $\tau\in [6/Q,0.33]$ default fit range. If they are included in the
  default dataset then for our final fit in \eq{asO1finalcor} the $\chi^2=439$
  increases by $+81$ and the central fit values show a slight decrease to
  $\alpha_s(m_Z)=0.1132$ and a slight increase to $\Omega_1=0.336\,{\rm GeV}$.}
Our default global data set contains a total of $487$ bins.  In the numerical
analysis performed in Sec.~\ref{sec:fit} we also examine alternative global data
sets with different $\tau$-ranges.

The data sets were corrected by the experiments to eliminate the QED effects from
initial state radiation using bin-by-bin correction factors determined from
Monte Carlo simulations. The primary aim of these corrections was to eliminate
the effective reduction of the c.m.\ energy available for the production of the
hadronic final state. In addition, in the data sets from the TASSO, L3 and ALEPH
collaborations the effects from final state radiation of photons were
eliminated, while they have been fully included in the data sets from the AMY,
JADE, SLC, DELPHI and OPAL collaborations. It should also be noted that the
approaches used by the experiments to treat photon radiation were dependent on
the c.m.\ energy $Q$. For the $Q=m_Z$ data any radiation of initial state
photons is naturally suppressed as the effective c.m.\ energy for the hadronic
final state gets shifted away from the Z pole. Therefore no specific photon cuts
were applied for the $Q=m_Z$ data prior to the application of the bin-by-bin
correction factors. For the data taken off the Z pole for either $Q<m_Z$ or
$Q>m_Z$ the effects of initial state radiation are substantial and explicit hard
photon cuts were applied in the data taking prior to the application of the
bin-by-bin correction procedure. We therefore consider the $Q=m_Z$ data sets as
more reliable concerning the treatment of QED effects.

Since the size of the QED effects we find in the measurements of $\alpha_s$ and
the soft function moment $\Omega_1$ is comparable to the experimental
uncertainties (see the results and discussions in Sec.~\ref{sec:fit}), a less
Monte Carlo dependent treatment of QED radiation would be certainly warranted.
(See Ref.~\cite{Denner:2010ia} for a recent discussion of QED radiation based on
full one-loop matrix elements.) However, given that the impact of QED
corrections we find for $\alpha_s$ and $\Omega_1$ is still smaller than the
current theoretical uncertainties from QCD, we use for our default numerical
analysis the theory code with QED effects switched on, as described in
Sec.~\ref{subsec:QED}. In Sec.~\ref{sec:fit} we also present results when QED
corrections are neglected for all data sets, and for the case when they are
neglected only for the TASSO, L3 and ALEPH data sets.

For the fitting procedure we use a $\chi^2$-analysis, where we combine the
statistical and the systematic experimental errors into the correlation matrix.
We treat the statistical errors of all bins as independent. The systematic
errors of the bins are correlated, but - unfortunately - practically no
information on the correlation is given in the experimental publications. We
therefore have to rely on a correlation model. For our analysis we assume as the
default that within one thrust data set, i.e.\ for the set of thrust bins
obtained by one experiment at one $Q$ value, the systematic experimental errors
are correlated in the minimal overlap model used by the LEP QCD working
group~\cite{Heister:2003aj,Abbiendi:2004qz}.  In the minimal overlap model the
off-diagonal entries of the experimental covariance matrix for the bins $i$ and
$j$ within one data set are equal to $[{\rm min}(\Delta_i^{\rm
  sys},\Delta_j^{\rm sys})]^2$, where $\Delta_{i,j}^{\rm sys}$ are the
systematic errors of the bins $i$ and $j$. This model implies a positive
correlation of systematic uncertainties within each thrust data set.  As a cross
check that our default correlation model does not introduce a strong bias we
also carry out fits were the experimental systematic errors are assumed to be
uncorrelated. Details are given in Sec.~\ref{sec:fit}.

\begin{table}[t] 
\begin{tabular}{ccc}
   parameter\ & \ default value\ & \ range of values \ \\
  \hline 
  $\mu_0$ & 2\,{\rm GeV} & 1.5 to 2.5\, {\rm GeV}\\ 
  $n_1$ & 5 & 2 to 8\\
  $t_2$ & 0.25 & 0.20 to 0.30\\
  $e_J$ & 0 & -1,0,1\\
  $e_H$ & 1 & 0.5 to 2.0\\
  $n_s$ & 0 & -1,0,1\\
\hline
  $s_2$ & -39.1 & $-36.6$ to $-41.6$ \\
  $\Gamma^{\rm cusp}_3$ & 1553.06 & $-1553.06$ to $+4569.18$ \\
  $j_3$ & 0 & $-3000$ to $+3000$ \\
  $s_3$ & 0 & $-500$ to $+500$ \\
\hline
  $\epsilon_2$ & 0 & -1,0,1 \\
  $\epsilon_3$ & 0 & -1,0,1 \\ 
\end{tabular}
%}
\caption{Theory parameters relevant for estimating the theory uncertainty, their
default values and range of values used for the theory scan during the fit
procedure.}
\label{tab:theoryerr}
\end{table}
\begin{figure}[t!]
\subfigure[]{
\includegraphics[width=0.95\columnwidth]{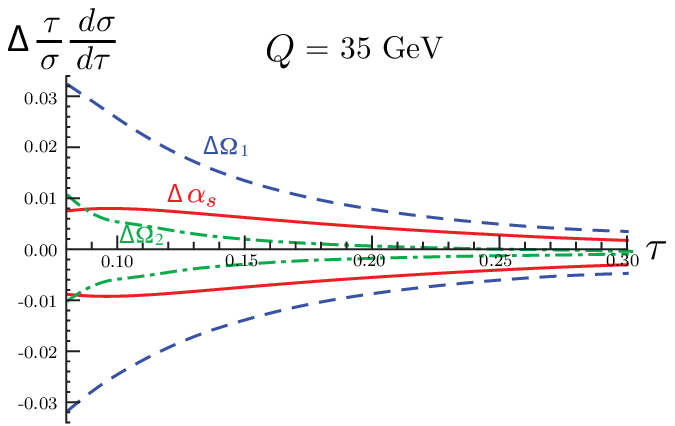}
\hspace{-0.3cm}
}\\
\subfigure[]{
\includegraphics[width=0.95\columnwidth]{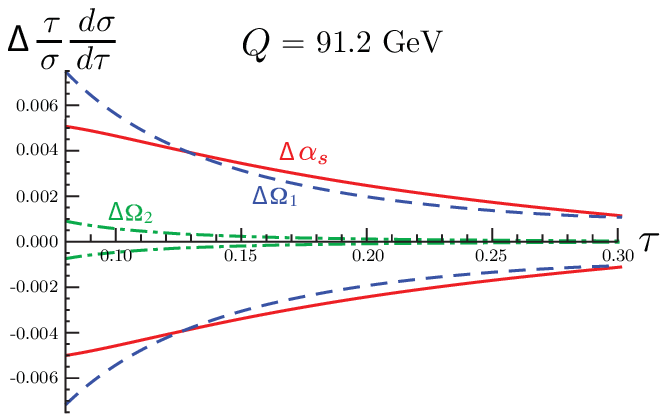}
\hspace{-0.3cm}
}\\
\subfigure[]{
\includegraphics[width=0.95\columnwidth]{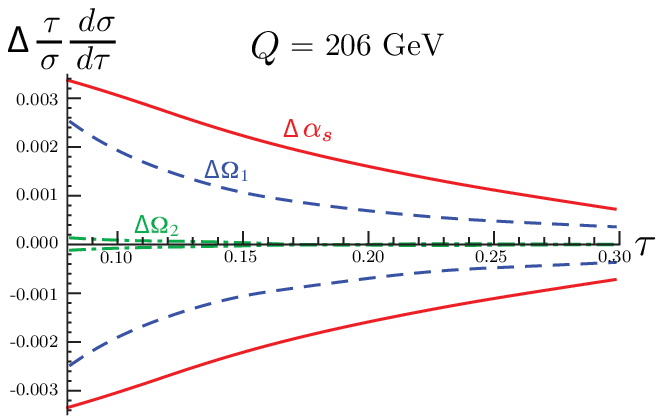}
\hspace{-0.3cm}
}
\caption{Difference between default cross section and the cross section varying
  only one parameter as a function of $\tau$. We  vary $\alpha_s(m_Z)$ 
  by $\pm \,0.001$ (solid red curves), $2\Omega_1$ by $\pm \,0.1$ (dashed blue
  curves) and $c_2$ by $\pm \,0.5$ (dash dotted green curves). The plot is shown
  for three different values of 
  the center of mass energy: (a) $Q=35 \GeV$, (b) $Q=91.2 \GeV$, (c) $Q=206
  \GeV$.\label{fig:diffplots} } 
\end{figure}

To estimate the theoretical errors in the $\alpha_s$-$\Omega_1$ plane at any order
and for any approximation used for the factorization
formula~(\ref{eq:masterformula}), we carry out independent fits for $500$
different sets of theory parameters which are randomly chosen in the ranges
discussed in the previous sections and summarized in Tab.~\ref{tab:theoryerr}.
We take the area covered by the points of the best fits in the
$\alpha_s$-$\Omega_1$ plane as the theory uncertainty treated like
$1$-sigma.\footnote{
This corresponds to a $1$-sigma error ($68\%$~CL) in $\alpha_s$ as well as in
$\Omega_1$.   
}
We emphasize that this method to estimate theoretical errors is more conservative
than the error band method~\cite{Jones:2003yv} employed for example in
Refs.~\cite{Becher:2008cf,Dissertori:2007xa}. However, our method required
considerably more computer power and it was necessary to use the Tier-2 centers
at Garching and MIT, as well as clusters at the MPI and the University of
Arizona. In Sec.~\ref{sec:fit} we also present the outcome of other ways to
estimate the theoretical error.

It is an important element of our analysis that we carry out global fits to the
data from all values of $Q\ge 35$ (and all experiments). This is motivated by
the strong degeneracy between $\alpha_s$ and $\Omega_1$ in the tail region which
can only be lifted when data from different $Q$ values are simultaneously
included in the fits.\footnote{The presence of this degeneracy is presumably
  also related to why Monte Carlos that are tuned to LEP data tend to have
  smaller hadronization corrections at $Q=m_Z$ than at larger $Q$ values. See
  Sec.~\ref{sec:comparison}.} In Fig.~\ref{fig:diffplots} the difference
${\rm d}\sigma/{\rm d}\tau-({\rm d}\sigma/{\rm d}\tau)_{\rm default}$ is
displayed for $0.08\le\tau\le 0.30$ and $Q=35$, $91.2$ and $206$~GeV. Here
$({\rm d}\sigma/{\rm d}\tau)_{\rm default}$ is the cross section for the default
setting of the theory parameters with $\alpha_s(m_Z)=0.114$ and
$\Omega_1=0.35\,{\rm GeV}$ and for ${\rm d}\sigma/{\rm d}\tau$ we vary either
$\alpha_s(m_Z)$ by $\pm\, 0.001$ (solid red curves) or $2\Omega_1$ by $\pm
\,0.1\,{\rm GeV}$ (dashed blue curves) from their default values. The figures
show that in the tail region changes in $\alpha_s$ can be compensated by changes
in $\Omega_1$.  This degeneracy makes it impossible to determine $\alpha_s$ and
$\Omega_1$ simultaneously with small uncertainties from tail fits that use data
from one $Q$ value (or from a narrow range of $Q$ values).  On the other hand,
we see that the correlation is $Q$ dependent when considering a large enough
range of $Q$ values. In our fits it is particularly important to include, apart
from the data from $Q=m_Z$, the low-energy data from JADE, TASSO, and AMY, and
the high energy data from the LEP-II experiments. Although the errors in these
analyses are larger than from the high-statistics $Q=m_Z$ run at LEP-I these
data sets are essential for breaking the degeneracy and simultaneously
extracting $\alpha_s$ and $\Omega_1$.

\section{Numerical Analysis}
\label{sec:fit}

\begin{figure*}[t!]
\subfigure[]{
\includegraphics[width=0.485\textwidth]{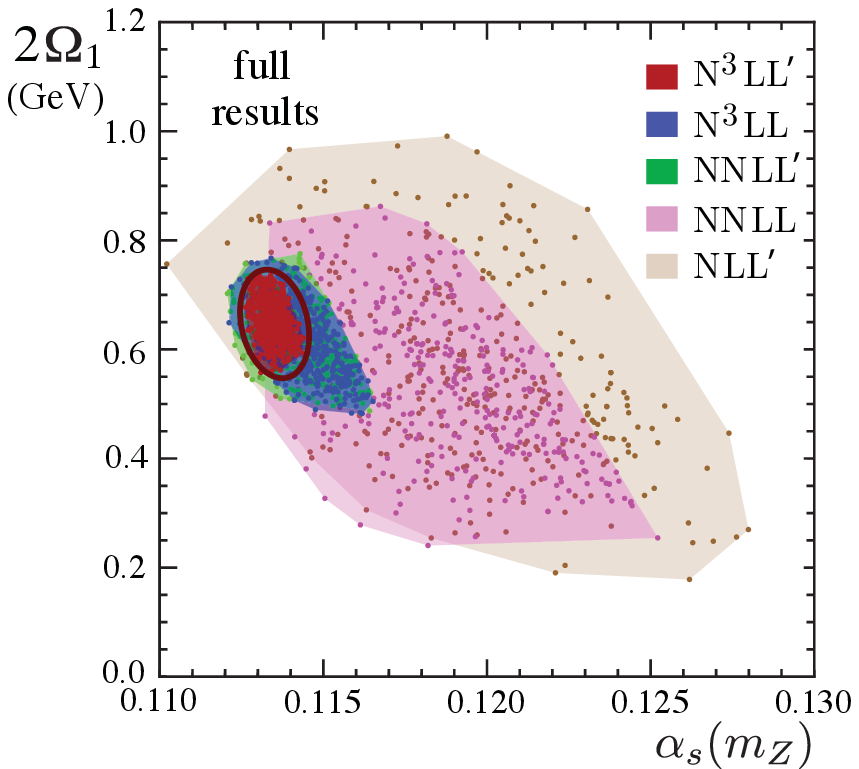}
\label{fig:M1alphagap}
}
\subfigure[]{
\includegraphics[width=0.485\textwidth]{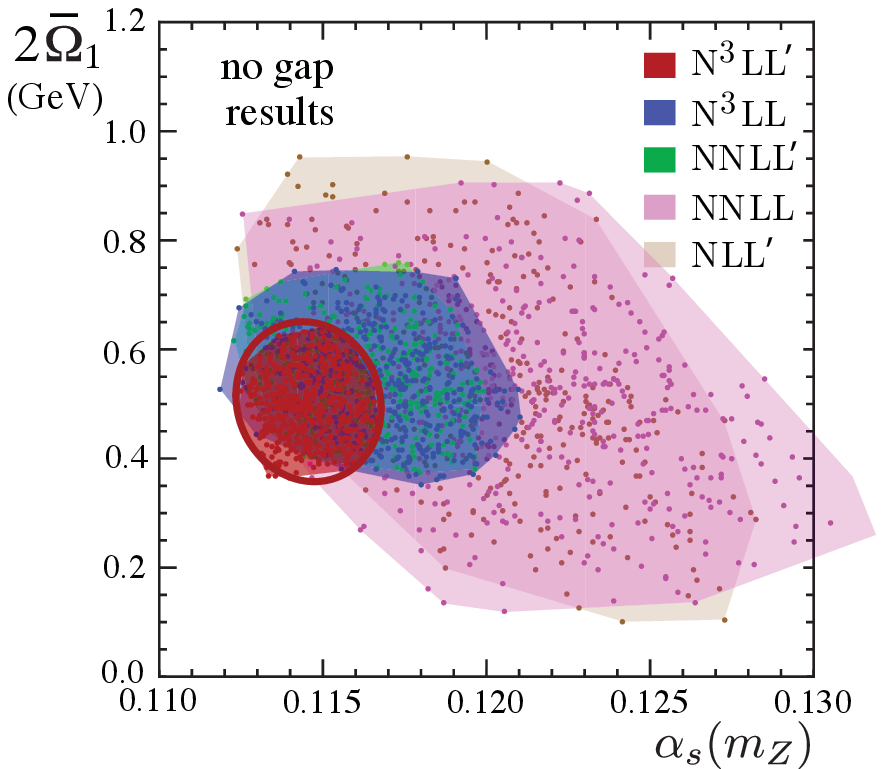}
\label{fig:M1alphanogap}
}
\vspace{-0.2cm}
\caption{Distribution of best fit points in the $\alpha_s(m_Z)$-$2\Omega_1$ and
  $\alpha_s(m_Z)$-$2\bar\Omega_1$
  planes. Panel~(a) shows results including perturbation theory, resummation of the
  logs, the soft nonperturbative 
  function and $\Omega_1$ defined in the R-gap scheme with renormalon
  subtractions. Panel~(b) shows the results as in panel~a, but with $\bar\Omega_1$
  defined in the $\msbar$ scheme without renormalon subtractions. In both panels
  the respective total (experimental+theoretical) 39\% CL standard error
  ellipses are displayed (thick dark red lines), which correspond to $1$-sigma
  (68\% CL) for either one-dimensional projection.
  \label{fig:M1alpha} }
\end{figure*}

\begin{figure*}[t!]
\subfigure[]{
\includegraphics[width=0.485\textwidth]{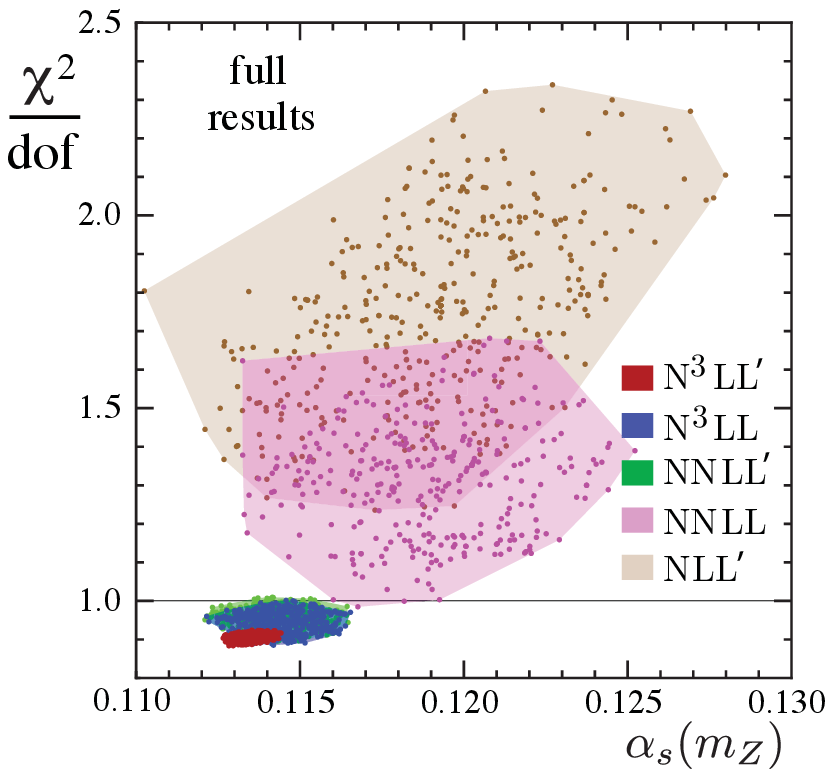}
\label{fig:chi2alphagap}
}
\subfigure[]{
\includegraphics[width=0.485\textwidth]{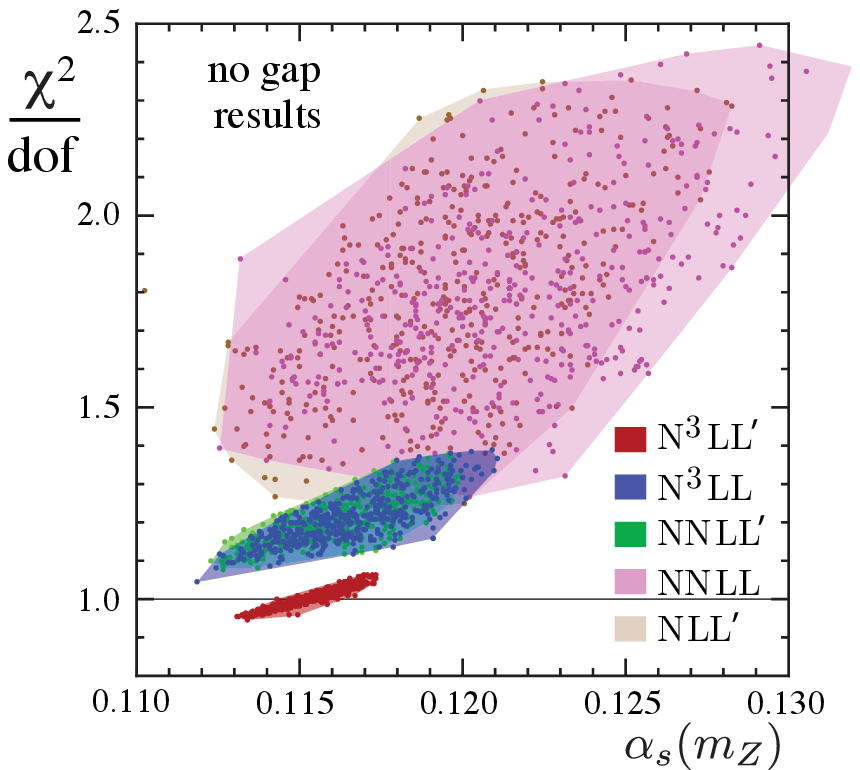}
\label{fig:chi2alphanogap}
}
\vspace{-0.2cm}
\caption{Distribution of best fit points in the $\alpha_s(m_Z)$-$\chi^2/{\rm dof}$
  plane. Panel~(a) shows the $\chi^2/{\rm dof}$ values of the points given in 
  Fig.~\ref{fig:M1alpha}a. Panel~(b) shows the $\chi^2/{\rm dof}$ values of the
  points given in Fig.~\ref{fig:M1alpha}b.
  \label{fig:chi2alpha} }
\end{figure*}

Having explained all ingredients of the factorization
formula~(\ref{eq:masterformula}) and the fit procedure we are now in the position to
discuss the numerical results of our analysis based on a global fit to the
experimental data for $Q\ge 35$~GeV in the tail region.  In the tail region the
dominant power corrections are encoded in the first moment $\Omega_1$, see
\eq{O1def}, so we can determine $\alpha_s(m_Z)$ and $\Omega_1$ from a
simultaneous fit. In this section we examine in detail the numerical results of
our fits concerning the treatment of the perturbative, hadronization and
experimental errors, QED and bottom mass corrections and their dependence on the
choice of the data set. We note that the values quoted for $\Omega_1$ in the
R-gap scheme are given for reference scales $R_\Delta=\mu_\Delta=2$~GeV, see
Sec.~\ref{subsec:gap}.

\begin{table}[t!]
\begin{tabular}{ccc}
order &$\alpha_s(m_Z)$ (with $\bar\Omega_1^{\msbar}$) & $\alpha_s(m_Z)$
(with $\Omega_1^{\rm Rgap}$)\\
\hline
NLL${}^\prime$ 
  & $0.1203\pm 0.0079$ & $0.1191\pm 0.0089$ \\
NNLL
  & $0.1222\pm 0.0097$ & $0.1192\pm 0.0060$ \\
NNLL${}^\prime$ 
  & $0.1161\pm 0.0038$ & $0.1143\pm 0.0022$ \\
\ntll
  & $0.1165\pm 0.0046$ & $0.1143\pm 0.0022$ \\
\ntllp(full)
  &  $0.1146\pm 0.0021$ & $\mathbf{0.1135\pm 0.0009}$ \\
\hline
\ntllp$\!\!${\tiny(QCD+$m_b$)}
  &  $0.1153\pm 0.0022 $ & $0.1141\pm 0.0009$ \\
\ntllp$\!\!${\tiny (pure QCD)}
  & $0.1152\pm 0.0021$ & $0.1140\pm 0.0008$ \\
\end{tabular}
%}
\caption{Theory errors from the parameter scan and central values for
  $\alpha_s(m_Z)$ at various 
  orders. The \ntllp value above the horizontal line is our final scan result,
  while the \ntllp values below the horizontal line show the effect of leaving
  out the QED corrections, and leaving out both the $b$-mass and QED 
  respectively. The central values are the average of the maximal and minimal
  values reached from the scan.}
\label{tab:results}
\end{table}
\begin{table}[t!]
\begin{tabular}{ccc}
order & \hspace{8mm}$\bar\Omega_1$ ($\msbar$)\hspace{8mm} 
 & \hspace{8mm}$\Omega_1$ (R-gap)\hspace{8mm} \\
\hline
NLL${}^\prime$ 
  & $0.264\pm 0.213$ & $0.293\pm 0.203$ \\
NNLL
  & $0.256\pm 0.197$ & $0.276\pm 0.155$ \\
NNLL${}^\prime$ 
  & $0.283\pm 0.097$ & $0.316\pm 0.072$ \\
\ntll
  & $0.274\pm 0.098$ & $0.313\pm 0.071$ \\
\ntllp(full)
  & $0.252\pm 0.069$ & $\mathbf{0.323\pm 0.045}$ \\
\hline
\ntllp$\!\!${\tiny (QCD+$m_b$)}
  & $0.238\pm 0.070$ & $0.310\pm 0.049$ \\
\ntllp$\!\!${\tiny (pure QCD)}
  & $0.254\pm 0.070$ & $0.332\pm 0.045$ \\
\end{tabular}
%}
\caption{Theory errors from the parameter scan and central values for $\Omega_1$
  defined at the 
  reference scales $R_\Delta=\mu_\Delta=2$~GeV in units of GeV at various
  orders. The \ntllp value above the horizontal line is our final scan result,
  while the \ntllp values below the horizontal line show the effect of leaving
  out the QED corrections, and leaving out both the $b$-mass and QED 
  respectively. The central values are the average of the maximal and minimal
  values reached from the scan.}
\label{tab:O1results}
\end{table}

\head{Theory Scan}

\noindent
In Fig.~\ref{fig:M1alpha} the best fit points of the theory parameters scan in
the $\alpha_s$-$2\Omega_1$ plane are displayed at NLL${}^\prime$ (brown), NNLL
(magenta), NNLL${}^\prime$ (green), N${}^3$LL (blue) and N${}^3$LL${}^\prime$
(red) order. The fit results at N${}^3$LL${}^\prime$ order include bottom mass
and QED corrections.  In Fig.~\ref{fig:M1alpha}a the results in the R-gap scheme
with renormalon subtractions are shown, and in Fig.~\ref{fig:M1alpha}b the
results in the $\msbar$ scheme without gap subtractions are given.

At each order 500~fits were carried out with the theory parameters randomly
chosen in the ranges given in Tab.~\ref{tab:theoryerr}. As described in
Sec.~\ref{sec:expedata}, we take the size of the area in the
$\alpha_s$-$2\Omega_1$ plane covered by the best fit points as a measure for the
theoretical uncertainties. To visualize the theoretical uncertainties we have
colored the respective areas according to the orders. The fit results clearly
show a substantial reduction of the theoretical uncertainties with increasing
orders. Explicit numerical results for the respective central values (determined
by the mean of the respective maximal and minimal values) and the theory errors
(determined by half of the difference between maximal and minimal values) for
$\alpha_s$ and $\Omega_1$ are given in Tabs.~\ref{tab:results} and
\ref{tab:O1results}, respectively.  We will consider these theory errors as
$1$-sigma.  At \ntllp order with $\Omega_1$ in the R-gap scheme the theory error
for $\alpha_s(m_Z)$ is $\pm 0.0009$ compared to $\pm 0.0021$ with $\bar\Omega_1$
in the $\msbar$ scheme. Also at NNLL${}^\prime$ and \ntll we see that the removal
of the ${\cal O}(\Lambda_{\rm QCD})$ renormalon leads to a reduction of the
theoretical uncertainties by about a factor of two in comparison to the results
with $\bar\Omega_1$ in the $\msbar$ scheme without renormalon subtraction. The
proper treatment of the renormalon subtraction is thus a substantial part of a
high-precision analysis for $\Omega_1$ as well as for $\alpha_s$.

\begin{figure}[t!]
\includegraphics[width=1.0\columnwidth]{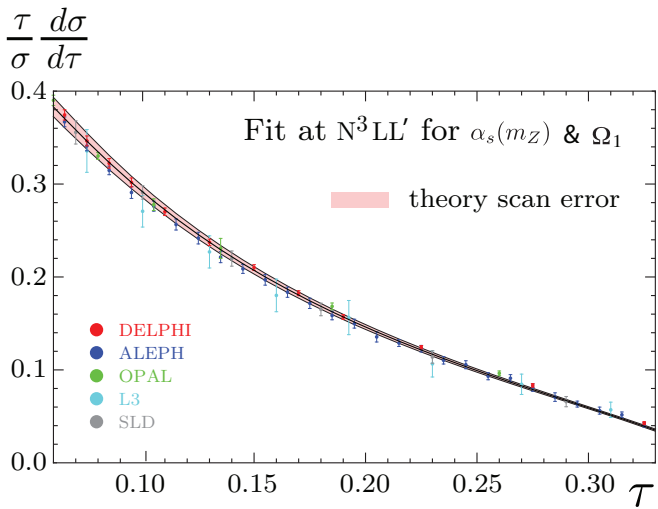}
\caption{Thrust distribution at $\mathrm{N^3LL'}$ order and $Q=m_Z$ including
  QED and $m_b$ corrections using the best fit values for $\alpha_s(m_Z)$ and
  $\Omega_1$ in the R-gap scheme given in Eq.~(\ref{eq:asO1finalcor}). The pink
  band represents the perturbative error determined from the scan method
  described in Sec.~\ref{sec:expedata}. Data from DELPHI, ALEPH, OPAL, L3, and
  SLD are also shown.}
\label{fig:tailfit}%
\end{figure}

It is instructive to analyze the minimal $\chi^2$ values for the best fit points
shown in Fig.~\ref{fig:M1alpha}. In Fig.~\ref{fig:chi2alpha} the distributions
of the best fits in the $\alpha_s$-$\chi^2_{\rm min}/{\rm dof}$ plane are shown
using the color scheme of Fig.~\ref{fig:M1alpha}.  Figure~\ref{fig:chi2alpha}a
displays the results in R-gap scheme, and Fig.~\ref{fig:chi2alpha}b the ones in
the $\msbar$ scheme. For both schemes we find that the $\chi^2_{\rm min}$ values
and the size of the covered area in the $\alpha_s$-$\chi^2_{\rm min}/{\rm dof}$
plane systematically decrease with increasing order. While the analysis in the
$\msbar$ scheme for $\bar\Omega_1$ leads to $\chi^2_{\rm min}/{\rm dof}$ values
around unity and thus an adequate description of the entire global data set at
N${}^3$LL${}^\prime$ order, we see that accounting for the renormalon
subtraction in the R-gap scheme leads to a substantially improved theoretical
description having $\chi^2_{\rm min}/{\rm dof}$ values below unity already at
NNLL${}^\prime$ and \ntll orders, with the \ntllp order result slightly lower at
$\chi^2_{\rm min}/{\rm dof}\simeq 0.91$. This demonstrates the excellent
description of the experimental data contained in our global data set. It also
validates the smaller theoretical uncertainties we obtain for $\alpha_s$ and
$\Omega_1$ at \ntllp order in the R-gap scheme.

As an illustration of the accuracy of the fit, in Fig.~\ref{fig:tailfit} we show
the theory thrust distributions at $Q=m_Z$ for the full \ntllp order with the
R-gap scheme for $\Omega_1$, for the default theory parameters and the
corresponding best fit values shown in bold in Tabs.~\ref{tab:results}
and~\ref{tab:O1results}. The pink band displays the theoretical uncertainty from
the scan method. The fit result is shown in comparison with data from DELPHI,
ALEPH, OPAL, L3, and SLD, and agrees very well. (Note that the theory values
displayed are actually binned according to the ALEPH data set and then joined by
a smooth interpolation.)

\head{Band Method}

\begin{table}[t]
\begin{tabular}{l|c c c}
&
Band & Band & \,\,Our scan\,\,
\tabularnewline[-3pt]
&\,\, method\! 1\,\, & \,\,method\! 2 \,\,& method 
\tabularnewline
\hline
\ntllp with $\Omega_1^{\rm Rgap}$&
$0.0004$&
$0.0008$&
$0.0009$\tabularnewline
\ntllp with $\bar\Omega_1^{\msbar}$&
$0.0016$&
$0.0019$&
$0.0021$\tabularnewline
\ntllp without $S_\tau^{\rm mod}$\,\,&
$0.0018$&
$0.0021$&
$0.0034$\tabularnewline
${\cal O}(\alpha_s^3)$ fixed-order &
$0.0018$&
$0.0026$&
$0.0046$\tabularnewline
\end{tabular}
\caption{Theoretical uncertainties for $\alpha_s(m_Z)$ obtained at \ntllp order
  from two versions of the error band method, and from our theory scan
  method. The uncertainties in the R-gap scheme (first line) include renormalon
  subtractions,  while the ones in the $\msbar$ scheme (second line) do not and
  are therefore larger. The same uncertainties are obtained in the analysis
  without nonperturbative function (third line). Larger uncertainties are
  obtained from a pure ${\cal O}(\alpha_s^3)$ fixed-order analysis (lowest
  line). Our theory scan method is more conservative than the error band method.   
  \label{tab:errorband}} 
\end{table}

\noindent
It is useful to compare our scan method to determine the perturbative errors
with the error band method ~\cite{Jones:2003yv} that was employed in the
analyses of Refs.~\cite{Dissertori:2007xa,Becher:2008cf,Dissertori:2009ik}.  In
the error band method first each theory parameter is varied separately in the
respective ranges specified in Tab.~\ref{tab:theoryerr} while the rest are kept
fixed at their default values. The resulting envelope of all these separate
variations with the fit parameters $\alpha_s(m_Z)$ and $\Omega_1$ held at their
best fit values determines the error bands for the thrust distribution at the
different $Q$ values.  Then, the perturbative error is determined by varying
$\alpha_s(m_Z)$ keeping all theory parameters to their default values and the
value of the moment $\Omega_1$ to its best fit value.  The resulting
perturbative errors of $\alpha_s(m_Z)$ for our full \ntllp analysis in the R-gap
scheme are given in the first line of Tab.~\ref{tab:errorband}. In the second
line the corresponding errors for $\alpha_s(m_Z)$ in the $\msbar$ scheme for
$\bar\Omega_1$ are displayed.  The left column gives the error when the band
method is applied such that the $\alpha_s(m_Z)$ variation leads to curves
strictly inside the error bands for all $Q$ values. For this method it turns out
that the band for the highest $Q$ value is the most restrictive and sets the
size of the error. The resulting error for the \ntllp analysis in the R-gap
scheme is more than a factor of two smaller than the error obtained from our
theory scan method, which is shown in the right column. Since the high $Q$ data
has a much lower statistical weight than the data from $Q=m_Z$, we do not
consider this method to be sufficiently conservative and conclude that it should
not be used.  The middle column gives the perturbative error when the band
method is applied such that the $\alpha_s(m_Z)$ variation minimizes a $\chi^2$
function which puts equal weight to all $Q$ and thrust values. This second band
method is more conservative, and for the \ntllp analyses in the R-gap and the
$\msbar$ schemes the resulting errors are only $10\%$ smaller than in the scan
method that we have adopted. The advantage of the scan method we use is that the
fit takes into account theory uncertainties including correlations.

\head{Effects of QED and the bottom mass}

\begin{figure}[t!]
\includegraphics[scale=0.95]{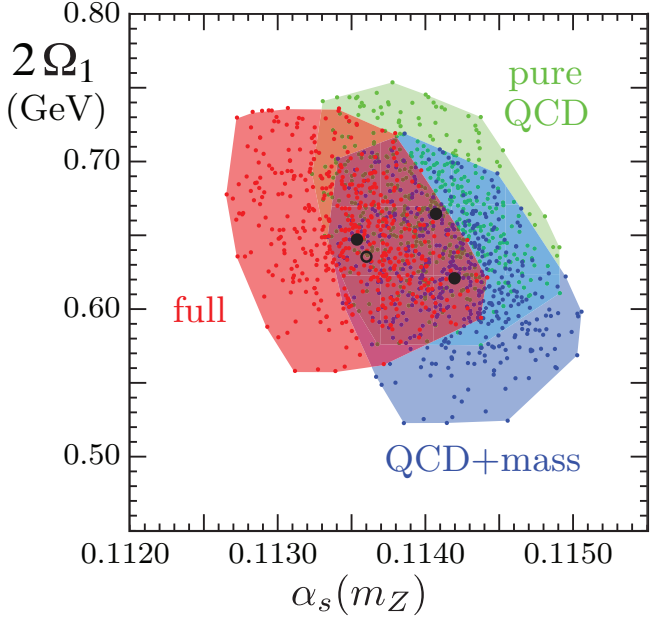}
\caption{Distribution of best fit points at \ntllp order with $\Omega_1$ in the
  R-gap scheme in pure QCD (light green), including $m_b$ effects (medium blue)
  and including $m_b$ effects and QED corrections (dark red). Solid circles
  indicate the central points for these three cases. The hollow circle
  represents the central point from the global fit with QED corrections
  neglected for the data from TASSO, L3 and ALEPH, but included for all other
  data sets.}
\label{fig:globalthrustanalysis}%
\end{figure}

\noindent
Given the high-precision we can achieve at \ntllp order in the R-gap scheme for
$\Omega_1$, it is a useful exercise to examine also the numerical impact of the
corrections arising from the nonzero bottom quark mass and the QED corrections.
In Fig.~\ref{fig:globalthrustanalysis} the distributions of the best fit points
in the $\alpha_s$-$2\Omega_1$ plane at \ntllp in the R-gap scheme is displayed
for pure massless QCD (light green points), including the bottom mass
corrections (medium blue points) and the bottom mass as well as the QED
corrections (dark red points).  The distribution of the best fit points with
bottom mass and QED corrections (dark red points) was already shown in
Fig.~\ref{fig:M1alpha}a.  The large black dots represent the corresponding central
values.  The corresponding numerical results are shown at the bottom of
Tabs.~\ref{tab:results} and~\ref{tab:O1results}.

We see that the QED and bottom
quark mass effects are somewhat smaller than the theoretical errors of the
\ntllp analysis but not negligible. Moreover we find that the qualitative impact
of the QED and the bottom quark mass effects is quite intuitive: The nonzero
bottom quark mass primarily causes a horizontal shift of the thrust distribution
towards larger $\tau$ values, since the small-$\tau$ threshold for massive quark
production is moved to a finite $\tau$ value. Here this is compensated primarily
by a reduced value of $\Omega_1$. Concerning QED effects, they cause an
effective increase of the coupling strength in the final state interactions
leading primarily to a decrease of $\alpha_s$ in the fit. 

As explained in Sec.~\ref{sec:expedata} the experimental correction procedures
applied to the AMY, JADE, SLC, DELPHI and OPAL data sets were designed to
eliminate initial state photon radiation, while those of the TASSO, L3 and ALEPH
collaborations eliminated initial and final state radiation.  It is
straightforward to test for the effect of these differences in the fits by using
our theory code with QED effects turned on or off depending on the data set.
Since our $\chi^2$ procedure treats data from different experiments as
uncorrelated it is also easy to implement this technically. Using our \ntllp
order code in the R-gap scheme we obtain the central values
$\alpha_s(m_Z)=0.1136$ and $\Omega_1=0.318$~GeV, indicated by the hollow circle
in Fig.~\ref{fig:globalthrustanalysis}. Comparing to our default results given
in Tabs.~\ref{tab:results} and \ref{tab:O1results}, which are based on the
theory code were QED effects are included for all data sets, we see that the
central value for $\alpha_s$ is larger by $0.0001$ and the one for $\Omega_1$ is
smaller by $0.006$~GeV. This shift is substantially smaller than our
perturbative error, and justifies our choice to use the theory code with QED
effects included as the default code for our analysis.

\head{Hadronization and Experimental Error}

\begin{figure}[t!]
\includegraphics[scale=1.2]{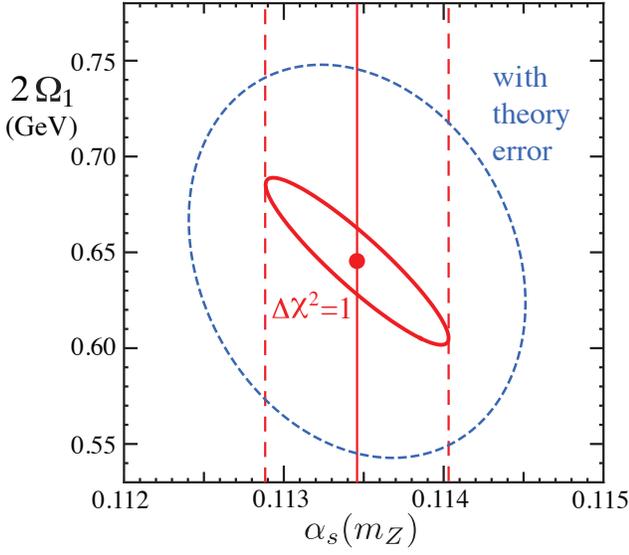}
\caption{Experimental $1$-sigma standard error ellipse (red solid) in the
  $\alpha_s$-$2\Omega_1$ plane. The larger ellipse shows the total uncertainty
  including theory errors (blue dashed). The fit is at \ntllp order in the R-gap
  scheme for $\Omega_1$ using the central values of the correlation matrix given
  in Eq.~(\ref{Vijresult}). The center of the ellipse are the central values of
  our final result given in Eq.~(\ref{eq:asO1finalcor}).}
\label{fig:redellipse}%
\end{figure}

\noindent
An important element in the construction of the $\chi^2$ function used for our
fit procedure is the correlation model for the systematic uncertainties given
for the experimental thrust bins. The results discussed above rely on the
minimal overlap model for the systematic experimental errors explained in
Sec.~\ref{sec:expedata}. The $1$-sigma ellipse based on the central values of
Eq.~(\ref{Vijresult}) and centered around
$(\alpha_s,2\Omega_1)=(0.1135,0.647~\mbox{GeV})$ is shown in
Fig.~\ref{fig:redellipse} by the red solid ellipse. This ellipse yields the
experimental errors and hadronization uncertainty related to $\Omega_1$ in our
analysis. We find that the size and correlation coefficients of the $1$-sigma
error ellipses at \ntllp order of all fits made in our theory scan are very
similar, and hence we can treat the theory error and these
hadronization/experimental errors as independent. 

The correlation matrix of the red solid error ellipses is ($i,j=\alpha_s,
2\Omega_1$)
\begin{align} \label{Vijresult} 
V_{ij}&  =\, 
\left( \begin{array}{cc}
\sigma_{\alpha_s}^2 
   & \,\, \sigma_{\alpha_s} \sigma_{2\Omega_1}\rho_{\alpha\Omega}\\
\sigma_{\alpha_s} \sigma_{2\Omega_1}\rho_{\alpha\Omega} 
   & \,\, \sigma_{2\Omega_1}^2
\end{array}\right) 
 \\
& =
\left( \begin{array}{cc}
3.29(16)\cdot 10^{-7}  & \,\, -2.30(12)\cdot 10^{-5}~\mbox{GeV}\\ 
-2.30(12)\cdot 10^{-5}~\mbox{GeV} & \,\, 1.90(18)\cdot 10^{-3}~\mbox{GeV}^2
\end{array}\right), \nn
\end{align} 
where the correlation coefficient is significant and reads
\begin{align} \label{eq:rhoaO}
  \rho_{\alpha\Omega}=-0.9176(60) \,.
\end{align} The numbers in the parentheses represent the variance from the
theory scan.  From Eq.~(\ref{Vijresult}) it is straightforward to extract the
experimental error for $\alpha_s$ and $\Omega_1$ and the error due to variations
of $\Omega_1$ and $\alpha_s$, respectively:
\begin{align}
\sigma_{\alpha_s}^{\rm exp} 
  & = \,\sigma_{\alpha_s}\,\sqrt{1-\rho^2_{\alpha\Omega}}
  =  \,0.0002 \,,
\nonumber\\
\sigma_{\Omega_1}^{\rm exp} 
  & = \,\sigma_{\Omega_1}\,\sqrt{1-\rho^2_{\alpha\Omega}}
  =  \,0.009~\mbox{GeV} \,,
\nonumber\\
\sigma_{\alpha_s}^{\rm \Omega_1} 
  & = \,\sigma_{\alpha_s}\, |\rho_{\alpha\Omega}|\,
  =  \,0.0005 \,,
\nonumber\\
\sigma_{\Omega_1}^{\rm \alpha_s} 
  & = \,\sigma_{\Omega_1}\, |\rho_{\alpha\Omega}|\,
  =  \,0.020~\mbox{GeV}
\,.
\end{align}
For $\alpha_s$, the error due to $\Omega_1$ variations is the dominant part of
the hadronization uncertainty. The blue dashed ellipse in
Fig.~\ref{fig:redellipse} shows the total error in our final result quoted in
\eq{asO1finalcor} below.

The correlation exhibited by the red solid error ellipse in
Fig.~\ref{fig:redellipse} is indicated by the line describing the semimajor axis
\begin{align}
  \frac{\Omega_1}{41.5\,{\rm GeV}} = 0.1213 - \alpha_s(m_Z) \,.
\end{align}
Note that extrapolating this correlation to the extreme case where we neglect
the nonperturbative corrections ($\Omega_1=0$) gives $\alpha_s(m_Z)\to 0.1213$.
This value is consistent with the fits in
Refs.~\cite{Dissertori:2007xa,Dissertori:2009ik} shown in
Tab.~\ref{tab:aseventshapes}, which are dominated by $Q=m_Z$ where the Monte
Carlo hadronization uncertainties are smallest.

\begin{figure}[t!]
\includegraphics[scale=1.25]{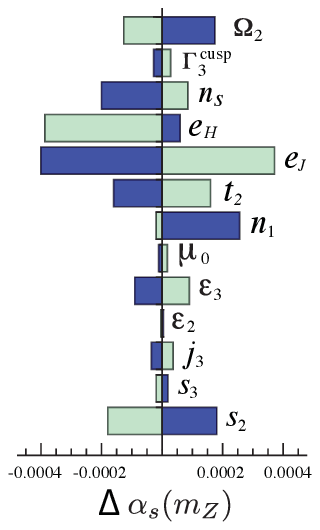}
\includegraphics[scale=1.25]{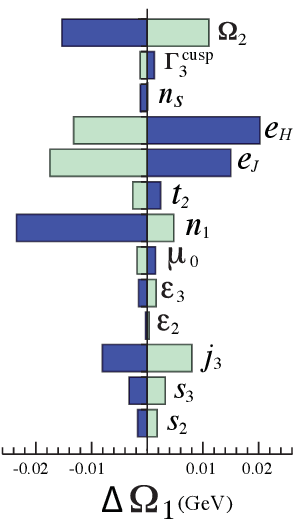} 
\caption{Variations of the best fit values for $\alpha(m_Z)$ and $\Omega_1$
  from up (dark shaded blue) and down (light shaded green) variations for the
  theory parameters with 
  respect to the default values and in the ranges given in
  Tab.~\ref{tab:theoryerr}. For the variation of the moment $\Omega_2$ we use
  $\Omega_2/\Omega_1^2=1.18^{+0.32}_{-0.18}$ as explained in the text.   
\label{fig:asO1updown}}
\end{figure}

\head{Individual Theory Scan Errors}

\noindent
It is a useful exercise to have a closer look at the size of the theory
uncertainties caused by the variation of each of the theory parameters we vary
in our fit procedure in order to assess the dominant sources of theory errors.
In Fig.~\ref{fig:asO1updown} two bar charts are shown for the variation of the
best fit values for $\alpha_s(m_Z)$ and $\Omega_1(R_\Delta,\mu_\Delta)$ at
\ntllp order in the R-gap scheme with our default theory parameters. The bars
show individual up-down variations of each of the theory parameters in the
ranges given in Tab.~\ref{tab:theoryerr}.  The changes of the best fit values
related to up variations of the theory parameters are given in dark blue and
those related to down variations are given in light green.

We see that the dominant theory uncertainties are related to variations of the
profile functions ($n_1, t_2, e_J, e_H$) and the renormalization scale parameter
($n_s$) for the nonsingular partonic distribution ${\rm d}\hat\sigma_{\rm
  ns}/{\rm d}\tau$ . The uncertainties related to the numerical errors of the
perturbative constants ($s_2$, $s_3$, $j_3$) as well as the numerical errors in
the extraction of the nonsingular distribution for small $\tau$ values,
($\epsilon_2$, $\epsilon_3$) are -- with the exception of $s_2$ -- much smaller
and do not play an important role. The theory error related to the unknown
4-loop contribution to the cusp anomalous dimension is negligible.  Adding
quadratically the symmetrized individual errors shown in
Fig.~\ref{fig:asO1updown} for each parameter, we find $0.0006$ for $\alpha_s$
and $0.029$ for $\Omega_1$. This is about $2/3$ of the theoretical uncertainty
we have obtained by the theory parameter scan, and it demonstrates that the
theory parameter scan represents a more conservative method to estimate the
theory error.

In Fig.~\ref{fig:asO1updown} we have also shown the variation of the best fit
values for $\alpha_s(m_Z)$ and $\Omega_1(R_\Delta,\mu_\Delta)$ due to variations
of the second soft function moment parameter $\Omega_2$. Our default choice for
the parametrization of the soft function $S_\tau^{\rm mod}$ uses $c_0=1$
and $c_{n>0}=0$ with $\bar\Delta(R_\Delta,\mu_\Delta)=0.05$~GeV. In this case
$\lambda$ is the only variable parameter of the soft model function $S_\tau^{\rm
  mod}$, and $\Omega_2$ is predetermined by \eq{Omega12} with $c_2=0$. As
explained in Sec.~\ref{sec:model} we modify $\Omega_2$ by setting $c_2$ to
nonzero values.  It is instructive to discuss the $\Omega_2$ values one should
consider.  From the Cauchy-Schwarz inequality one can show that
$\Omega_2/\Omega_1^2\ge 1$, giving a strict lower bound on $\Omega_2$.  This
bound can only be reached if $S_\tau^{\rm mod}$ is a delta-function.  Moreover,
if $S_\tau^{\rm mod}$ is positive definite, vanishing at $k=0$, has a width of
order $\Lambda_{\rm QCD}$, has its maximum at a $k$ value of order $\Lambda_{\rm
  QCD}$, and has an exponential fall-off for large $k$, then one finds
$\Omega_2/\Omega_1^2 < 1.5$.  We therefore adopt the range $1\le
\Omega_2/\Omega_1^2\le 1.5$ as a conservative $\Omega_2$ variation to carry out
an error estimate. For our default parametrization we have
$\Omega_2/\Omega_1^2=1.18$ and changing $c_2$ between $\pm 0.5$ gives a
variation of $\Omega_2/\Omega_1^2$ between $1.05$ and $1.35$.  We find that the
best fit values for $\alpha_s$ and $\Omega_1$ are smooth linear functions of
$\Omega_2/\Omega_1^2$ which allows for a straightforward extrapolation to the
conservative range between $1.0$ and $1.5$.  The results for the variations of
the best fit values for $\alpha_s(m_Z)$ and $\Omega_1$ for
$\Omega_2/\Omega_1^2=1.18^{+0.32}_{-0.18}$ read
$(\delta\alpha_s(m_Z))_{\Omega_2}=^{+0.00017}_{-0.00013}$ and
$(\delta\Omega_1)_{\Omega_2}=^{+0.011}_{-0.015}$ and are also shown in
Fig.~\ref{fig:asO1updown}. The symmetrized version of these errors are included
in our final results. For our final results for $\alpha_s(m_Z)$ we add the
uncertainties from $\Omega_1$ and the one from $\Omega_2$ quadratically giving
the total hadronization error. For $\Omega_1(R_\Delta,\mu_\Delta)$ we quote the
error due to $\Omega_2$ separately.

\head{Final Results}

\noindent
As our final result for $\alpha_s(m_Z)$ and $\Omega_1(R_\Delta,\mu_\Delta)$,
obtained at \ntllp order in the R-gap scheme for $\Omega_1$, including bottom
quark mass and QED corrections we obtain 
\begin{align} \label{eq:asO1finalcor}
\alpha_s(m_Z) & \, = \, 
 0.1135 \,\pm\, (0.0002)_{\rm exp} 
\nonumber\\[2mm] &
        \,\pm\, (0.0005)_{\rm hadr} 
        \,\pm \, (0.0009)_{\rm pert},
\nonumber\\[4mm]
\Omega_1(R_\Delta,\mu_\Delta) & \, = \,
 0.323 \,\pm\, (0.009)_{\rm exp} 
        \,\pm\, (0.013)_{\rm \Omega_2} 
\nonumber\\[2mm] &        \,\pm\, (0.020)_{\rm \alpha_s(m_Z)} 
\,\pm \, (0.045)_{\rm pert}~\mbox{GeV},
\end{align} 
where $R_\Delta=\mu_\Delta=2$~GeV and we quote individual $1$-sigma errors for
each parameter.  Eq.~(\ref{eq:asO1finalcor}) is the main result of this work.
In Fig.~\ref{fig:redellipse} (blue dashed line) and Fig.~\ref{fig:M1alpha}a
(thick dark red line) we have displayed the corresponding combined total
(experimental+theoretical) standard error ellipse.  To obtain the combined
ellipse we take the theory uncertainties given in Tabs.~\ref{tab:results}
and~\ref{tab:O1results} together with the $\Omega_2$ uncertainties, adding them
in quadrature.  The central values in Eq.~(\ref{eq:asO1finalcor}) are determined
by the average of the respective maximal and minimal values of the theory scan,
and are very close to the central values obtained when running with our default
theory parameters. The fit has $\chi^2/dof=0.91$ with a variation of $\pm 0.03$
for the displayed scan points.  Having added the theory scan and $\Omega_2$
uncertainties reduces the correlation coefficient in \eq{rhoaO} to
$\rho_{\alpha\Omega}^{\rm total} = -0.212$. As a comparison we have also shown
in Fig.~\ref{fig:M1alpha}b the combined total (experimental+theoretical) error
ellipse at \ntllp in the $\msbar$ scheme for $\bar\Omega_1$ where the ${\cal
  O}(\Lambda_{\rm QCD})$ renormalon is not subtracted.

Since our treatment of the correlation of the systematic experimental errors is
based on the minimal overlap model, it is instructive to also examine the
results treating all the systematic experimental errors as uncorrelated. 
At \ntllp order in the R-gap scheme the results that are
analogous to Eqs.~(\ref{eq:asO1finalcor}) read
$\alpha_s(m_Z)=0.1141 \pm (0.0002)_{\rm exp} \pm (0.0005)_{\rm hadr} \pm
(0.0010)_{\rm pert}$ and $\Omega_1(R_\Delta,\mu_\Delta)=0.303 \pm (0.006)_{\rm exp}
\pm (0.013)_{\Omega_2} \pm (0.022)_{\alpha_s} \pm (0.055)_{\rm pert}$~GeV 
with a combined correlation
coefficient of $\rho_{\alpha\Omega}^{\rm total}=-0.180$. 
The results are compatible with the results
of Eqs.~(\ref{eq:asO1finalcor}) and indicate that the ignorance of the exact
correlation of the systematic experimental errors does not crucially affect
the outcome of the fit.

\head{Data Set Choice}

\noindent
We now address the question to which extent the results of
Eqs.~(\ref{eq:asO1finalcor}) depend on the thrust ranges contained in the global
data set used for the fits. Our default global data set accounts for all
experimental thrust bins for $Q\ge 35$ in the intervals $[\tau_{\rm
  min},\tau_{\rm max}]=[6/Q,0.33]$. (See Sec.~\ref{sec:expedata} for more
details.) This default global data set is the outcome of a compromise that (i)
keeps the $\tau$ interval large to increase statistics, (ii) sets $\tau_{\rm
  min}$ sufficiently large such that the impact of the soft function moments
$\Omega_i$ with $i\ge 2$ is small and (iii) takes $\tau_{\rm max}$ sufficiently
low to exclude the far-tail region where the missing order $\alpha_s\Lambda_{\rm
  QCD}/Q$ corrections potentially become important.

\begin{figure}[t!]
\includegraphics[scale=1.05]{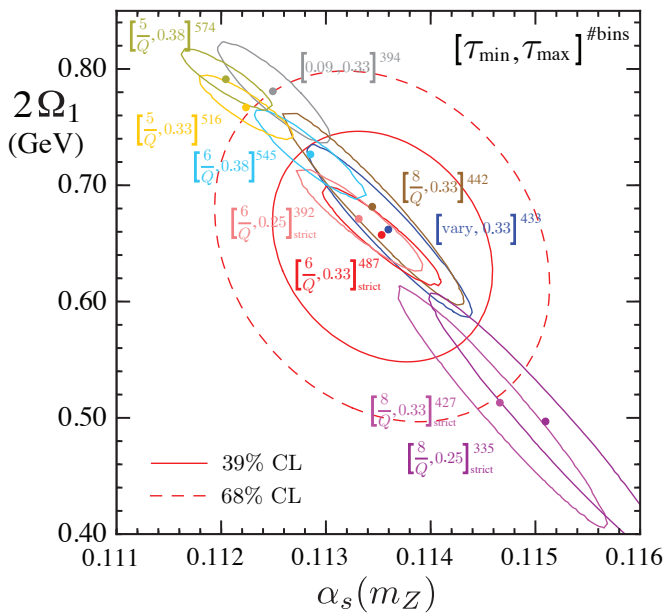}
\caption{The smaller elongated ellipses show the experimental 39\% CL error
  ($1$-sigma for $\alpha_s$) and best fit points for different global data sets
  at \ntllp order in the R-gap scheme and including bottom quark mass and QED
  effects. The default theory parameters given in Tab.~\ref{tab:theoryerr} are
  employed. The larger ellipses show the combined theoretical plus experimental
  error for our default data set with 39\% CL (solid, $1$-sigma for one
  dimension) and 68\% CL (dashed).
  \label{fig:alldataset}}
\end{figure}

In Fig.~\ref{fig:alldataset} the best fits and the respective experimental 39\%
and 68\% CL error ellipses for the default values of the theory parameters given
in Tab.~\ref{tab:theoryerr} are shown for global data sets based on different
$\tau$ intervals. The results for the various $\tau$ intervals are each given in
different colors. The results for our default global data set is given in red
color, and the subscript ``strict'' for some intervals means that bins are
included in the data set if more than half their range is contained within the
interval. For intervals without a subscript the criterion for selecting bins
close to the boundaries of the $\tau$ interval is less strict and generically,
if the $\tau_{\rm min}$ and $\tau_{\rm max}$ values fall in such bins, these
bins are included. The numbers in superscript for each of the $\tau$ intervals
given in the figure refers to the total number of bins contained in the global
data set. We observe that the main effect on the outcome of the fit is related
to the choice of $\tau_{\rm min}$ and to the total number of bins.
Interestingly all error ellipses have very similar correlation and are lined up
approximately along the line
\begin{align}
\frac{\Omega_1}{50.2 \,{\rm GeV}} =0.1200-\alpha_s(m_Z) \,.
\end{align}
Lowering $\tau_{\rm min}$ increases the dependence on $\Omega_2$ and leads to
smaller $\alpha_s$ and larger $\Omega_1$ values. On the other hand, increasing
$\tau_{\rm min}$ leads to a smaller data set and to larger experimental error
ellipses, hence to larger uncertainties.

It is an interesting but expected outcome of the fits that the pure experimental
error for $\alpha_s$ (the uncertainty of $\alpha_s$ for fixed central
$\Omega_1$) depends fairly weakly on the $\tau$ range and the size of the global
data sets shown in Fig.~\ref{fig:alldataset}.  If we had a perfect theory
description then we would expect that the centers and the sizes of the error
ellipses would be statistically compatible.  Here this is not the case, and one
should interpret the spread of the ellipses shown in Fig.~\ref{fig:alldataset}
as being related to the theoretical uncertainty contained in our \ntllp order
predictions. In Fig.~\ref{fig:alldataset} we have also displayed the combined
(experimental and theoretical) 39\% CL standard error ellipse from our default
global data set which was already shown in Fig.~\ref{fig:M1alpha}a (and is
$1$-sigma, 68\% CL, for either one dimensional projection). We also show the
68\% CL error ellipse by a dashed red line, which corresponds to $1$-sigma
knowledge for both parameters. As we have shown above, the error in both the
dashed and solid larger ellipses is dominated by the theory scan uncertainties,
see Eqs.~(\ref{eq:asO1finalcor}).  The spread of the error ellipses from the
different global data sets is compatible with the $1$-sigma interpretation of
our theoretical error estimate, and hence is already represented in our final
results.

\head{Analysis without Power Corrections}

\begin{table}[t]
\begin{tabular}{l|c c}
& $\alpha_s(m_Z)\pm $(pert. error) & $\chi^2/({\rm dof})$ \tabularnewline
\hline
\ntllp with $\Omega_1^{\rm Rgap}$&
$0.1135\pm 0.0009$&
$0.91$\tabularnewline
\ntllp with $\bar\Omega_1^{\msbar}$&
$0.1146\pm 0.0021$&
$1.00$\tabularnewline
\ntllp without $S_\tau^{\rm mod}$\,\,&
$0.1241\pm 0.0034$&
$1.26$\tabularnewline
\!\!\parbox{20ex}{${\cal O}(\alpha_s^3)$ fixed-order\\[2pt]
 without $S_\tau^{\rm mod}$}
 &
$0.1295\pm 0.0046$&
$1.12$ 
\end{tabular}
\caption{Comparison of global fit results for our full analysis to a fit
  where the renormalon is not canceled with $\bar\Omega_1$, a fit without
  $S_{\tau}^{\rm mod}$ (meaning without power corrections with $S_\tau^{\rm
    mod}(k)=\delta(k)$), and a fit at fixed order without power corrections 
  and log resummation. All results include bottom mass and QED corrections.
  \label{tab:nomodel}} 
\end{table}

\noindent
Using the simple assumption that the thrust distribution in the tail region is
proportional to $\alpha_s$ and that the main effect of power corrections is a
shift of the distribution in $\tau$, we have estimated in Sec.~\ref{sec:intro}
that a $300\,{\rm MeV}$ power correction will lead to an extraction of
$\alpha_s$ from $Q=m_Z$ data that is $\delta\alpha_s/\alpha_s \simeq (-9\pm
3)\%$ lower than an analysis without power corrections. In our theory code we
can easily eliminate all nonperturbative effects by setting $S_\tau^{\rm
  mod}(k)=\delta(k)$ and $\bar\Delta=\delta=0$. At \ntllp order and using our
scan method to determine the perturbative uncertainty a global fit to our
default data set yields $\alpha_s(m_Z)=0.1241\pm (0.0034)_{\rm pert}$ which is
indeed $9\%$ larger than our main result in \eq{asO1finalcor} which accounts for
nonperturbative effects. It is also interesting to do the same fit with a purely
fixed-order code, which we can do by setting $\mus=\muj=\muh$ to eliminate the
summation of logarithms. The corresponding fit yields $\alpha_s(m_Z)=0.1295\pm
(0.0046)_{\rm pert}$, where the displayed error has again been determined from
the theory scan which in this case accounts for variations of $\muh$ and the
numerical uncertainties associated with $\epsilon_2$ and $\epsilon_3$. (A
comparison with Ref.~\cite{Dissertori:2007xa} is given below in
Sec.~\ref{sec:comparison}.)

These results have been collected in Tab.~\ref{tab:nomodel} together with the
$\alpha_s$ results of our analyses with power corrections in the R-gap and the
$\msbar$ schemes. For completeness we have also displayed the respective
$\chi^2/{\rm dof}$ values which were determined by the average of the maximal
and the minimum values obtained in the scan.

\section{Far-tail and Peak Predictions}
\label{sec:fartailpeak}

\begin{figure}[t!]
\includegraphics[width=1.0\columnwidth]{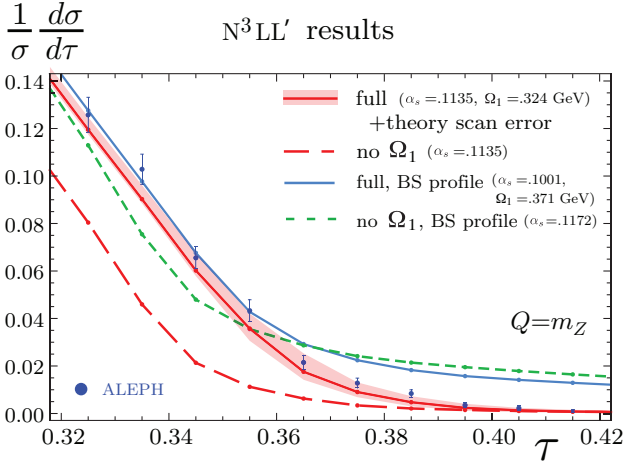}
\caption{Thrust distributions in the far-tail region at $\mathrm{N^3LL'}$ order
  with QED and $m_b$ corrections included at $Q=m_Z$ together with data from
  ALEPH. The red solid line is the cross section in the R-gap scheme using
  $\alpha_s(m_Z)$ and $\Omega_1$ obtained from fits using our full code, see
  Eq.~(\ref{eq:asO1finalcor}). The light red band is the perturbative
  uncertainty obtained from the theory scan method. The red dashed line shows
  the distribution with the same $\alpha_s$ but without power corrections. The
  light solid blue line shows the result of a full $\mathrm{N^3LL'}$ fit with
  the BS profile that does not properly treat the multijet thresholds. The short
  dashed green line shows predictions at $\ntllp$ with the BS profile, without
  power corrections, and with the value of $\alpha_s(m_Z)$ obtained from the fit
  in Ref.~\cite{Becher:2008cf}. All theory
  results are binned in the same manner as the experimental data, and then
  connected by lines.}
\label{fig:fartailfit}%
\end{figure}

\begin{figure}[t!]
\includegraphics[width=1.0\columnwidth]{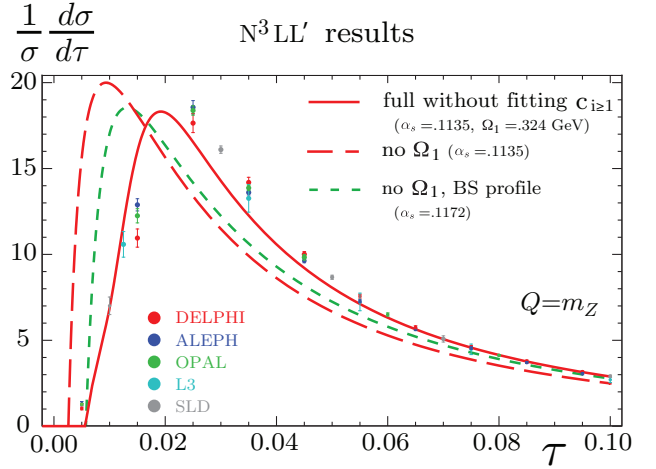}
\caption{Thrust cross section for the result of the $\mathrm{N^3LL'}$ fit, with
  QED and $m_b$ corrections included at $Q=m_Z$.  The red solid line is the
  cross section in the R-gap scheme using $\alpha_s(m_Z)$ and $\Omega_1$
  obtained from fits using our full code, see Eq.~(\ref{eq:asO1finalcor}). The
  red dashed line shows the distribution with the same $\alpha_s$ but without
  power corrections. The short-dashed green line shows predictions at $\ntllp$
  with the BS profile, without power corrections, and with the value of
  $\alpha_s(m_Z)$ obtained from the fit in Ref.~\cite{Becher:2008cf}. Data from
  ALEPH, DELPHI, L3, SLD, and OPAL are also shown.}
\label{fig:peakfit}%
\end{figure}

\noindent
The factorization formula~(\ref{eq:masterformula}) can be simultaneously
used in the peak, tail, and far-tail regions. To conclude the discussion of the
numerical results of our global analysis in the tail region, we use the results
obtained from this tail fit to make predictions in the peak and the far-tail
regions.

In Fig.~\ref{fig:fartailfit} we compare predictions from our full $\ntllp$ code
in the R-gap scheme (solid red line) to the accurate ALEPH data at $Q=m_Z$ in
the far-tail region.  As input for $\alpha_s(m_Z)$ and $\Omega_1$ we use our
main result of Eq.~(\ref{eq:asO1finalcor}) and all other theory parameters are
set to their default values (see Tab.~\ref{tab:theoryerr}).  We find excellent
agreement within the theoretical uncertainties (pink band).  Key features of our
theoretical result in Eq.~(\ref{eq:masterformula}) that are important in this
far-tail region are i) the nonperturbative correction from $\Omega_1$, and ii)
the merging of $\mu_S(\tau)$, $\mu_J(\tau)$, and $\mu_H$ toward
$\mus=\muj=\mu_H$ at $\tau=0.5$ in the profile functions, which properly treats
the cancellations occurring at multijet thresholds. To illustrate the importance
of $\Omega_1$ we show the long-dashed red line in Fig.~\ref{fig:fartailfit}
which has the same value of $\alpha_s(m_Z)$, but turns off the nonperturbative
corrections. To illustrate the importance of the treatment of multijet
thresholds in our profile function, we take the BS profile which does not
account for the thresholds (the BS profile is defined and discussed below in
Sec.~\ref{sec:comparison}), and use the smaller $\alpha_s(m_Z)$ and larger
$\Omega_1$ that are obtained from the global fit in this case. The result is
shown by the solid light blue line in Fig.~\ref{fig:fartailfit}, which begins to
deviate from the data for $\tau>0.36$ and gives a cross section that does not
fall to zero at $\tau=0.5$.  The fact that $\alpha_s(m_Z)$ is smaller by
$0.0034$ for the light blue line, relative to the solid red line, indicates that
the proper theoretical description of the cross section in the far-tail region
has an important impact on the fit done in the tail region.  The final curve
shown in Fig.~\ref{fig:fartailfit} is the short-dashed green line, which is the
result at the level of precision of the analysis by Becher and Schwartz in
Ref.~\cite{Becher:2008cf}. It uses the BS profile, has no power corrections, and
has the value of $\alpha_s$ obtained from the fit in Ref.~\cite{Becher:2008cf}.
It also misses the $Q=m_Z$ data in this region.  The results of other ${\cal
  O}(\alpha_s^3)$ thrust analyses, such as Davison and
Webber~\cite{Davison:2008vx} and Dissertori et
al.~\cite{Dissertori:2007xa,Dissertori:2009ik}, significantly undershoot the
data in this far-tail region.\footnote{ See the top panel of Fig.~9 in
  Ref.~\cite{Davison:2008vx}, the top left panel of Fig.~4 in
  Ref.~\cite{Dissertori:2007xa}, and the left panel of Fig.~2 in
  Ref.~\cite{Dissertori:2009ik}.}  To the best of our knowledge, the theoretical
cross section presented here is the first to obtain predictions in this far-tail
region that agree with the data. Note that our analysis does include some ${\cal
  O}(\alpha_s^k \Lambda_{\rm QCD}/Q)$ power corrections through the use of
\eq{ns1}.  It does not account for the full set of ${\cal O}(\alpha_s
\Lambda_{\rm QCD}/Q)$ power corrections as indicated in
Eq.~(\ref{eq:masterformula}) (see also Tab.~\ref{tab:orders}b), but the
agreement with the experimental data seems to indicate that missing power
corrections may be smaller than expected.

Unbinned predictions for the thrust cross section at $Q=m_Z$ in the peak region
are shown in Fig.~\ref{fig:peakfit}.  The green dashed curve shows the result at
the level of precision in Becher and Schwartz, that is $\ntllp$, with the BS
profile, without power corrections, and with the value of $\alpha_s(m_Z)=0.1172$
obtained from their fit. This purely perturbative result peaks to the left of
the data. With the smaller value of $\alpha_s(m_Z)$ obtained from our fit, the
result with no power corrections peaks even slightly further to the left, as
shown by the long-dashed red curve. In contrast, the red solid curve shows the
prediction from our full $\ntllp$ code in the R-gap scheme with our central
fit values of $\alpha_s(m_Z)$ and $\Omega_1$ given in Eq.~(\ref{eq:asO1finalcor}).
It clearly indicates that the value of $\Omega_1$ obtained from the fit in the
tail region shifts the theory prediction in the peak region much closer to the
experimental data. The residual difference between the solid red theory curve
and the experimental data can be attributed to the fact that the peak is
sensitive to power corrections from higher moments, $\Omega_{k\ge 2}$, which
have not been fit in our analysis. In our theoretical cross section result this
would correspond to fitting $\bar\Delta(R_\Delta,\mu_\Delta)$, and a subset of
the higher coefficients $c_{i\ge 1}$.  The $c_{i\ge 1}$ were all set to zero in
the curves shown here, and we leave the presentation of results of this extended
fit to a future publication.

\section{Cross checks and Comparisons}
\label{sec:comparison}

The result for $\alpha_s(m_Z)$ we obtain from our global \ntllp analysis in the
R-gap scheme with 487 bins given in Eq.~(\ref{eq:asO1finalcor}) is consistent at
$1$-sigma with the result of Davison and Webber~\cite{Davison:2008vx}
($\alpha_s(m_Z)=0.1164\pm(0.0022)_{\rm hadr+exp}\pm(0.0017)_{\rm pert}$). They
also carried out a global thrust analysis with a total of 430 experimental bins.
As explained in Sec.\ \ref{sec:intro}, in their theory formula nonperturbative
effects were included as a power correction in the effective coupling model
which was fit from the experimental data, and their approach also accounts for a
renormalon subtraction of the perturbative distribution. In these respects their
analysis is similar to ours. However, it differs as their theory formula
contains only resummation of logarithms at NLL order, and it also uses a
different renormalon subtraction scheme which is based on the running coupling
approximation for the subtraction corrections and does not account for the
resummation of large logarithms. Moreover the separation of singular and
nonsingular perturbative contributions and method to turn off the
log resummation at large $\tau$ is not equivalent to the one we employ.  The
difference between their central value and perturbative error and our
Eq.~(\ref{eq:asO1finalcor}) can be attributed to these items.  Their combined
hadronization and experimental uncertainty utilizes an error rescaling using the
value $\chi^2_{\rm min}/{\rm dof}=1.09$ obtained for their best fit.

On the other hand, our main result for $\alpha_s(m_Z)$ given in
Eq.~(\ref{eq:asO1finalcor}) is smaller than the results of Dissertori et
al.~\cite{Dissertori:2007xa} by $2.9$-sigma, of Dissertori et
al.~\cite{Dissertori:2009ik} by $2.2$-sigma, and of Becher and
Schwartz~\cite{Becher:2008cf} by $1.6$-sigma. (These results are displayed in
Table~\ref{tab:aseventshapes}.) In these analyses $\alpha_s(m_Z)$ was determined
from fits to data for individual $Q$ values and, as explained in
Sec.~\ref{sec:intro}, nonperturbative corrections and their associated
uncertainty were taken from Monte Carlo generators in Dissertori et al., or left
out from the fit and used to assign the hadronization uncertainty for the final
result in Becher and Schwartz. It is possible to turn off pieces of our
theoretical code to reproduce the perturbative precision of the codes used in
Refs.~\cite{Dissertori:2007xa}\footnote{ We do not attempt to reproduce the
  NLL/${\cal O}(\alpha_s^3)$ code of Ref.~\cite{Dissertori:2009ik} as the final
  outcome is similar to Ref.~\cite{Dissertori:2007xa}.  } and
\cite{Becher:2008cf}.  It is the main purpose of the remainder of this section
to show the outcome of the fits based on these modified theory codes.  We show
in particular, that the main reason why the above results for $\alpha_s(m_Z)$
are higher than our result of Eq.~(\ref{eq:asO1finalcor}) is related to the fact
that the nonperturbative corrections extracted from Monte Carlo generators at
$Q=m_Z$ are substantially smaller than and incompatible with the ones obtained
from our fit of the field theory power correction parameter $\Omega_1$.  In
Sec.~\ref{sec:intro} we already discussed why the use of Monte Carlo generators
to estimate nonperturbative corrections in high-precision perturbative
predictions is problematic.

\begin{table}[t]
\begin{tabular}{c c|c c c}
&& BS &
our BS &
default 
\\[-4pt]
\raisebox{0.1cm}{Experiment}&\raisebox{0.1cm}{\,\,\,Energy\,\,\,} 
&\,\,\,\,results~\cite{Becher:2008cf}\,\,\,\, 
& \,\,\,\,profile\,\,\,\, & \,\,\,\,profile\,\,\,\, 
\tabularnewline
\hline 
ALEPH &$91.2\,$GeV&
$0.1168(1)$
& $0.1170$ & $0.1223$
% $0.1175$ & $0.1226$
\tabularnewline
ALEPH &$133\,$GeV&
$0.1183(37)$
& $0.1187$ & $0.1235$
%$0.1192$ & $0.1239$
\tabularnewline
ALEPH &$161\,$GeV&
$0.1263(70)$
& $0.1270$ & $0.1328$
% $0.1275$ & $0.1332$
\tabularnewline
ALEPH &$172\,$GeV&
$0.1059(80)$
& $0.1060$ & $0.1088$
% $0.1064$ & $0.1091$
\tabularnewline
ALEPH &$183\,$GeV&
$0.1160(43)$
& $0.1166$ & $0.1205$
% $0.1170$ & $0.1208$
\tabularnewline
ALEPH &$189\,$GeV&
$0.1203(22)$
& $0.1214$ & $0.1260$
% $0.1219$ & $0.1264$
\tabularnewline
ALEPH &$200\,$GeV&
$0.1175(23)$
& $0.1182$ & $0.1224$
% $0.1186$ & $0.1227$
\tabularnewline
ALEPH &$206\,$GeV&
$0.1140(23)$
& $0.1149$ & $0.1185$
% $0.1153$ & $0.1188$
\tabularnewline
OPAL &$91\,$GeV&
$0.1189(1)$
& $0.1198$ & $0.1251$
% $0.1203$ & $0.1255$
\tabularnewline
OPAL &$133\,$GeV&
$0.1165(38)$
& $0.1175$ & $0.1218$
% $0.1180$ & $0.1222$
\tabularnewline
OPAL &$177\,$GeV&
$0.1153(33)$
& $0.1160$ & $0.1200$
% $0.1164$ & $0.1203$
\tabularnewline
OPAL &$197\,$GeV&
$0.1189(14)$
& $0.1197$ & $0.1241$
% $0.1202$ & $0.1244$
\tabularnewline
average &
& $0.1172(10)$
& $0.1180$& $0.1221$
% $0.1184$& $0.1224$
\tabularnewline
\hline 
\parbox{12ex}{global fit\\[0pt] (stat)}
 & all $Q$ &
& $0.1188$ & $0.1242$
% $0.1193$ & $0.1246$
\\[4pt]
\parbox{12ex}{global fit\\[0pt] (stat+syst)}
 & all $Q$ &
& $0.1192$ & $0.1245$
% $0.1198$ & $0.1249$
\tabularnewline
\end{tabular}
%}
\caption{Comparison of the results for $\alpha_s(m_Z)$ quoted by Becher and Schwartz
  in Ref.~\cite{Becher:2008cf} with results we obtain from our adapted code
  where power corrections, the $m_b$ and QED corrections, the 
  ${\cal O}(\alpha_s^2)$ axial singlet corrections are neglected. The ${\cal
    O}(\alpha_s^3)$ nonlogarithmic constants $h_3$ and $s_3$ are set to the
  values used in Ref.~\cite{Becher:2008cf} as described in the text.
  We follow the fit approach of Ref.~\cite{Becher:2008cf} and employ their profile
  functions for the nonsingular, hard, jet and soft scales, with results shown
  in the column labeled ``our BS profile''. In the last column we show results
  with this same code, but using our default profile functions. The errors in
  the third column are the statistical experimental uncertainty.}
\label{tab:comparisonBS}
\end{table}

We start with an examination related to the code used by Becher and
Schwartz~\cite{Becher:2008cf}, which has $\ntllp$ accuracy but does not include
power corrections or renormalon subtractions.  This treatment can be reproduced
in our factorization formula by turning off the nonperturbative soft
nonperturbative function by setting $S_\tau^{\rm mod}(k)=\delta(k)$ and
$\bar\Delta=\delta=0$.  Moreover they used the central scale setting $\mu_H=Q$,
$\mu_J=Q\sqrt{\tau}$ and $\mu_S=Q\tau$.  We can reproduce this from our profile
functions for $\mu_0=n_1=e_J=0$, $t_2=3/2$ and $e_H=n_s=1$, which we call the BS
profile setting. The BS profile functions for $\muj(\tau)$ and $\mus(\tau)$ are
shown by dashed curves in Fig.~\ref{fig:Profile}. (Note that the BS profile
setting does not cause $\mu_S$, $\mu_J$, and $\mu_H$ to merge in the far-tail
region and become equal at $\tau=0.5$, which is needed to switch off the SCET
resummation of logarithms in the multijet region to satisfy the constraints from
multijet thresholds.)  Becher and Schwartz set the ${\cal O}(\alpha_s^3)$
nonlogarithmic correction in the Euclidean hard factor $C(-q^2)$ to zero (with
$H_Q=|C(q^2)|^2$ for $q^2=Q^2>0$), which in our
notation corresponds to $h_3=11771.50$ (somewhat larger than the now known
$h_3$).  We also set $s_2=-40.1$ (see Ref.~\cite{Hoang:2008fs,Becher:2008cf}) and
$s_3=-324.631$ for the non-logarithmic 
${\cal O}(\alpha_s^2)$ and ${\cal O}(\alpha_s^3)$ constants in the soft function
(both within our range of uncertainties). 
The value for $s_3$ corresponds to setting the  ${\cal O}(\alpha_s^3)$
nonlogarithmic corrections in the 
expanded position space soft function to zero.
Finally, we also turn off our QED and
bottom quark mass corrections and the ${\cal O}(\alpha_s^2)$ axial singlet
corrections, and use the fixed-order normalization from \eq{Rhad}.  For the fit
procedure we follow Becher and Schwartz and analyze all ALEPH and OPAL data for
individual $Q$ values in the $\tau$ ranges given in their work and account only
for statistical experimental errors in the $\chi^2$ functions. The outcome of
the fits for $\alpha_s(m_Z)$ at \ntllp order is given in the fourth column of
Tab.~\ref{tab:comparisonBS}.  The third column shows their central values and
the respective statistical experimental errors as given in
Ref.~\cite{Becher:2008cf}.  The numbers we obtain are $0.0001$ to $0.0011$
%$0.0007$ to $0.0016$
higher than their central values, and we attribute this discrepancy to the
nonsingular contributions.\footnote{Becher and Schwartz uncovered a numerical
  problem with the original EERAD3 code at very small $\tau$, which
  correspondingly had an impact on the nonsingular function used in their
  analysis which was extracted from EERAD3.  When their nonsingular distribution
  is updated to results from the new EERAD3 code they become significantly
  closer to ours, differing by $\lesssim 0.0002$.  We thank M.~Schwartz for
  correspondence about this and for providing us with their new fit values.}
(Becher and Schwartz also used a difference of cumulants for their fits, as in
\eq{sigdiff} with the choice $\tilde\tau_1=\tau_1$ and $\tilde\tau_2=\tau_2$,
rather than integrating $\df\sigma/\df\tau$ as we do for the table.  The
spurious contribution induced by this choice has a significant effect on the
$\chi^2$ values, but a small effect on $\alpha_s(m_Z)$, changing the values shown
in the table by $\le 0.0003$.  For cumulants that use $\tilde
\tau_1=\tilde\tau_2 =(\tau_1+\tau_2)/2$ with no spurious contribution, the
difference from our integrated distribution results is reduced to $\le 0.0001$
for $\alpha_s(m_Z)$, and $\chi^2$ values are almost unaffected.)

The numbers obtained at $\ntllp$ above are significantly larger than our central
fit result $\alpha_s(m_Z)=0.1135$ shown in Eq.~(\ref{eq:asO1finalcor}) obtained
from our full code. These differences are mainly related to the nonperturbative
power correction and partly due to the BS profile setting. To distinguish these
two and other effects we can take the purely perturbative code described above
and turn back to our default setting for the profile functions with the
parameters given in Tab.~\ref{tab:theoryerr}. The results are shown in the fifth
column in Tab.~\ref{tab:comparisonBS} using again only statistical experimental
errors in the $\chi^2$ functions.  The $\alpha_s(m_Z)$ values using our default
profile functions are by $0.0028$ to $0.0058$ larger than for the BS profile
setting in the fourth column.\footnote{With our full code, which accounts in
  particular for power corrections and renormalon subtractions, the shift due to
  the modified profile functions becomes smaller; shifts in $\alpha_s(m_Z)$ of
  $0.005$ become $0.003$.} (The fifth column results again integrate the
distribution over each bin rather than using differences of cumulants, which for
our profile is important for the reasons discussed in
Sec.~\ref{sec:convergence}.\footnote{Using the cumulant method with
  $\tilde\tau_1=\tau_1$ and $\tilde\tau_2=\tau_2$ in \eq{sigdiff}, which has a
  spurious contribution, changes the values in the fifth column of
  Tab.~\ref{tab:theoryerr} by about $-0.003$ to $-0.005$. On the other hand,
  using the cumulant method without a spurious contribution,
  $\tilde\tau_1=\tilde\tau_2=(\tau_1+\tau_2)/2$, changes the values in the fifth
  column by $\le 0.0001$. }) A similar difference arises from a global fit to our
default data set of Sec.~\ref{sec:expedata} using the same fit procedure: For
the BS profile setting we obtain $\alpha_s(m_Z)=0.1189$, while the default
profile setting gives $\alpha_s(m_Z)=0.1242$ (second to last line of
Tab.~\ref{tab:comparisonBS}).  Using instead the $\chi^2$-analysis of our main
analysis which includes the experimental systematical errors we obtain
$\alpha_s(m_Z)=0.1192$ for the BS profile setting and $\alpha_s(m_Z)=0.1245$ for
the default profile setting (last line of Tab.~\ref{tab:comparisonBS}). The
latter result is by $0.0110$ larger than our $0.1135$ central fit result in
Eq.~(\ref{eq:asO1finalcor}).  This 10\% effect is almost entirely coming from
the power correction $\Omega_1$.  The difference of $0.3\%$ to the full
perturbative result of $\alpha_s(m_Z)=0.1241$ given in Table~\ref{tab:nomodel}
illustrates the combined effect of the QED, the bottom quark mass and the ${\cal
  O}(\alpha_s^2)$ axial singlet corrections and the ${\cal O}(\alpha_s^3)$ hard
constant $h_3$.
 
Finally, let us examine the results related to the code used by Dissertori~et
al.~in Ref.~\cite{Dissertori:2007xa}, which uses the fixed-order ${\cal
  O}(\alpha_s^3)$ results without a resummation of logarithms, but accounts for
nonperturbative corrections determined from the difference of running Monte
Carlo generators in parton and hadron level modes. Since in this work we are not
concerned with extracting the parton-hadron level transfer matrix from Monte
Carlo generators, we use in the following our code neglecting power corrections
by setting $S_\tau^{\rm mod}(k)=\delta(k)$, setting $\bar\Delta=\delta=0$, and
setting $\mu_H=\mu_J=\mu_S$. The latter switches off the log resummation factors
in Eq.~(\ref{eq:masterformula}) such that only the ${\cal O}(\alpha_s^3)$ fixed
order expression remains. We also include the $m_b$ corrections, but neglect QED
effects. Since these modifications give us a code that does not contain
nonperturbative corrections, the differences to Ref.~\cite{Dissertori:2007xa} we
obtain will serve as a quantitative illustration for the size of the
hadronization corrections obtained by a transfer matrix from the Monte Carlo
generators PYTHIA, HERWIG, and ARIADNE, tuned to global hadronic observables
at $m_Z$.

\begin{table}[t]
\begin{tabular}{c c|c c}
 & &
\,\,Dissertori et al.\,\,&
Our fixed 
\\[-4pt]
\raisebox{0.1cm}{Experiment} & \raisebox{0.1cm}{Energy} &
\,\,\,\, results~\cite{Dissertori:2007xa} \,\,\,\, & order code \\
\hline 
ALEPH &$91.2\,$GeV&
$0.1274(3)$&
$0.1281$\tabularnewline
ALEPH &$133\,$GeV&
$0.1197(35)$&
$0.1289$\tabularnewline
ALEPH &$161\,$GeV&
$0.1239(54)$&
$0.1391$\tabularnewline
ALEPH &$172\,$GeV&
$0.1101(72)$&
$0.1117$\tabularnewline
ALEPH &$183\,$GeV&
$0.1132(32)$&
$0.1247$\tabularnewline
ALEPH &$189\,$GeV&
$0.1140(20)$&
$0.1295$\tabularnewline
ALEPH &$200\,$GeV&
$0.1094(22)$&
$0.1260$\tabularnewline
ALEPH &$206\,$GeV&
$0.1075(21)$&
$0.1214$\tabularnewline
\end{tabular}
%}
\caption{Comparison of the thrust results quoted in
  Ref.~\cite{Dissertori:2007xa} with
  our numerical reproduction. For this numerical exercise we have used their
  procedure to get the error matrix for the experimental data.  This amounts to
  considering only the statistical errors in an uncorrelated way, with the
  resulting experimental error shown in the third column. Whereas in
  the code of Ref.~\cite{Dissertori:2007xa} hadronization corrections are
  included determined from Monte Carlo simulations our numbers are based on a pure
  partonic code neglecting nonperturbative effects. We use the
  default value for the scale setting, i.e. $\mu=Q$. } 
\label{tab:comparison-FO}
\end{table}

For the fits for $\alpha_s(m_Z)$ we follow Dissertori~et
al.~\cite{Dissertori:2007xa} analyzing ALEPH data for individual $Q$ values in
the $\tau$ ranges given in their work and accounting only for statistical
experimental errors in the $\chi^2$ functions.  The results of Dissertori~et
al.\ and the outcome for our best fits are given in the third and fourth column
of Tab.~\ref{tab:comparison-FO}, respectively. We have also quoted the
respective statistical errors from Ref.~\cite{Dissertori:2007xa}.  For the high
statistics data at $Q=m_Z$ our $\alpha_s(m_Z)$ result is larger than theirs, but
the discrepancy amounts to only $0.0007$ which is a $0.5\%$ shift in
$\alpha_s(m_Z)$. This illustrates the small size of the nonperturbative
hadronization corrections encoded in the Monte Carlo transfer matrix at $Q=m_Z$.
This is clearly incompatible with the size of the nonperturbative correction we
have obtained from simultaneous fits of $\alpha_s$ and $\Omega_1$, confirming
the concerns on Monte Carlo hadronization corrections explained in
Sec.~\ref{sec:intro}.  Interestingly, with the exception of $Q=172\,{\rm GeV}$,
our fixed-order results for all $Q$ are relatively stable and close to the
result at $Q=m_Z$, while their $\alpha_s(m_Z)$ values, which use the transfer
matrix for nonperturbative effects, are systematically lower for $Q>m_Z$ by $7$
to $13\%$. Thus the nonperturbative effects from the Monte Carlo transfer matrix
are substantially larger for $Q>m_Z$.\footnote{Note that the weighted average of
  the $Q>m_Z$ thrust results of Dissertori~et al.\ is $\alpha_s(m_Z)=0.1121$ and
  is consistent with our result in \eq{asO1finalcor} within the larger
  uncertainties. Also note that the $Q$ dependence of our $\Omega_1(R,R)/Q$
  power correction is affected by its anomalous dimension, cf.
  Fig.~\ref{fig:RunningO1}. }  The same behavior is also visible in the results
of Ref.~\cite{Dissertori:2009ik}, which includes NLL resummation of logarithms.
Since the transfer matrix is obtained from Monte Carlo tuned to the more
accurate $Q=m_Z$ data, we believe that this issue deserves further
investigation.  To complete the discussion we use the same fixed-order theory
code to quote results for a global fit to our default data set. Using the fit
procedure as described in Sec.~\ref{sec:expedata} we obtain
$\alpha_s(m_Z)=0.1300\pm (0.0047)_{\rm pert}$.  (The corresponding errors
obtained from the error band method are given in the fourth line of
Tab.~\ref{tab:errorband}.)

\begin{figure*}[t!]
\includegraphics[scale=0.9]{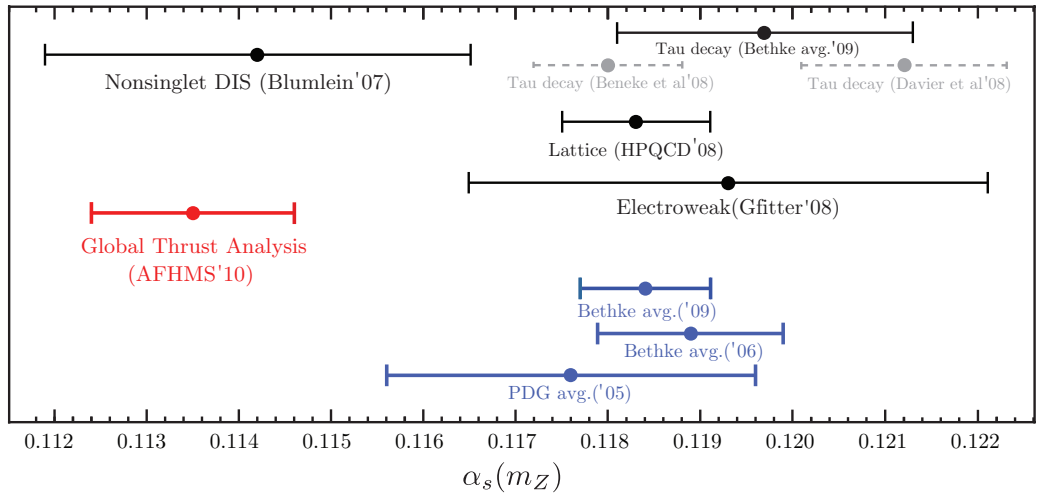}
\caption{Comparison of selected determinations of $\alpha_s(m_Z)$ defined in
  the $\msbar$ scheme.}
\label{fig:ascomparison}%
\end{figure*}

\section{Conclusions}
\label{sec:conc}

In this work we have provided a factorization formula for the thrust
distribution in $e^+e^-$ annihilation which incorporates the previously known
${\cal O}(\alpha_s^2)$ and ${\cal O}(\alpha_s^3)$ perturbative QCD corrections
and summation of large logarithms at N${}^3$LL order for the singular terms in
the dijet limit where the thrust variable $\tau=1-T$ is small. The factorization
formula used here incorporates a systematic description of nonperturbative
effects with a soft function defined in field theory. The soft function
describes the dynamics of soft particle radiation at large angles. We have also
accounted for 
bottom mass and QED photon effects for fixed-order contributions as well as for
the summation of QED logarithms. With specifically designed $\tau$-dependent
profile functions for the renormalization scales the
factorization formula can be applied in the peak, tail and far-tail regions of
the thrust distribution. It has all nonperturbative effects accounted for up to
terms of ${\cal 
  O}(\alpha_s \Lambda_{\rm QCD}/Q)$, which is parametrically smaller than the
remaining perturbative uncertainty ($<2\%$ for $Q=m_Z$) of the thrust
distribution predictions in the tail region where we carried out the fits to the
experimental data.

In the tail region, $2\Lambda_{\rm QCD}/Q \ll \tau \lesssim 1/3$, the dominant
effects of the nonperturbative soft function are encoded in its first moment
$\Omega_1$, which is a power correction to the cross section.  
Fitting to tail data at multiple $Q$s as we
did in this work, the strong coupling $\alpha_s(m_Z)$ and the moment $\Omega_1$
can be 
simultaneously determined.  An essential ingredient to reduce the theoretical
uncertainties to the level of $<2\%$ in the thrust distribution is our use of a
short-distance scheme for $\Omega_1$, called the R-gap scheme, that induces
subtractions 
related to an ${\cal O}(\Lambda_{\rm QCD})$ renormalon contained in the $\msbar$
perturbative thrust cross section from large angle soft gluon radiation. The
R-gap scheme introduces an additional scale that leads to large logarithms in
the subtractions, and we carry out a summation of these additional logarithms
with renormalization group equations in the variable $R$. The R-gap scheme reduces
the perturbative uncertainties in our best highest order theory code by roughly
a factor of two compared to the pure $\msbar$ definition, $\bar\Omega_1$, where
renormalon effects are not treated.

The code we use in this analysis represents the most complete theoretical
treatment of thrust existing at this time. As our final result we obtain
\begin{align}
\label{eq:finalasO1}
\alpha_s(m_Z) & \, = \, 
 0.1135 \,\pm\, 0.0011,
\nonumber\\[4mm]
\Omega_1(R_\Delta,\mu_\Delta) & \, = \,
 0.323 \,\pm\, 0.051~\mbox{GeV},
\end{align} 
where $\alpha_s$ is defined in the $\msbar$ scheme, and $\Omega_1$ in the R-gap
scheme at the reference scales $R_\Delta=\mu_\Delta=2$~GeV.
Here the respective total $1$-sigma errors are shown.  The results with individual
$1$-sigma errors quoted separately for the different sources of uncertainties
are given in Eq.~(\ref{eq:asO1finalcor}). Neglecting 
the nonperturbative effects incorporated in the soft function, and in particular
$\Omega_1$, from the fits gives $\alpha_s(m_Z)=0.1241$ which exceeds the result
in Eq.~(\ref{eq:finalasO1}) by $9\%$. This is consistent with a simple scaling
argument one can derive from experimental data, see Eq.~(\ref{eq:fracalphas}) in
Sec.~\ref{sec:intro}.

Analyses of event shapes with a simultaneous fit of $\alpha_s$ and a power
correction have been carried out earlier with the effective coupling model.
Davison and Webber~\cite{Davison:2008vx} analyzed the thrust distribution and
determined $\alpha_s(m_Z)=0.1164\pm 0.0028$ also using ${\cal O}(\alpha_s^3)$
fixed-order input, but implementing the summation of logarithms only at NLL
order (for further discussion see Sec.~\ref{sec:comparison}). Recently Gehrmann
et al.~\cite{Gehrmann:2009eh} analyzed moments of different event shape
distributions, also with the effective coupling model, and obtained
$\alpha_s(m_Z)=0.1153\pm 0.0029$ using fixed-order perturbation theory at ${\cal
  O}(\alpha_s^3)$.  Both analyses neglected bottom mass and QED corrections.
Our result in Eq.~(\ref{eq:finalasO1}) is compatible with these analyses at
$1$-sigma, but has smaller uncertainties.

These results and our result for $\alpha_s(m_Z)$ in \eq{finalasO1} are
substantially smaller than the results of event shape analyses employing input
from Monte Carlo generators to determine nonperturbative effects. We emphasize
that using parton-to-hadron level transfer matrices obtained from Monte Carlo
generators to incorporate nonperturbative effects is not compatible with a
high-order theoretical analysis such as ours, and thus analyses relying on such
Monte Carlo input contain systematic errors in the determination of $\alpha_s$
from thrust data. The small effect of hadronization corrections on thrust
observed in Monte Carlo generators at $Q=m_Z$ and the corresponding small shift
in $\alpha_s(m_Z)$ do not agree with the $9\%$ shift we have obtained from our
fits as mentioned above. For the reasons discussed earlier, we believe Monte
Carlo should not be used for hadronization uncertainties in higher order
analyses.

Although our theoretical approach represents the most complete treatment of
thrust at this time, and all sources of uncertainties known to us have been
incorporated in our error budget, there are a number of theoretical issues
related to subleading contributions that deserve further investigation. These
issues include (i) the summation of logarithms for the nonsingular partonic
cross section, (ii) the structure of the ${\cal O}(\alpha_s \Lambda_{\rm
  QCD}/Q)$ power corrections, (iii) analytic perturbative computations of the
${\cal O}(\alpha_s^2)$ and ${\cal O}(\alpha_s^3)$ nonlogarithmic coefficients
$s_2$ and $s_3$ in the partonic soft function, the ${\cal O}(\alpha_s^3)$
nonlogarithmic coefficient $j_3$ in the partonic jet function, and the 4-loop
QCD cusp anomalous dimension $\Gamma^{\rm cusp}_3$.  Concerning issue (i) we
have incorporated in our analysis the nonsingular contributions in fixed-order
perturbation theory and estimated the uncertainty related to the higher order
logarithms through the usual renormalization scale variation. Further
theoretical work is needed to derive the renormalization group structure of
subleading jet, soft, and hard functions in the nonsingular contributions and to
use these results to sum the corresponding logarithms.  Concerning issue (ii) we
have shown that our theoretical description for the thrust distribution contains
a remaining theoretical uncertainty from nonperturbative effects of order ${\cal
  O}(\alpha_s\Lambda_{\rm QCD}/Q)$.  Parametrically, this uncertainty is
substantially smaller than the perturbative error of about $1.7\%$ for the
thrust distribution in the tail region at LEP-I energies that is contained in
our best theory code. Furthermore, our predictions in the far-tail region at
$Q=m_Z$ appear to indicate that the dominant corrections of this order are
already captured in our setup. Nevertheless a systematic analysis of these
subleading effects is certainly warranted.

Apart from investigating these theoretical issues, it is also warranted to apply
the high-precision approach using soft-collinear effective theory to other event
shape distributions in order to validate the result in Eq.~(\ref{eq:finalasO1}).
Event shapes that can be clearly treated with similar techniques are: heavy and
light jet masses, the C-parameter, and the
angularities~\cite{Berger:2003iw,Hornig:2009vb}.  For many of these event shapes
it has been proven field theoretically that the same parameter $\Omega_1$
describes the leading power corrections in the tail region~\cite{Lee:2006fn},
although there might be caveats related to the experimental treatment of hadron
masses~\cite{Salam:2001bd,Gardi:2002bg}. Thus, one has the potential to extend
the analysis done here to include additional data without additional parameters.
An analysis for the heavy jet mass accounting for perturbative contributions at
$\ntll$ in $\msbar$ with different profile functions and a simple soft
function model for 
power corrections without renormalon subtractions, was recently carried out in
Ref.~\cite{Chien:2010kc}, providing a first step in this direction.

To conclude this work we cannot resist comparing our result for $\alpha_s(m_Z)$
with the results of a selection of analyses using other techniques and
observables, as shown in Fig.~\ref{fig:ascomparison}. We include a N$^3$LO
analysis of data from deep inelastic scattering in the nonsinglet
channel~\cite{Blumlein:2006be}\footnote{Analyses studying $\alpha_s$ with data
  that depends also on the gluon PDF have been carried out in
  Refs.~\cite{Martin:2009bu,Alekhin:2009ni,Demartin:2010er,Lai:2010nw}.}, the
recent HPQCD lattice determination based on fitting Wilson loops and the
$\Upsilon$-$\Upsilon^\prime$ mass difference~\cite{Davies:2008sw}, the result
from fits to electroweak precision observables based on the Gfitter
package~\cite{Flacher:2008zq}, analyses of $\tau$-decay data using
fixed-order~\cite{Beneke:2008ad} and contour-improved perturbation
theory~\cite{Davier:2008sk}, together with an average of $\tau$ results from
Ref.~\cite{Bethke:2009jm}. Finally we also show a collection of
$\alpha_s$-averages from Refs.~\cite{Bethke:2009jm,Bethke:2006ac,Yao:2006px}.
The DIS result is consistent with our fit result, whereas the deviation from
HPQCD is $3.5\sigma$. It is interesting to note that the high energy extractions
from thrust and DIS appear to be smaller than the low energy extractions from
Lattice and $\tau$ decays.

\begin{acknowledgments}

  This work was supported in part by the Director, Office of Science, Office of
  Nuclear Physics of the U.S.\ Department of Energy under the Contracts
  DE-FG02-94ER40818, DE-FG02-06ER41449, and the European Community's Marie-Curie
  Research Networks under contract MRTN-CT-2006-035482 (FLAVIAnet) and
  MRTN-CT-2006-035505 (HEPTOOLS). IS was also supported in part by the DOE OJI
  program, the Sloan Foundation, and by a Friedrich Wilhelm Bessel award from
  the Alexander von Humboldt foundation.  VM has been partially supported by a
  DFG ``Eigenen Stelle'' under contract MA 4882/1-1.  MF was supported in part
  by the German Academic Exchange Service (DAAD) D/07/44491. RA, MF and IS thank
  the Max-Planck Institute for Physics for hospitality while parts of this work
  were completed.  We thank Thomas Gehrmann for useful conversations on the
  experimental treatment of photon radiation and in particular for providing us
  with the most up-to-date numerical results from the program EERAD3 that made
  our \ntllp analysis possible.  We are particularly indebted to Stefan Kluth
  for countless valuable discussions and information on experimental data and
  the usage of the program EVENT2. We are grateful to Daniel Wicke for useful
  discussions concerning the treatment of experimental data.  We thank Thomas
  Hahn for computing support.  MF thanks Sean Fleming for scientific advice. We
  thank T.~Gehrmann, S.~Kluth, M.~Schwartz, and F.~Tackmann for comments on the
  manuscript. We thank the Aspen Center for Physics for a stimulating
  environment when this work was finalized.
\end{acknowledgments}
\vspace*{2mm}

\appendix
\section{Formulae}
\label{app:appendix}

In this appendix we collect all the remaining formulas used in our analysis for
the case of massless quarks. The total hadronic cross section at tree level at
the energies we are considering is
\begin{equation}
\sigma_0(Q)=\sum_{q\neq {\rm top}} 
 \left[\, \sigma^q_{ax}(Q)+\sigma^q_{vec}(Q)\,\right] ,
\end{equation}
where Q is the c.m. energy. For a quark of flavor $q$ the tree level
axial-vector and vector cross sections are
\begin{eqnarray}
\sigma^{qa}_{0} &=&N_c\dfrac{4\pi\alpha^2}{3Q^2}\frac{Q^{4}(v_{e}^{2}+a_{e}^{2})a_{q}^2}{(m_{Z}^{2}-Q^{2})^{2}+
\frac{Q^{4}}{m_{Z}^{2}}\,\Gamma_{Z}^{2}}\,,\\
\sigma^{qv}_{0} &=&N_c\dfrac{4\pi\alpha^2}{3Q^2}\bigg[e_{q}^{2}
  -\frac{2\,e_{q}v_{q}v_{e}\,Q^{2}(Q^{2}-m_{Z}^{2})}{(m_{Z}^{2}-Q^{2})^{2}+
\frac{Q^{4}}{m_{Z}^{2}}\,\Gamma_{Z}^{2}}\nonumber\\
&&+\,\frac{Q^{4}(v_{e}^{2}+a_{e}^{2})v_{q}^{2}}{(m_{Z}^{2}-Q^{2})^{2}+
\frac{Q^{4}}{m_{Z}^{2}}\,\Gamma_{Z}^{2}}\bigg]\,,\nn
\end{eqnarray}
where  $e_{q}$ is the electric charge of the quark, and
\begin{equation}
v_{q}=\frac{T_{3}^{q}-2\,e_{q}\sin^{2}\theta_{W}}{\sin(2\,\theta_{W})}\,,\quad
a_{q}=\frac{T_{3}^{q}}{\sin(2\,\theta_{W})}\,.
\end{equation}
Here $T_{3}^{q}$ is the third component of the weak isospin, and $\theta_{W}$
is the weak mixing angle. For our numerics we use the following values:
\begin{align}\label{eq:numerical_constants}
\sin^{2}\theta_{W}&=0.23119\,,      &m_{Z}&=91.187\,\mathrm{GeV}\,,\nonumber\\
\Gamma_{Z}&=2.4952\,\mathrm{GeV}\,, &m_t&=172\,\mathrm{GeV}\,,\nonumber\\
m_b&=4.2\,\mathrm{GeV}\,,           &\alpha(m_Z)&=1/127.925\,.
\end{align}

\head{Singular Cross Section Formula}

\noindent
To simplify the numerical evaluation of the singular part of the differential
cross section given in \eq{singular} we take $\mu=\mu_J$ so that
$U_J^\tau(s-s',\mu_J,\mu_J)=\delta(s-s')$ and express the result in the
following form
\begin{align} \label{eq:dswithP}
&\int\!\! {\rm d}k \frac{{\rm d}\hat\sigma_s}{{\rm d}\tau}\Big(\tau-\frac{k}{Q}\Big) 
  S_{\tau}^{\rm mod}\Big(k-2\bar\Delta(R,\mu_S)\Big) \nn\\
&= Q\, \sum_I \sigma_0^I\,
H_Q^I(Q,\mi_H)\,U_H\big(Q,\mi_H,\mi_J\big)\,\nonumber
\\&\times
\int {\rm d}k\,P\big(Q,Q\tau-k,\mi_J\big)\,\nn\\
&\times
     e^{-{2\delta(R,\mu_S)}\frac{{\rm d}}{{\rm
      d}k}}\, S_{\tau}^{\rm mod}\Big(k-2\bar{\Delta}(R,\mu_S)\Big),
\end{align}
where the perturbative corrections from the partonic soft function, jet
function, and soft evolution factor are contained in $P(Q,k,\mu_J)=\int {\rm d}s\int
{\rm d}k'\, J_\tau(s,\mu_J)U_S^\tau(k',\mu_J,\mu_S) S_\tau^{\rm
  part}(k-k'-s/Q,\mu_S)$. The integrals in $P$ can be carried out explicitly so
that it is given by a simple set of functions. The soft nonperturbative function
$S_{\tau}^{\rm mod}(k-2\bar{\Delta})$ is discussed in
Sec.~\ref{sec:model}, and in \eq{dswithP} we have integrated by parts so the
derivative in the exponential with the $\delta(R,\mu_S)$ acts on this nonperturbative
function.  $H_Q^I$, $J_{\tau}$, $S_\tau^{\rm part}$ and $\exp(-2\delta(R,\mu_S)
{\rm d}/{\rm d}k)$ (cf. \eq{deltaseries}) involve series in $\alpha_s(\mu_h)$,
$\alpha_s(\mu_J)$, and $\alpha_s(\mu_S)$ with no large logs, and in our
numerical analysis we expand the product of these series out, order-by-order in
$\alpha_s$. This expansion is crucial for $S_\tau^{\rm part}(k,\mu_S)$ and
$\exp(-2\delta(R,\mu_S) {\rm d}/{\rm d}k)$ since it is needed to allow the renormalon in the
two series to cancel.

For simplicity where possible we give ingredients in a numerical form for SU(3)
color with $n_f=5$ active flavors. The vector hard function to ${\cal
  O}(\alpha_s^3)$
is~\cite{Matsuura:1987wt,Matsuura:1988sm,Gehrmann:2005pd,Moch:2005id,Lee:2010cg,Baikov:2009bg}
\begin{align} \label{eq:Hardnumeric}
& H^v_Q(Q,\mu) \nn\\
& =
  1+\alpha_{s}(\mu_h)\Big(\!0.745808\!-\!1.27324 L_{Q}\!-\!0.848826L_{Q}^{2}\Big)
\notag\\&
+\alpha_{s}^{2}(\mu_h)\Big(2.27587- 0.0251035\, L_{Q}- 1.06592\, L_{Q}^{2}
\notag\\&
+0.735517L_{Q}^{3}+0.360253L_{Q}^{4}\Big)
\notag\\&
+\alpha_{s}^{3}(\mu_h)\Big(0.00050393 \,h_{3}+2.78092 L_{Q}-2.85654 L_{Q}^{2}
\notag\\&
-0.147051 L_{Q}^{3}+0.865045L_{Q}^{4}-0.165638 L_{Q}^{5}
\notag\\&
-0.101931\, L_{Q}^{6}\Big)\,,
\end{align}
where $L_Q=\ln\frac{\mi_h}{Q}$ and from \eq{h3} we have $h_3=8998.080$.  Our
axial-vector hard function for $b$ quarks has an extra two-loop singlet piece
from the large top-bottom mass splitting, $H_Q^{ba}=H_Q^v + H_Q^{\rm singlet}$.
$H_Q^{\rm singlet}$ was given in \eq{Hsinglet} and involves the real
function~\cite{Kniehl:1989qu}
\begin{align} \label{eq:I2fn}
&I_{2}(r_t)
 =10\,\Phi(r_t)^{2}+6\gamma(r_t)+\frac{\pi^{2}}{3}
-\frac{1}{r_t^2} \Big\{\mathrm{Cl}_{2}[2\,\Phi(r_t)]\Phi(r_t)
\nn\\&
 +\mathrm{Cl}_{3}[2\,\Phi(r_t)]-\Phi(r_t)^{2}  -\zeta(3)\Big\}
-\frac{2}{r_t}
\Big\{2\,\Phi(r_t)\,\mathrm{Cl}_{2}[4\,\Phi(r_t)]
\nn\\&
-2\mathrm{Cl}_{3}[2\Phi(r_t)]
+\mathrm{Cl}_{3}[4\Phi(r_t)]
+[4\gamma(r_t)+3]\Phi(r_t)^{2}
+\zeta(3) \Big\} 
\nn\\&
+\sqrt{\frac{1}{r_t}-1} \bigg\{  4(4h(r_t)+ \gamma(r_t))\Phi(r_t) 
+4\mathrm{Cl}_{2}[4\,\Phi(r_t)]
\nn\\&
-6\Phi(r_t)-6\mathrm{Cl}_{2}[2\Phi(r_t)]
-\frac{\mathrm{Cl}_{2}[2\Phi(r_t)]+2\gamma(r_t)\Phi(r_t)}{r_t}\bigg\}\,,
\end{align}
where $r_t=Q^2/(4m_t^2)$ and 
\begin{align} \label{eq:specialf}
\Phi(r_t) &= \arcsin(\sqrt{r_t})\,,\qquad
 \gamma(r_t) =\ln(2)+\frac{1}{2}\ln(r_t)
\,, \nn \\
\mathrm{Cl}_{2}(x) &= \mathrm{Im}[\mathrm{Li}_{2}(e^{ix})] \,,
 \qquad
 \mathrm{Cl}_{3}(x) = \mathrm{Re}[\mathrm{Li}_{3}(e^{ix})]
 \,, \nn\\
 h(r_t) &=\ln(2)+\frac{1}{2}\ln(1-r_t)\, .
\end{align}

The resummation of large logs from $\mu_H$ to $\mu_J$ is given by
$U_H(Q,\mi_H,\mi_J)$ in \eq{dswithP} which is the solution of the RGE for the
square of the SCET Wilson coefficient~\cite{Bauer:2000yr}
\begin{align} \label{appendix:U_H}
U_H(Q,\mi_H,\mi)=
e^{2 K(\Gamma_H,\gamma_H,\mi,\mi_H)}
\Bigg(\frac{\mi_H^2}{Q^2}\Bigg)^{\omega(\Gamma_H\!,\,\mi,\mi_H)},
\end{align}
and the functions $\omega$ and $K$ are given in \eqs{w}{K} below.  

Finally using results for the convolution of plus-functions from
Ref.~\cite{Ligeti:2008ac} we have the momentum space formula
\begin{align} \label{appendix:P}
&P\big(Q,k,\mi_J\big)=
  \frac{1}{\xi}\,E_S^{(\tau)}\big(\xi,\mi_J,\mi_S\big)\nonumber\\
&\times\sum_{\substack{n,m,k,l=-1\\m+n+1\geq k\\k+1\geq l}}^{\infty}
  V_{k}^{mn}\,\,J_m\Big[\as(\mi_J),\frac{\xi Q}{\mi^2_{J}}\Big]
  \,S_{n}\Big[\as(\mi_S),\frac{\xi}{\mi_S}\Big]\,
\notag\\
&\times
  V_l^k\big[-2\omega(\Gamma_S,\mi_J,\mi_S)\big]\,\mathcal
  L_l^{-2\omega(\Gamma_S,\mi_J,\mi_S)}\Big(\frac{k}{\xi}\Big) \,.
\end{align}
This result is independent of the dummy variable $\xi$.\footnote{When convoluted
  with $S_\tau^{\rm mod}$ we evaluate the right-hand side of
  Eq.~(\ref{appendix:P}) for $\xi=Q\tau-2\bar\Delta(R,\mu_S)$ which simplifies
  the final numerical integration.} Here $E_S^{(\tau)}(\xi,\mu_J,\mu_S)$ encodes
part of the running between the jet and the soft
scale~\cite{Balzereit:1998yf,Neubert:2004dd},
\begin{align} \label{appendix:E}
&E_{S}^{(\tau)}(\xi,\mi_J,\mi_S)=
  \exp\big[2 K(\Gamma_S,\gamma_S,\mi_J,\mi_S)\big]
\\&\times
  \Big(\frac{\xi}{\mi_S}\Big)^{-2\omega(\Gamma_S,\mi_J,\mi_S)}\ 
\frac{\exp\big[2\gamma_E\, \omega(\Gamma_S,\mi_J,\mi_S) \big]}
 {\Gamma\big[1-2\,\omega(\Gamma_S,\mi_J,\mi_S)\big]} \,.
\notag
\end{align}
The sum in Eq.~(\ref{appendix:P}) contains coefficients of the momentum space
soft and jet functions. Shifting the plus-functions so that they have common
arguments gives 
\begin{align}
  J(p^-k,\mu_J) &= \frac{1}{p^-\xi} \sum_{m=-1}^\infty
  J_m\Big[\alpha_s(\mu_J),\frac{p^-\xi}{\mu_J^2}\Big] {\cal
    L}_m\Big(\frac{k}{\xi}\Big) , \nn\\
  S(k,\mu_S) &= \frac{1}{\xi} \sum_{n=-1}^\infty
  S_n\Big[\alpha_s(\mu_S),\frac{\xi}{\mu_S}\Big] {\cal
    L}_n\Big(\frac{k}{\xi}\Big) .
\end{align} 
Here the thrust soft function coefficients are
\begin{align}
S_{-1}[\alpha_s,x] 
  &= S_{-1}(\alpha_s) + \sum_{n=0}^\infty S_n(\alpha_s)
   \frac{\ln^{n+1} x}{n+1} \,,\nn \\
S_n[\alpha_s,x] 
 &= \sum_{k=0}^\infty \frac{(n+k)!}{n!\, k!} S_{n+k}(\alpha_s) \ln^k x
 \,.
\end{align}
The soft function is known to ${\cal O}(\alpha_s^3)$ except for the constant
$s_3$ term~\cite{Schwartz:2007ib,Fleming:2007qr,Becher:2008cf,Hoang:2008fs}
\begin{align} \label{eq:Sncoeff}
S_{-1}(\alpha_s) &= 1 + 0.349066 \alpha _s +
 (1.26859 + 0.0126651\, s_2) \alpha _s^2 
  \notag\\
&+ \left (
  1.54284 + 0.00442097 \, s_2 
 + 0.00100786\, s_3 \right) \alpha_s^3 
  \nn \,,\\
S_0(\alpha_s) &= 2.07321 \alpha _s^2 + 
(4.80020- 0.0309077\, s_2) \alpha _s^3 \nn\,,\\
S_1(\alpha_s) &= -1.69765\, \alpha_s - 6.26659\, \alpha_s^2
  \notag\\
 &
 -(16.4676+0.021501\, s_2)\, \alpha _s^3 \,,\nn \\
S_2(\alpha_s) &= 1.03573\, \alpha_s^2 - 0.567799\, \alpha_s^3 \,,\nn\\
S_3(\alpha_s) &= 1.44101\, \alpha_s^2 + 9.29297\, \alpha_s^3 \,,\nn\\
S_4(\alpha_s) &= -1.46525\, \alpha_s^3 \,,\nn\\
S_5(\alpha_s) &= -0.611585\, \alpha_s^3 \,.
\end{align}
Note that $s_2$ and $s_3$ are the ${\cal O}(\alpha_s^{2,3})$ coefficients of the
non-logarithmic terms in the series expansion of the logarithm of the position
space thrust soft function. The coefficients appearing in the shifted
thrust jet function are
\begin{align}
 J_{-1}[\as,x]=&
 J_{-1}(\as)+\sum_{n=0}^{\infty}\,J_{n}(\as)\,\frac{\ln^{n+1}x}{n+1} 
 \,, \nn\\
J_{n}[\as,x]=& \sum_{k=0}^{\infty}\,\frac{(n+k)!}{n!\, k!}
 \,J_{n+k}(\as)\,\ln^{k}x\,,
\end{align}
and are known up to ${\cal O}(\alpha_s^3)$ except for the constant $j_3$
term~\cite{Lunghi:2002ju,
Bauer:2003pi,Bosch:2004th,
Becher:2006qw,Moch:2004pa,Becher:2008cf}
\begin{align} \label{eq:Jncoeff}
J_{-1}(\as) &= 1 - 0.608949 \as - 2.26795 \as^2
  \notag\\&\qquad + (2.21087 + 0.00100786\, j_3)\,\as^3  \,,\nn \\
J_{0}(\as) &= -0.63662 \as + 3.00401 \as^2
  \notag\\&\qquad + 4.45566 \as^3 \,,\nn\\
J_{1}(\as) &= 0.848826 \as - 0.441765 \as^2 - 11.905 \as^3\,,\nn \\
J_{2}(\as) &= -1.0695 \as^2 + 5.36297 \as^3\,,\nn\\
J_{3}(\as) &= 0.360253 \as^2 + 0.169497 \as^3\,,\nn\\
J_{4}(\as) &= -0.469837 \as^3\,,\nn \\
J_{5}(\as) &= 0.0764481\as^3.
\end{align} 
The $\cL$ distributions are defined as [$n\ge 0$]
\begin{equation} \label{eq:cLna_def}
\cL_n^a(x) = \biggl[\frac{\theta(x)\ln^n x}{x^{1-a}}\biggr]_+
  = \frac{\df^n}{\df a^n}\, \cL^{a}(x)
\,,\end{equation}
${\cal L}_{-1}^a(x) = {\cal L}_{-1}(x) = \delta(x)$, and for $a> -1$
\begin{equation}
\label{eq:cLa_def}
\cL^a(x) = \biggl[\frac{\theta(x)}{x^{1-a}} \biggr]_+
= \lim_{\e\to 0}\, \frac{\df}{\df x}
\biggl[ \theta(x - \e)\, \frac{x^a - 1}{a} \biggr] \,.
\end{equation}
\begin{widetext}
In Eq.~(\ref{appendix:P}) we use the coefficients~\cite{Ligeti:2008ac} 
\begin{align} \label{eq:Vkna_def}
\V_k^n(a) &= \begin{cases}
   \displaystyle a\, \frac{\df^n}{\df b^n}\,\frac{\V(a,b)}{a+b}\bigg\vert_{b = 0}\,,
  &   k=-1\,, \\[10pt]
   \displaystyle  a\, \binom{n}{k}   \frac{\df^{n-k}}{\df b^{n-k}}\, \V(a,b)
   \bigg\vert_{b = 0} + \delta_{kn} \,, \quad
  & 0\le k\le n \,,  \\[10pt]
  \displaystyle  \frac{a}{n+1} \,,
  & k=n+1  \,,
\end{cases}
\end{align}
and the coefficients
\begin{align} \label{eq:Vkmn_def}
\V_k^{mn} &= \begin{cases}
 \displaystyle \frac{\df^m}{\df a^m}\, \frac{\df^n}{\df b^n}\,\frac{\V(a,b)}{a+b}\bigg\vert_{a = b = 0} \,,
   & k=-1\,, \\[10pt]
 \displaystyle  \sum_{p=0}^m\sum_{q=0}^n\delta_{p+q,k}\,\binom{m}{p} \binom{n}{q}
\frac{\df^{m-p}}{\df a^{m-p}}\, \frac{\df^{n-q}}{\df b^{n-q}} \ \V(a,b)
  \bigg\vert_{a = b = 0}\,, \quad
   & 0\le k \le m+n \,,\\[15pt]
 \displaystyle  \frac{1}{m+1} + \frac{1}{n+1}\,,
   & k=m+n+1 \,,
\end{cases}
\end{align}
where 
\begin{equation}
\V(a,b) = \frac{\Gamma(a)\,\Gamma(b)}{\Gamma(a+b)} - \frac{1}{a} - \frac{1}{b}
\,.
\end{equation}
Special cases not covered by the general formulae in \eqs{Vkna_def}{Vkmn_def} include
\begin{align}
&\V_{-1}^{-1}(a) = 1
\,,
&\V_0^{-1}(a) &= a
\,,
&\V_{k \geq 1}^{-1}(a) &= 0
\,,\quad
&\V^{-1,n}_k &= \V^{n,-1}_k = \delta_{nk}
\,.\end{align}

\head{Evolution factors and Anomalous Dimensions}

The evolution factors appearing in Eqs.~(\ref{appendix:U_H}),
(\ref{appendix:P}), and (\ref{appendix:E}) are
\begin{align} \label{eq:w}
\omega(\Gamma,\mi,\mi_0) &=2\int_{\alpha_s(\mu_0)}^{\alpha_s(\mu)}\frac{{\rm
    d}\,\alpha}{\beta(\alpha)}\,\Gamma(\alpha) \notag\\
 &=  -\frac{\Gamma_0}{\beta_0}\bigg\{\ln r
    +\frac{\alpha_s(\mu_0)}{4\pi}\Big(\frac{\Gamma_1}{\Gamma_0}
    -\frac{\beta_1}{\beta_0}\Big)(r-1) 
    +\frac{1}{2}
    \frac{\alpha_s^2(\mu_0)}{(4\pi)^2}\Big(\frac{\beta_1^2}{\beta_0^2}
    -\frac{\beta_2}{\beta_0}+\frac{\Gamma_2}{\Gamma_0}
    -\frac{\Gamma_1\beta_1}{\Gamma_0\beta_0}\Big)(r^2-1) \notag\\& +\frac{1}{3}
    \frac{\alpha_s^3(\mu_0)}{(4\pi)^3}
    \Big[\frac{\Gamma_3}{\Gamma_0}-\frac{\beta_3}{\beta_0}
    +\frac{\Gamma_1}{\Gamma_0}\Big(\frac{\beta_1^2}{\beta_0^2}-\frac{\beta_2}{\beta_0}\Big)
      -\frac{\beta_1}{\beta_0}\Big(\frac{\beta_1^2}{\beta_0^2}-
2\,\frac{\beta_2}{\beta_0}+\frac{\Gamma_2}{\Gamma_0}\Big) \Big](r^3-1)\bigg\},
\end{align}
and 
\begin{align} \label{eq:K}
&K(\Gamma,\gamma,\mi,\mi_0)-\omega\Big(\frac{\gamma}{2},\mi,\mi_0\Big)
= 2\int_{\alpha_s(\mu_0)}^{\alpha_s(\mu)}
  \frac{{\rm d}\,\alpha}{\beta(\alpha)}\,\Gamma(\alpha)
   \int_{\alpha_s(\mu_0)}^{\alpha}\frac{{\rm d}\alpha'}{\beta(\alpha')}
\notag\\[3pt]
&\ =\frac{\Gamma_0}{2\beta_0^2}\Bigg\{\frac{4\pi}{\alpha_s(\mu_0)}
  \Big(\ln r+\frac{1}{r}-1\Big) +
  \Big(\frac{\Gamma_1}{\Gamma_0}-\frac{\beta_1}{\beta_0}\Big)(r-1-\ln r)
  -\frac{\beta_1}{2\beta_0}\ln^2 r
 +\frac{\alpha_s(\mu_0)}{4\pi}\bigg[
 \Big(\frac{\Gamma_1\beta_1}{\Gamma_0\beta_0}-\frac{\beta_1^2}{\beta_0^2}\Big)
  (r-1-r \ln r)
\nn\\[3pt]
&\ \ -B_2 \ln r
    +\Big( \frac{\Gamma_2}{\Gamma_0}-\frac{\Gamma_1\beta_1}{\Gamma_0\beta_0} +
  B_2 \Big)\frac{(r^2\!-\!1)}{2}
  +\Big(\frac{\Gamma_2}{\Gamma_0}-\frac{\Gamma_1\beta_1}{\Gamma_0\beta_0}\Big) 
  (1\!-\!r)\bigg]
  +\frac{\alpha_s^2(\mu_0)}{(4\pi)^2}
  \bigg[ \Big[\Big(\frac{\Gamma_1}{\Gamma_0}-\frac{\beta_1}{\beta_0} \Big)B_2
  +\frac{B_3}{2}\Big]\frac{(r^2\!-\!1)}{2}
\nn\\
&  \ \ 
+ \Big(\frac{\Gamma_3}{\Gamma_0} -
  \frac{\Gamma_2\beta_1}{\Gamma_0\beta_0}
  +\frac{B_2\Gamma_1}{\Gamma_0}+B_3\Big) \Big(\frac{r^3-1}{3}-\frac{r^2-1}{2}\Big)
 -\frac{\beta_1}{2\beta_0}
  \Big(\frac{\Gamma_2}{\Gamma_0}-\frac{\Gamma_1\beta_1}{\Gamma_0\beta_0}+B_2\Big)
    \Big(r^2\ln r-\frac{r^2-1}{2}\Big) -\frac{B_3}{2}\ln r
\notag\\
  &\ \ 
  -B_2\Big(\frac{\Gamma_1}{\Gamma_0}-\frac{\beta_1}{\beta_0} \Big)(r-1)
   \bigg]
 \Bigg\},
\end{align}
where $r=\alpha_s(\mu)/\alpha_s(\mu_0)$ depends on 4-loop running couplings, and
the coefficients are $B_2=\beta_1^2/\beta_0^2-\beta_2/\beta_0$ and $B_3=-\beta_1^3/\beta_0^3+2\beta_1\beta_2/\beta_0^2-\beta_3/\beta_0$.
\end{widetext}
These results are expressed in terms of series expansion coefficients of
the QCD $\beta$ function $\beta[\alpha_s]$, of $\Gamma[\alpha_s]$ which is given
by a constant of proportionality times the QCD cusp anomalous dimension, and of
a non-cusp anomalous dimension $\gamma[\alpha_s]$,
\begin{align}
 &\beta(\as)=-2\,\as\,\sum_{n=0}^{\infty}\beta_n\Big(\frac{\as}{4\pi}\Big)^{n+1}\!\!,\;\\
 &\Gamma(\as)=\sum_{n=0}^{\infty}\Gamma_n\Big(\frac{\as}{4\pi}\Big)^{n+1}\!\!,\;
 \gamma(\as)=\sum_{n=0}^{\infty}\gamma_n\Big(\frac{\as}{4\pi}\Big)^{n+1}\!\!.\notag
\end{align}
The coefficients for $n_f=5$ are~\cite{Tarasov:1980au, Larin:1993tp,
  vanRitbergen:1997va, Korchemsky:1987wg, Moch:2004pa,Czakon:2004bu}
\begin{align}\label{eq:betaCusp}
\beta_0 &= 23/3\,,\quad
 \beta_1 = 116/3\,,\quad
\beta_2 = 180.907,
\\
 \beta_3 &= 4826.16,
\nn \\
\Gamma^{\rm{cusp}}_0 &= 16/3,\quad
 \Gamma^{\rm{cusp}}_1 = 36.8436,\quad
 \Gamma^{\rm{cusp}}_2 = 239.208
 \,.
\nn
\end{align}
For the unknown four-loop cusp anomalous dimension we use the Pad\`e
approximation assigning 200\% uncertainty:
\begin{align}%08
  \Gamma_{3}^{\rm cusp} = (1 \pm 2) \frac{(\Gamma _2^{\rm
      cusp}){}^2}{\Gamma_{1}^{\rm cusp}}.
\end{align}
The anomalous dimensions for the hard, jet, and soft functions
are~\cite{Catani:1992ua,Vogt:2000ci,Moch:2004pa,Neubert:2004dd,Moch:2005id,Idilbi:2006dg,Becher:2006mr}
\begin{align}
\Gamma^H_n &= -\,\Gamma^{\rm{cusp}}_n,\quad
\Gamma^J_n = 2\,\Gamma^{\rm{cusp}}_n,\quad
\Gamma^S_n = -\,\Gamma^{\rm{cusp}}_n,
\nn \\[4pt]
\gamma^H_0 &= -\,8,\quad
\gamma^H_1 = 1.14194,\quad
\gamma^H_2 = -\,249.388,
\nn \\[4pt]
\gamma^J_0 &= 8,\quad
\gamma^J_1 = -\,77.3527,\quad
\gamma^J_2 = -\,409.631,
\nn \\
\gamma^S_n &= -\,\gamma^H_n-\gamma^J_n.
\end{align}

To determine the strong coupling $\alpha_s(\mu)$ in terms of $\alpha_s(m_Z)$ at
$4$-loops with 5 light flavors we use
\begin{align} \label{alphas}
  \frac{1}{\alpha_s(\mu)} &= \frac{X}{\alpha_s(m_Z)} + 0.401347248\, \ln X 
  \\
 &+ \frac{\alpha_s(m_Z)}{X} \big[ 0.01165228\, (1-X) + 0.16107961\,
 {\ln X}\big]
  \nn\\
 &+ \frac{\alpha_s^2(m_Z)}{X^2}
  \big[ 0.1586117\, (X^2-1) +0.0599722\,(X
  \nn\\
 &+\ln X-X^2)
  +0.0323244\, \{(1-X)^2-\ln^2 X\} \big] ,
 \nn
\end{align}
where $X=1+\alpha_s(m_Z) \ln(\mu/m_Z) \beta_0/(2\pi)$ and the displayed numbers
are determined from the $\beta_i$ in \eq{betaCusp}. The form in
Eq.~(\ref{alphas}) agrees very well with the numerical solution of the
beta function equation.

\begin{widetext}
\head{Nonsingular Cross Section Formula }

\noindent
At ${\cal O}(\alpha_s^2)$ there is an axial singlet contribution to the nonsingular
terms through the three-parton cut of Fig.~\ref{fig:Haxial}, which is given by
the function $f_{\rm singlet}$ appearing in Eq.~(\ref{eq:nonsingular}) for
$f_{\rm qcd}^{ba}$. The result for this function can be extracted from results
in Ref.~\cite{Hagiwara:1990dx} and reads: [$r_t=Q^2/(4m_t^2)$]
\begin{align}  \label{eq:three-partons-cut}
&f_{\rm singlet}\big(\tau,r_t)=-\,\frac{64}{3}\,\theta\left( \frac{1}{3}-\tau\right) \left[\int_{1+\tau}^{2(1-\tau)}\mathrm{d}y\, y\, g(y-1,r_t)+(1-3\,\tau)\,\dfrac{(1+\tau)}{2}\, g(\tau,r_t)\right],\\
&g(\tau,r_t)=\frac{2\,r_t\left[\sqrt{\frac{1-r_t\,\tau}{r_t\,\tau}}\sin^{-1}\left(\sqrt{r_t\,\tau}\right)-\sqrt{\frac{1-r_t}{r_t}}\sin^{-1}\!\left(\sqrt{r_t}\right)\right]
+ \big[\sin^{-1}\!\left(\sqrt{r_t\,\tau}\right)\big]^{2}
-\big[\sin^{-1}\!\left(\sqrt{r_t}\right)\big]^{2}-r_t\log(\tau)}{4\,r_t\,(1-\tau)^{2}}\,.\nn
\end{align}
\end{widetext}

\head{R-evolution}

\noindent
Finally we display here the function $D^{(k)}$~\cite{Hoang:2008yj}
which appears in the solution in Eq.~(\ref{eq:DeltaSoln}) of the R-RGE equation
for ${\bar \Delta}(R,R)$:
\begin{align} \label{eq:Dk}
D^{(k)}(\alpha_1,\alpha_2)&=e^{i\pi{\hat b}_1}\sum_{j=0}^{k}(-1)^j S_j\times\nonumber\\
&[\Gamma(-{\hat b}_1-j,t_1)-\Gamma(-{\hat b}_1-j,t_2)]\,,
\end{align}
which is real since the complex phase $e^{i\pi{\hat b}_1}$ cancels the imaginary
part coming from the incomplete Gamma functions, defined as
\begin{equation}
\Gamma(c,t)=\int_{t}^{\infty}\:\mathrm{d}x\, x^{c-1}e^{-x}\,.
\end{equation}
Here $k$ is the order of the
matrix elements (that is $k=0$ for NLL$^\prime$ and NNLL, $k=1$ for
NNLL$^\prime$ and $\ntll$, and $k=2$ for $\ntllp$. For lower orders
$D^{(k)}=0$). In \eq{Dk} we have defined
\begin{align}
t_i&=-\dfrac{2\pi}{\beta_{0}\alpha_i}\,,\quad{\hat b}_1=\dfrac{\beta_1}{2\beta_0^2}\,,\quad S_0=0\,,\quad S_1=\dfrac{\gamma_1^R}{(2\beta_0)^2}\,,\nonumber\\
S_2&=\dfrac{\gamma_2^R}{(2\beta_0)^3}-
\dfrac{2\beta_0^2\beta_1+\beta_1^2-\beta_0 \beta_2}{16\beta_0^6}\,\gamma_1^R\,,
\end{align}
where the R-anomalous dimensions $\gamma_i^R$ were given in \eq{gammaR}.

\head{Total Hadronic Cross Section}

\noindent
The total hadronic QCD cross section, can be evaluated in fixed-order
perturbation theory with $\mu\simeq Q$, and was given in \eq{sigtot} with the
vector QCD results given in \eq{Rhad}. The function appearing in the singlet
contribution in \eq{sigtot} at ${\cal O}(\alpha_s^2)$ is~\cite{Kniehl:1989qu}
\begin{align}
&I(r_t)=\frac{-\,\Phi(r_t)\,\mathrm{Cl}_{2}[2\,\Phi(r_t)]-\,\mathrm{Cl}_{3}[2\,\Phi(r_t)]+\zeta_3}{r_t^{2}}\nonumber\\
&\qquad\quad+\sqrt{\frac{1}{r_t}-1}\,\bigg(\frac{2\,[\,1-\gamma(r_t)\,]\Phi(r_t)-{\rm Cl}_2[2\,\Phi(r_t)]}{r_t}\nonumber\\
&\qquad\quad+2\,{\rm Cl}_2[2\,\Phi(r_t)]+2\,[\,2\,\gamma(r_t)-3\,]\Phi(r_t)\bigg)\nonumber\\
&\qquad\quad+6\,\gamma(r_t)+2\,\Phi(r_t)^{2}-\frac{4\,\Phi(r_t)^{2}+1}{r_t}-\frac{\pi^{2}}{3}\,,
\end{align}
where the necessary functions appear in \eq{specialf}.  Note that we have
dropped the four particle cut contribution $I_4=\pi^2/3-15/4$ since we have not
accounted for it in the ${\cal O}(\alpha_s^2)$ nonsingular distribution.

\section{Soft Function OPE Matching}
\label{app:SoftOPE}

To derive \eq{Sope} we must demonstrate uniqueness of the power correction
$\Omega_1$ and derive its perturbative Wilson coefficient to all orders in
$\alpha_s$. We carry out these two parts of the proof in turn.

Since the operator appearing in the matrix element $\Omega_1$ is non-local, the
proof of uniqueness is more involved than for a typical OPE where we could just
enumerate all local operators of the appropriate dimension. Here we are
integrating out perturbative soft gluons in $S_{\tau}(k,\mu)$, while retaining
nonperturbative soft gluons. The hierarchy between these soft gluons is in their
invariant masses, $k^2\gg\Lambda_{\rm QCD}^2$. This process can not introduce
Wilson lines in new light-like directions, nor additional Wilson lines following
paths in $n$ and $\bar n$. Thus the Wilson lines will be the same as those in
the full theory operator, \eq{Sdef}. Additional Wilson lines could only be
induced by integrating out collinear or hard gluons, which would yield power
corrections suppressed by the hard or jet scales.  The second point to
demonstrate is that dimension one combinations of derivatives other than
$i\widehat\partial$ do not lead to new nonperturbative matrix elements at this
order. The key is that for derivative operators inside our vacuum matrix element
involving Wilson lines, boost invariance along the thrust axis relates all
matrix elements to $\Omega_1$~\cite{Lee:2006fn}.  The proof relies on boost
invariance along the thrust axis of derivative operators inside the vacuum
matrix element.  To see this one defines the transverse energy flow operator
${\cal E}_T(\eta)$ by its action on
states~\cite{Korchemsky:1999kt,Belitsky:2001ij}
\begin{align}
  {\cal E}_T(\eta) | X \rangle 
  = \sum_{i\in X} \,| \vec k_i^\perp | \,\delta(\eta-\eta_i) \,|X\rangle \,.
\end{align}
Any dimension one derivative operator we might wish to consider, such as
$n\cdot\partial$, $\bar n\cdot \partial$, $\partial_t$, $\partial_z$, $\ldots$,
or combinations thereof, are given by an integral $\int\df\eta\: h(\eta)\: {\cal
  E}_T(\eta)$ for an appropriate rapidity function $h(\eta)$. For example, for
the thrust derivative $i\widehat\partial$ we have $h(\eta) = e^{-|\eta|}$. Boost
invariance implies~\cite{Lee:2006fn}
\begin{align}
 &\bar\Omega_1 = \frac1{2N_c} \big\langle 0 \big| {\rm tr}\ \overline Y_{\bar
    n}^T(0) Y_n(0)\, i\widehat\partial\, Y_n^\dagger(0) \overline Y_{\bar
    n}^*(0) \big| 0 \big\rangle
   \nn\\
 &= \int\!\!\df\eta\, \frac{h(\eta)}{2N_c}\langle 0 \big| {\rm tr}\ \overline Y_{\bar
    n}^T(0) Y_n(0)\, {\cal E}_T(\eta\!+\!\eta') Y_n^\dagger(0) \overline Y_{\bar
    n}^*(0) \big| 0 \big\rangle
   \nn\\
 &= \frac{1}{N_c} \langle 0 \big| {\rm tr}\ \overline Y_{\bar
    n}^T(0) Y_n(0)\, {\cal E}_T(\eta') Y_n^\dagger(0) \overline Y_{\bar
    n}^*(0) \big| 0 \big\rangle,
\end{align}
for arbitrary $\eta'$. The same steps hold for any other derivative operator and
function $h(\eta)$, and different choices only affects the constant calculable
prefactor. This suffices to show the second point.

To derive an all orders expression for the Wilson coefficient of $\Omega_1$ we
construct an analog of the OPE matching done for the soft function in $B\to
X_s\gamma$~\cite{Bauer:2003pi}. The proof is considerably simpler for $B\to
X_s\gamma$ because the OPE in that case yields local HQET operators.
Nevertheless the thrust soft function can be manipulated such that a similar
strategy can be used.  Using the thrust axis we define hemisphere $a$ where $p^+
< p^-$ and hemisphere $b$ where $p^- < p^+$. Consider the soft function written
as a matrix element squared
\begin{align} \label{eq:SwithX}
  S_\tau(k,\mu) &= \frac{1}{N_c} 
  \sum_X \delta(k-\!k_s^{a+}\!-\!k_s^{b-} )\, {\rm tr}\:
  \langle 0 | \overline Y_{\bar n}^T(0) Y_n(0)  | X \rangle \nn\\
 & \times
  \langle X | Y_n^\dagger(0) \overline Y_{\bar n}^*(0)   | 0 \rangle 
  \,,
\end{align}
where the trace is over color, $k_s^{a+}=n\cdot p_X^a$ is the total
plus-momentum of the particles in state $X$ in hemisphere $a$ and $k_s^{b-}=\bar
n\cdot p_X^b$ is the minus-momenta of particles in $X$ in hemisphere $b$. To
carry out the OPE we need to consider a state that has overlap with the operator
in \eq{O1bar}. Thus we could replace the vacuum by very soft nonperturbative
gluons with momenta of ${\cal O}(\Lambda_{\rm QCD})$ and then consider matrix
elements with perturbative gluons having momenta $\sim k \gg \Lambda_{\rm QCD}$.
Since the OPE is independent of the particular states we choose, we will instead
consider a simpler alternative in the following.

First we write the matrix element in \eq{SwithX} as
\begin{align}
  & {\rm tr}\:
  \big\langle 0 \big| \overline Y_{\bar n}^T Y_n \big| X \big\rangle
  \big\langle X \big| Y_n^\dagger \overline Y_{\bar n}^*
  \big| 0 \big\rangle  
  \\
 & = \big\langle 0 \big|
  \bar\zeta_{\bar n} \overline Y_{\bar n}^T Y_n \zeta_n 
  \big| X u_n v_{\bar n} \big\rangle
  \big\langle X u_n  v_{\bar n} \big| 
  \bar\zeta_n Y_n^\dagger \overline Y_{\bar n}^* \zeta_{\bar n}  
 \big| 0 \big\rangle  
 \nn \,,
\end{align}
where $\zeta_n$ and $\zeta_{\bar n}$ are non-interacting collinear fields whose
contractions with the sterile quark $u_n$ and anti-quark $v_{\bar n}$ are chosen
with a normalization to reproduce the original matrix element (and a sum over
their color correctly reproduces the trace). Here $u_n$ should be thought of as
a very energetic collinear quark in hemisphere $a$ with large label momentum
$p_n^-$, and zero residual momentum. The large momentum is conserved by soft
interactions from the Wilson lines due to the SCET multipole expansion. Here the
plus-momentum of $u_n$ is included into $k_s^{a+}$, but is zero and does not
contribute to the $\delta$-function. The same is true for $v_{\bar n}$ which has
zero minus-momentum, large label $p_{\bar n}^+$ momentum, and is always in
hemisphere $b$. We introduced $u_n$ and $v_{\bar n}$ so that we can use them to
systematically add a very soft momentum to the end of the Wilson lines (at
$\infty$). They provide a convenient state with which to carry out the OPE,
because there is nonzero overlap taking only the $1$ out of the Wilson lines,
$Y$.  In particular they allow us to perform the OPE and pick out the
$i\widehat\partial$ present in $\bar \Omega_1$ at tree level, without the
necessity to add explicit soft gluons with momenta $\ll k$.

To carry out the OPE we now give $u_n$ a very small soft momentum $\ell^+$ and
$v_{\bar n}$ a very small soft momentum $\ell^-$, and denote them by $u_n^\ell$
and $v_{\bar n}^\ell$ respectively. These particles are kept on-shell by
adjusting their large label $\perp$-momenta so that $\ell^+ = p_{n\perp}^2 /
p_n^-$ and $\ell^- = p_{\bar n\perp}^2/p_{\bar n}^+$. Due to the multipole
expansion these $\perp$-momenta have no influence on diagrams with perturbative
soft gluons having momenta $k \ll p_{n\perp} = -p_{\bar n \perp}$. The Wilson
line propagators reduce to the same as before, such as
\begin{align}
  \frac{p_n^-}{k^+ p_n^- + \ell^+ p_n^- + p_{n\perp}^2} 
  = \frac{1}{k^+}  \,.
\end{align}
This property is familiar in SCET where soft couplings to energetic collinear
quarks in SCET remain eikonal for any values of the quark's large momenta by
using the equations of motion, as long as the final particles are on-shell. Thus
at any order in perturbation theory, with any number of soft gluons and soft
quarks of momenta $\sim k$ in the matrix elements, the only change caused by
$\ell^\pm$ is on the $\delta(k-k_s^{a+}-k_s^{b-})$ in \eq{SwithX} which is
shifted to $\delta(k-\ell -k_s^{a+} - k_s^{b-})$, where $\ell\equiv \ell^+ +
\ell^-$.  Expanding with $\ell\ll k$ the matrix element with this choice of
state evaluates to
\begin{align} \label{eq:SopeE}
  S_{\tau}^{\rm part}(k-\ell,\mu) = S_{\tau}^{\rm part}(k) -  \frac{{\rm d}S_{\tau}^{\rm
    part}(k)}{{\rm d}k} \ \ell + \ldots 
  \,.
\end{align}
At lowest order in emission of very soft gluons $\sim \Lambda_{\rm QCD}$ the
corresponding matrix element in the lower energy theory is
\begin{align}
 & \frac{1}{N_c} \langle 0 | \bar\zeta_{\bar n} \overline Y_{\bar n}^T Y_n \zeta_n 
  i \widehat\partial \big| u_n^\ell v_{\bar n}^\ell \big\rangle
  \big\langle u_n^\ell  v_{\bar n}^\ell \big| 
  \bar\zeta_n Y_n^\dagger \overline Y_{\bar n}^* \zeta_{\bar n}  
 \big| 0 \big\rangle 
  =  \ell \,.
\end{align}
Virtual radiative corrections do not correct this result since they are
scaleless and vanish in pure dimensional regularization.  Thus we can identify $\ell
\to 2\bar\Omega_1$ in \eq{SopeE}, and this then yields the stated result for the
OPE in \eq{Sope}.

\section{Operator Expansion for the First Thrust Moment}
\label{app:MomentOPE}

\begin{figure}[t]
      \vspace{0pt}
      \includegraphics[width=0.95\linewidth]{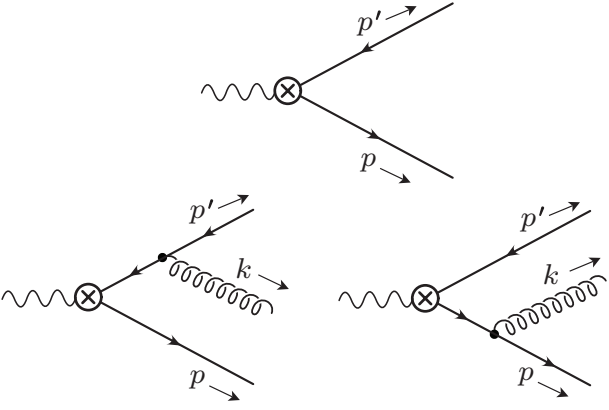}
      \caption{Amplitudes for zero and one soft gluon. }
      \label{fig:Tree_qqg}
\end{figure}

For moment integrals of the thrust distribution over $\tau\in [0, 1/2]$ there is
not a hierarchy of scales that induces large logs, and one may formulate the
theoretical result in terms of an expansion in $\alpha_s$ and $\Lambda_{\rm
  QCD}/Q$. The zero'th moment of thrust is just the total cross section for
$e^+e^- \to \text{hadrons}$, and the power corrections are formulated in terms
of the well known OPE~\cite{Shifman:1978bx}. For higher moments the fact that
thrust constrains a non-trivial combination of final state momenta makes
carrying out an OPE more difficult. For example, when we weigh the integral by
a power of thrust it is not possible to collapse all propagators to a point, so
the nonperturbative parameters are no longer constrained to be given by a basis
of local operators.  In the effective coupling model~\cite{Dokshitzer:1995qm}
the same nonperturbative parameter $\alpha_0$ that appears for the thrust
distribution, also occurs in the first moment. However it is not clear to what
level of accuracy this carries over to a field theoretical description of power
corrections derived from QCD. In this appendix we show how one can carry out an
OPE for the 1st moment of the thrust distribution, and demonstrate that at
leading order it only involves the same nonperturbative matrix element
$\Omega_1$ from \eq{O1def}.

To carry out an OPE for the thrust moment we can work order by order in the hard
$\alpha_s(Q)$ expansion, and analyze direct computations where we couple soft
nonperturbative gluons to hard partons in Feynman diagrams. The appropriate
non-local operator(s) appearing in the expansion will be identified by the
structure of the amplitudes in this computation. In the following discussion the
soft gluons will {\em not} be treated as final state particles for which there
is a phase space integral, but rather as a means of probing the structure of the
nonperturbative operator.  The lowest order graphs with zero or one soft gluon
and a virtual photon current (for simplicity) are shown in \fig{Tree_qqg}. Here
$k^\mu\sim \Lambda_{\rm QCD}$ is soft, and $p^\mu\sim Q$, $p^{\prime\mu}\sim Q$
are hard momenta. To carry out the OPE we calculate and square the on-shell
amplitude, ${\cal M}_h^{\mu} {\cal M}_h^{\nu}$, where $\mu$, $\nu$ are the
virtual photon current indices. We sum over the final quark/antiquark spins
since these particles are hard and are being integrated out. On the other hand
the gluon vector indices $\alpha$, $\alpha'$ are left uncontracted and are used
to help in identifying the operator for the nonperturbative matrix element. For
simplicity, the indices $\alpha$ and $\alpha'$ are suppressed in writing down
the amplitudes below. We start out without making restrictions on the number of
gluons coming from ${\cal M}_h^{\mu}$ and ${\cal M}_h^{\nu*}$, which corresponds
to directly matching onto the nonperturbative operator, without considering the
final vacuum matrix element which gives a nonperturbative parameter.  Since
$p^2=p^{\prime 2}=k^2=0$ the denominators of the propagators in the one gluon
graphs reduce to $2 p\cdot k$ and $2p'\cdot k$.  In the numerators we can drop
$\slash\!\!\!k$'s relative to the large $\slash\!\!\!  p$ and $\slash\!\!\!
p'$. The interference between the zero and one gluon amplitudes gives
\begin{align} \label{eq:As1}
 & {\cal M}_h^{\mu} {\cal M}_h^{\nu*}
  =  N_c {\rm tr}[ \slash\!\!\! p \gamma^\mu \slash\!\!\! p^\prime \gamma^\nu]
 \ \frac{2 g T^A}{N_c} \bigg[ 
   \frac{p^{\prime\alpha} }{p^\prime\cdot k} -\frac{p^\alpha}{p\cdot k} \bigg].
\end{align}
The interference with one gluon from each of ${\cal M}_h^{\mu}$ and ${\cal
  M}_h^{\nu*}$ is
\begin{align} \label{eq:As2}
 & {\cal M}_h^{\mu} {\cal M}_h^{\nu*}
  =  N_c \, {\rm tr}[ \slash\!\!\! p \gamma^\mu \slash\!\!\! p^\prime \gamma^\nu]
  \\
 & \times \frac{g^2 T^A T^B}{N_c} \bigg[ \frac{p^\alpha p^{\alpha'}}{(p\cdot k)^2} 
  + \frac{p^{\prime\alpha} p^{\prime\alpha'}}{(p^\prime\cdot k)^2} 
  - \frac{(p^\alpha p^{\prime\alpha'}+p^{\prime\alpha} p^{\alpha'})}
     {(p\cdot k)(p^\prime \cdot k)} \bigg].
  \nn
\end{align}
Continuing in this fashion with any number of gluons from ${\cal M}_h^{\mu}$ and
any number from ${\cal M}_h^{\nu*}$ we always find the tree level amplitude
squared with no soft gluons, \mbox{$N_c\, {\rm tr}[ \slash\!\!\! p \gamma^\mu
  \slash\!\!\! p^\prime \gamma^\nu]$}, times an amplitude from the soft gluons.

Since the hard quarks are on-shell and back-to-back their 
four-momenta are given by light-like vectors along the thrust axis, 
\begin{align} \label{eq:ppp}
 p^\mu &= n^\mu \frac{\bar n \cdot p}{2} \,, 
 &p^{\prime\mu} &= {\bar n}^\mu \frac{n\cdot p}{2} \,,
\end{align}
up to power corrections beyond those considered here.  Here $n^\mu=(1,\hat {\bf
  t})$ and $\bar n^\mu =(1,-\hat {\bf t})$ are identical to the $n$ and $\bar n$
appearing in \eq{O1def}. Using \eq{ppp} the soft gluon amplitudes in
\eqs{As1}{As2} are eikonal with precisely the right factors to come from the
$\overline Y_{\bar n}^T(0)$, $Y_n(0)$, $Y_n^\dagger(0)$, $\overline Y_{\bar
  n}^*(0)$ in the $\Omega_1$ matrix element in \eq{O1def}.

For the first moment observable we can focus on amplitudes that have the same
number of gluons in ${\cal M}_h^{\mu}$ and ${\cal M}_h^{\nu*}$, and at least one
gluon for the $i\widehat\partial$ operation in \eq{O1def} to act on. Since the
gluon is soft, the factor of $\tau$ in $\int {\rm d}\tau
(\tau/\sigma)(\df\sigma/\df\tau)$ is given by
\begin{align}
 \tau &= {\rm min}\Big[\frac{2 p\cdot k}{q^2}, \frac{2 p'\cdot k}{q^2} \Big]
   =  \frac{1}{Q} {\rm min} \Big[ n\cdot k , \bar n \cdot k \Big] 
  \\
 &= \frac{1}{Q} \Big\{ n\cdot k\, \theta(\bar n\cdot k \!-\! n\cdot k) + \bar n \cdot k\,
 \theta( n\cdot k \!-\! \bar n \cdot k) \Big\}\,,
  \nn
\end{align}
and is exactly equal to $i\widehat\partial$ given in \eq{ptau} acting on the
soft gluon in \fig{Tree_qqg}. Hence in the first moment of thrust we find that
$\tau$ together with the soft gluon amplitude give precisely $2\Omega_1/Q$, with
the vacuum matrix in \eq{O1def} (where the trace comes from the sum over color
for the final state quarks).  The remaining \mbox{$N_c\, {\rm tr}[ \slash\!\!\!
  p \gamma^\mu \slash\!\!\!  p^\prime \gamma^\nu]$} amplitude goes together with
the two-body phase space to yield the tree level cross section $\sigma_0^I$.
Together these results yield \eq{thrustmoment1} for the lowest order OPE for the
first moment of thrust.

\bibliography{thrust2}

\end{document}